\documentclass[%
 reprint,
superscriptaddress,
 amsmath,amssymb,
 aps,
]{revtex4-1}
\usepackage[version=3]{mhchem} 

\usepackage{balance} 
\usepackage{rotating}
\usepackage{float}
\usepackage{url}
\usepackage{tabularx}
\usepackage{multirow}
\usepackage{amsmath}
\usepackage{multirow}
\usepackage{wrapfig}
\usepackage{bm}
\usepackage{ulem}
\newcommand{\Lower}[1]{\smash{\lower 1.5ex \hbox{#1}}}

\newcommand{\down}{\sout{$\downarrow$}}
\newcommand{\up}{\sout{$\uparrow$}}
\newcommand{\zero}{\sout{\phantom{$\downarrow \negthickspace \uparrow$}}}
\newcommand{\double}{\sout{$\downarrow \negthickspace \uparrow$}}
\newcommand*\rot{\rotatebox{90}}
\setcitestyle{super}
\begin{document}
\title[Elucidating cation--cation interactions in neptunyl dications using multireference ab initio theory]
 {Elucidating cation--cation interactions in neptunyl dications using multireference ab initio theory}
\author{Aleksandra \L{}achma\'{n}ska}
\author{Pawe{\l} Tecmer}
\affiliation
{Institute of Physics, Faculty of Physics, Astronomy and Informatics, Nicolaus Copernicus University in Torun, Grudziadzka 5, 87-100 Torun, Poland}
\author{\"{O}rs Legeza}
\affiliation
{Strongly Correlated Systems ``Lend\"{u}let'' Research Group, Wigner Research Center for Physics, H-1525 Budapest, Hungary}
\author{Katharina Boguslawski}
\email{k.boguslawski@fizyka.umk.pl}
\altaffiliation[Also at ]{Faculty of Chemistry, Nicolaus Copernicus University in Torun, Gagarina 7, 87-100 Torun }
\affiliation
{Institute of Physics, Faculty of Physics, Astronomy and Informatics, Nicolaus Copernicus University in Torun, Grudziadzka 5, 87-100 Torun, Poland}

 
 
\begin{abstract}
Understanding the binding mechanism in neptunyl clusters formed due to cation--cation interactions is of crucial importance in nuclear waste reprocessing and related areas of research.
Since experimental manipulations with such species are often rather limited, we have to rely on quantum-chemical predictions of their electronic structures and spectroscopic parameters.
In this work, we present a state-of-the-art quantum chemical study of the T-shaped and diamond-shaped neptunyl(V) and neptunyl(VI) dimers.
Specifically, we scrutinize their molecular structures, (implicit and explicit) solvation effects, the interplay of static and dynamical correlation, and the influence of spin-orbit coupling on the ground state and lowest-lying excited states for different total spin states and total charges of the neptunyl dications.
Furthermore, we use the picture of interacting orbitals (quantum entanglement and correlation analysis) to identify strongly correlated orbitals in the cation--cation complexes that should be included in complete active space calculations.
Most importantly, our study highlights the complex interplay of correlation effects and relativistic corrections in the description of the ground and lowest-lying excited states of neptunyl dications.
\end{abstract}
\maketitle
\section{Introduction}

The neptunium atom is one by-product of nuclear fission reactions of nuclear power plants that is characterized by its high chemical reactivity and toxicity.
It is anticipated that neptunium compounds are the main source of radiation during sustained storage and thus should be sheltered by materials of special design, like, polydentate organic ligands.\mbox{~\cite{sessler2001, tian2005, tian2007}}
From an environmental point of view, we need to segregate heavy-element compounds from spent nuclear fuel.\mbox{~\cite{neptunium-important-1, neptunium-important-2, neptunium-important-3}}
Since the mid-20-th century, industrial reprocessing is based on the PUREX process that was designed to {separate} uranium and plutonium compounds.\mbox{~\cite{purex01}}
All transuranic actinides and lanthanides can be segregated in the GANEX (grouped actinide extraction) process.\mbox{~\cite{ganex01}}
However, the large variety and complexity of fission products and their broad range of oxidation states require a few stages of extraction, stripping, clean up, conversion to oxide compounds, and converting waste to the solid state.
Specifically, the management of neptunium is one of the key challenges in such processes.\mbox{~\cite{ganex02}}

Current research on novel nuclear fuel processing techniques of oxidized americium-containing compounds attracted a great deal of attention on neptunium oxides as {they can form complexes similar to americium} compounds.~\cite{ccis-in-solution-4} 
Examples are the linear NpO$_2^+$ and NpO$_2^{2+}$ species containing pentavalent (+5) and hexavalent (+6) neptunium,~\cite{Cotton_book, Actinides_bible, bursten_91, krot2004, denning2007, actinoid_rev_2012, gomes2015applied} also known as the neptunyl(V) and neptunyl(VI) cations, respectively.
These compounds have been immensely studied theoretically. \cite{swart2005, fromager2005, kovacs2011, laura2013, gendron2014}
Neptunyl cations have the 5f$^1$ and 5f$^2$ open-shell electron configuration and are characterized {by an equilibrium Np--O bond length} of about 1.7--1.8 \AA. The vibrational spectra of neptunyl(V) contains two characteristic (symmetric and asymmetric) stretching frequencies in the range from 870 cm$^{-1}$ to 1030 cm$^{-1}$.
For neptunyl(VI), {the corresponding vibrational} frequencies can be found at around 970 cm$^{-1}$ and 1080 cm$^{-1}$.\cite{kovacs2011}
Studies on neptunyl ions indicate that configurations, where the valence electrons are found in the 5f$_\delta$ and 5f$_\phi$ open-shell configurations, are favoured and are characteristic for the low-lying part of their electronic spectrum. Both experiments and calculations reveal low-intensity f--f transitions.\cite{infante2006, matsika2000, neptunyl-spectra-theory-1, neptunyl-ion-spectra-exp-1, neptunyl-ion-spectra-exp-2, neptunyl-ion-spectra-exp-3}

Furthermore, neptunyl ions can form small oligomeric or polymeric clusters that are known as cation--cation interactions (CCIs).~\cite{CCI_79, tecmer-song2016, baker2012, tetrameric-cci, neptunyl-ion-spectra-exp-4} 
{This characteristic, albeit peculiar behaviour} was first observed for neptunium(V) in uranyl perchlorate solution and aqueous chlorate media.~\cite{sullivan1960, sullivan1961, sullivan1962}
{The stability of CCI complexes depends on the composition of the solution.}~\cite{CCI_79} 
{Neptunyl dimers are formed if the concentration of Np exceeds 0.2 M. Their presence causes a change in the vibrational spectrum, where two new bands appear. The absorption spectra is only weakly affected by a red shift and peak broadening.}~\cite{guillaume1981} 
{The Np--Np distance is 4.2 \r{A} indicating no direct bonding between the actinide centers.}~\cite{guillaume1983}
New examples of CCIs in solution are still being reported.~\cite{ccis-in-solution-2, ccis-in-solution-3, ccis-in-solution-4}
Specifically, such inter-cationic interactions promote the formation of crystalline structures.~\cite{Arnold2009, Jin2011, Jin2011-2, wang2011, wang2012, diwu2012, laura2013, ccis-in-solids-1, ccis-in-solids-2, ccis-in-solids-3,ic-cci-neptunyl-2018}
CCIs are a technical difficulty for nuclear waste reprocessing systems as they facilitate the stabilization of actinide complexes in spent nuclear fuel.

The CCIs structures are stable primarily because of the bonding interaction between the oxygen and the neptunium atoms of two neighbouring complexes, where the effective charge of the oxygen atoms is negative \cite{vallet2004} in contrast to the effective positive charge localised on the actinide atoms.~\cite{choppin1984}
The stability of CCIs is strongly influenced by the \mbox{Np--O} bond distance, \cite{guillaume1981} which changes in different environments.~\cite{rao1979, gainar1983, halperin1983, roesch1990}
Since experimental studies on such compounds are impeded, theoretical modelling can provide much sought-after insights into the reactivity and stability of CCIs.
However, heavy elements are immensely difficult to describe theoretically because correlation effects and relativistic effects \cite{reiher-wolf,fromager2005,infante2006,pawel1,Dolg_rev_2012,CUO_DMRG,SO_in_actinides,pawel_saldien,Tecmer2016,boguslawski2017} have to be described on equal footing, which poses a challenge for present-day quantum chemistry, especially when the molecular system under investigation contains more than one heavy element.~\cite{actinide_oxides_1, computational-difficulties-casscf, computational-difficulties-casscf-2}

In this work we use state-of-art quantum chemistry methods to model the electronic structure of various neptunium-containing clusters in different environments. Furthermore, following the procedure outlined in ref.~\citenum{boguslawski2017}, we focus on constructing stable and reliable active spaces that properly describe static/nondynamic electron correlation effects in the investigated neptunyl clusters. 
As a quantitative criterion, we use the orbital-pair mutual information to select the most important orbitals of large and moderate orbital-pair correlations. 

This work is organized as follows. {In section 2}, the computational details are presented. Section 3 discusses the structural parameters and spin-state energetics of various neptunium dioxide clusters including their entropy-based active spaces, electronic spectra, and the influence of spin-orbit {coupling and solvation}. Finally, we conclude in section 4.

\section{Computational details}

\subsection{Geometry optimization}
The structures of [NpO$_2$]$_2^{2+}$, [NpO$_2$]$_2^{3+}$, and [NpO$_2$]$_2^{4+}$ were optimized with the TURBOMOLE 7.0 software package{\mbox{~\cite{ahlrichs_89, turbomole, turbomole-manual}} using the DFT module. The def2-TZVP\mbox{~\cite{def2-tzvp_o}}} basis set was chosen for the oxygen atoms,{while a small-core (relativistic) Effective Core Potential (ECP-60) in combination with the def-TZVP basis set\mbox{~\cite{ECP_Np_1, ECP_Np_2, def-tzvp_np}} was taken for the neptunium atoms. Previous studies suggest that a small-core ECP basis provides results that agree well with all-electron calculations for actinide compounds.\mbox{~\cite{pawel_saldien}}}
The BP86 exchange--correlation functional \cite{becke88, perdew86} was applied. Our choice was motivated by the good performance of the BP86 exchange--correlation functional in geometry optimizations of actinide species~\cite{schreckenbach_dft,actinoid_rev_2012,kovacs2011}. We should note that dispersion interactions (through, for instance, the D3 correction~\cite{Grimme-D3-2010}) have not been considered in our calculations. {The bonding between two monomers includes Np--O interactions where the distance between the atoms is similar to the sum of their corresponding single-bond covalent radii and all the Np--O distances are much shorter than the sum of their van der Waals radii. Thus, dispersion forces are expected to not affect the geometry of the optimized structures. Recent findings of Rotzinger\mbox{~\cite{neptunyl-solution}} confirm that a dispersion correction does not significantly affect the molecular structure of neptunyl monomers.}

All neptunyl clusters were studied both as free molecules in gas phase and in aqueous solution.
The aqueous solution was simulated by solvation effects using the Conductor-like Screening Model (COSMO) \cite{COSMO} module. 
$C_{2h}$ point group symmetry was used for the diamond-shaped clusters and $C_{2v}$ was used for all T-shaped clusters.
{The structures of [NpO$_2$]$_2^{2+}$ and [NpO$_2$]$_2^{4+}$ optimized for the states with zero, two, and four unpaired $\alpha$-electrons, while the structure of [NpO$_2$]$_2^{3+}$ was optimized for the states with one and three unpaired electrons. We should note that in all unrestricted Kohn--Sham DFT calculations each state is defined through the difference of $\alpha$ and $\beta$ electrons and hence does not correspond to a pure singlet, doublet, triplet, etc. state. To facilitate a direct comparison to wave function based methods, we will label those states through their number of unpaired electrons $N_{\textrm{un}}$.}
For each state, a vibrational frequency analysis was performed to verify that the optimized geometry corresponds to a (global or local) minimum. The geometry of the lowest energy state was chosen for subsequent electronic structure calculations. The ground-state structures of all studied compounds are visualized in Figure \ref{fig:geometry}.

In order to cross-check the reliability of the COSMO solvation model, we performed additional calculations, where the first solvation shell (nine water molecules for the T-shaped structure and eight water molecules for the diamond-shaped clusters) was explicitly included (using the same exchange--correlation functional and basis sets as described above).
{We should note that we considered only those clusters for which stable ground-state equilibrium structures could be found using the COSMO solvation model, \textit{i.e.}, where the cluster did not break apart.}
To assess the impact of the water ligands on the CCI structures, we optimized the molecular geometries for a different number of water ligands. Specifically, we included two, three, four, five, and seven \ce{H2O} molecules for the {state with $N_{\textrm{un}}$=4} of the T-shaped cluster. 
For the diamond-shaped [NpO$_2$]$_2^{2+}$ ({$N_{\textrm{un}}$=4}) and [NpO$_2$]$_2^{3+}$ ({$N_{\textrm{un}}$=3}), we recomputed the geometries of the neptunyl-neptunyl clusters in the presence of {four and six water molecules. Furthermore, we performed single-point calculations to obtain relative energies for different structures with the same number of electrons using different exchange--correlation functionals. The optimized structures and relative energies are provided in the ESI$\dag$.}

\subsection{Remarks on state and orbital labelling}

Although spin does not represent a good quantum number in heavy-element compounds{, due to spin-orbit coupling,} we will nonetheless label electronic structures according to their formal total spin quantum number.
This labeling is commonly used in the literature and allows us to straightforwardly identify the number of unpaired electrons in the complexes and clusters.
Furthermore, if not mentioned otherwise, we will label the orbitals according to their irreducible representation of the D$_{\infty h}$ point-group of the linear moieties ($\sigma_g$-, $\sigma_g^*$-, $\pi_g$-, $\pi_g^*$-, $\delta_u$-, $\phi_u$-type). This labeling will facilitate the direct comparison of the clusters to the individual subunits.

\subsection{CASSCF with ECP}

The Complete Active Space Self Consistent Field calculations with ECP (CASSCF+ECP) were performed with the MOLPRO 2012.1 software suite.~\cite{molpro2012,molpro2012_2,casscf-01,casscf-02} For the neptunium atoms, we used an ECP60MWB~ANO{\mbox{~\cite{ECP_Np_2, U_ECP}}} basis set introduced by the Stuttgart/Cologne group, with the core consisting of 60 electrons, and which features energy-consistent, semi-local pseudopotentials. For the oxygen atoms, Dunning's cc-pVDZ basis was used.~\cite{basis-Dunning-JCP1989-90-1007} The orbitals are visualized using Jmol 14.9.1.~\cite{jmol}
Furthermore, we used a simplified gas-phase model with no implicit or explicit solvent in all CASSCF+ECP calculations.
Explicit water molecules are considered only in our CASPT2/SO-RASSI calculations (\textit{vide infra}).
Since we obtained similar results in the electronic spectrum for CCIs with and without water molecules, the gas-phase models will suffice for assessing the influence of the relativistic Hamiltonian on the electronic structure of the investigated CCIs.
In all CASSCF calculations, we used the DFT structures that have been optimized in combination with the COSMO solvation model (cf. Figure\mbox{~\ref{fig:geometry}}). 
Moreover, we performed state-averaged calculations, where the average includes two states from each irreducible representation.
The {quintet state ($N_{\textrm{un}}$=4)} was optimized for the molecules with a total charge of 2+, while for the 3+ charged moiety we calculated the {quartet state ($N_{\textrm{un}}$=3)}.
Our choice of the active space orbitals is based on previous studies on CCIs presented by Vlaisavljevich~\textit{et al.}~\cite{laura2013}. We should note that in the CCI clusters the Np--O bond lengths are longer than in the isolated neptunyl cations. Recent work presented by some of us~\cite{uranyl-dissociation} highlights the importance of $\phi$ orbitals if the actinide--oxygen bond elongates. Furthermore, our latest study on plutonium oxides suggests that $\phi$ orbitals are important to reliably treat static electron correlation. Thus, in all active space calculations, $\phi$ orbitals have been included in the active space.
Specifically, the active space of all species contains orbitals corresponding to $\phi_u$- and $\delta_u$-type orbitals that are characteristic for linear moieties.
This resulted in CAS(4,8)SCF calculations for the T-shaped molecule.
Additional $\sigma_g$- and $\sigma_g^*$-type orbitals are included only in the diamond-shaped structures. 
These orbitals, built primarily from 5f neptunium atomic orbitals, are occupied by eight electrons (CAS(8,12)SCF) in the 2+ charged diamond-shaped molecule, while for the 3+ charged ion we perform analogous CAS(7,12)SCF calculations.
All active orbitals are presented in the ESI.

\subsection{CASSCF, CASPT2, and spin-orbit coupling calculations with the second order Douglas--Kroll--Hess Hamiltonian}

The CASSCF calculations and Complete Active Space Second-order Perturbation Theory (CASPT2)\mbox{~\cite{caspt21,caspt22,MS-CASPT2}} calculations were performed using the OpenMolcas (version 17.0) software package.\mbox{~\cite{Molcas6, Molcas7, Molcas-code, Molcas8}} Relativistic effects were accounted for by the second order Douglas--Kroll--Hess Hamiltonian (DKH2).\mbox{~\cite{dkh1, dkh2}}
We used the ANO-RCC-VDZP basis set for all atoms.\mbox{~\cite{basis-ano-rcc-main, basis-ano-rcc-actinides}}
Furthermore, we investigated two different models of CCIs: (i) the simplified gas-phase structures optimized by DFT/COSMO (as in all CASSCF+ECP calculations) and (ii) the hydrated CCIs with a complete first solvation shell (9 water ligands for the T-shaped cluster, 8 water ligands for the diamond compounds).
For the T-shaped CCI, we computed 12 CAS(4,8)SCF wave functions for the {quintet state ($N_{\textrm{un}}$=4)} (four in the A$_1$ and in the A$_2$ irreducible {representation}, two in the B$_1$ and in the B$_2$ irreducible representation) and eight CASSCF wave functions for the triplet state {($N_{\textrm{un}}$=2;} two in every irreducible representation). The state-averaged calculations include an average over states in the same irreducible representation.
For the diamond-shaped [NpO$_2$]$_2^{2+}$, we performed calculations for two active spaces. The minimal CAS(4,8) for the diamond-shaped dimers contains only $\phi_u$- and $\delta_u$-type orbitals as for the T-shaped species. The extended CAS(8,12) for the diamond-shaped complexes contains similar active space orbitals as chosen in CAS(8,12)SCF+ECP. {An average over eight states (two in every irreducible representation) was performed} for both the triplet ($N_{\textrm{un}}$=2) and quintet ($N_{\textrm{un}}$=4) states.
For the diamond-shaped [NpO$_2$]$_2^{3+}$, the calculations using both minimal CAS(3,8) and extended CAS(7,12) result in broken symmetry solutions and are thus not presented here.
{Furthermore, similar active spaces have been chosen for the gas-phase structures (i) and the hydrated clusters (ii).}
The CASSCF wave functions were used to calculate multistate CASPT2 energy corrections, where the ionization potential-electronic
affinity (IPEA) shifted H$_0$ Hamiltonian~\cite{caspt2-h0} was applied with an imaginary shift set to 0.1.
{The spin--orbit (SO) interaction effects (in the Atomic Mean Field Approximation\mbox{~\cite{AMFI_1,AMFI_2,AMFI_3}}) were calculated using the Restricted Active Space State Interaction (RASSI) approach,\mbox{~\cite{rassi}} where the energy correction due to dynamical correlation was included in an approximate manner by dressing (\textit{i.e.}, shifting) the diagonal elements of the spin-orbit Hamiltonian by the CASPT2 energies.
}

\subsection{DMRG}

Density Matrix Renormalization Group (DMRG) calculations~\cite{white,white2,white-qc,marti2010b,Ors_ijqc,wouters-review,yanai-review,dmrg-17,Stein2016,dmrg-21} were performed using the Budapest QC-DMRG program.~\cite{dmrg_ors} {In accordance with DFT} and CASSCF+ECP results, {we assume that the irreducible representation of the ground-state wave function} is A$_\mathrm{g}$ {for the diamond-shaped [NpO$_2$]$_2^+$ compound}, B$_\mathrm{u}$ {for the [NpO$_2$]$_2^{2+}$ molecule}, and A$_1$ {for the T-shaped CCIs}. The orbital basis set comprises the natural orbitals computed with the CASSCF+ECP method discussed in the previous subsection.

The DMRG active space comprises the CASSCF active space extended by including supplementary inactive and virtual orbitals. 
For all neptunyl clusters, the $\sigma_\mathrm{u}$-, $\sigma_\mathrm{u}^*$-, $\sigma_\mathrm{g}$-, $\sigma_\mathrm{g}^*$-, $\pi_\mathrm{u}$-, $\pi_\mathrm{u}^*$-, $\pi_\mathrm{g}$-, and $\pi_\mathrm{g}^*$-type orbitals were added to the active space.
For diamond-shaped [NpO$_2$]$_2^{2+}$, 14 occupied orbitals (4 in a$_\mathrm{g}$ and in b$_\mathrm{u}$, 3 in a$_\mathrm{u}$ and in b$_\mathrm{g}$) and 20 virtual orbitals (6 in a$_\mathrm{g}$ and in b$_\mathrm{u}$, 4 in a$_\mathrm{u}$ and in b$_\mathrm{g}$) were added to the CASSCF active space (4 in a$_\mathrm{g}$ and in b$_\mathrm{u}$, 2 in a$_\mathrm{u}$ and in b$_\mathrm{g}$) resulting in a total of 46 molecular orbitals and 36 electrons. For the CCI of charge 3+, similar orbitals were used, but the number of electrons equals 35 (DMRG(35,46)). 
The CASSCF active space of the T-shaped neptunyl cluster was extended by 13 occupied orbitals (5 in a$_1$, 5 in b$_1$, and 3 in b$_2$) and 26 virtual orbitals (9 in a$_1$, 3 in b$_1$, 6 in b$_2$, and 6 in a$_2$), yielding DMRG(30,45).

We should emphasize that our DMRG calculations will be used to elucidate the orbital correlations in the neptunium dications. As presented in ref.~\citenum{boguslawski2017}, this orbital-correlation analysis will allow us to dissect electron correlation effects and to determine which orbitals are hence important for non-dynamic, static, and dynamic correlation.
{Note that the DMRG algorithm approximates {a CASCI wave function in a large active space, but misses a large fraction of dynamical correlation}, which can be included \textit{a posteriori}. Since we are only interested in orbital correlations within a large active space, no dynamic energy correction on top of the DMRG wave function has been performed.
}

\section{Results}

In this work, we focus on the diamond-shaped and T-shaped dications containing neptunyl(V) and neptunyl(VI) as building blocks.
Thus, the corresponding supramolecular clusters are [NpO$_2$]$_2^{2+}$, [NpO$_2$]$_2^{3+}$, and [NpO$_2$]$_2^{4+}$.
For [NpO$_2$]$_2^{2+}$ and [NpO$_2$]$_2^{4+}$, we investigated {the corresponding quantum states with $N_{\textrm{un}}$=0, $N_{\textrm{un}}$=2, and $N_{\textrm{un}}$=4, while for [NpO$_2$]$_2^{3+}$ we performed calculations for $N_{\textrm{un}}$=1 and $N_{\textrm{un}}$=3.}
Our analysis suggests that [NpO$_2$]$_2^{2+}$ forms diamond-shaped and T-shaped clusters in both aqueous solution and the gas phase, while [NpO$_2$]$_2^{3+}$ is stable only as a diamond-shaped molecule in solution. 
For [NpO$_2$]$_2^{4+}$, no equilibrium geometries were found as the monomers diverge{, that is, the cluster breaks apart}. Thus, our calculations suggest that the CCIs cannot overcome the electrostatic repulsion between the neptunyl(VI) subunits. 
Furthermore, the BP86 exchange--correlation functional combined with the COSMO solvation model did not yield a stable T-shaped [NpO$_2$]$_2^{3+}$ complex.
The explicit inclusion of water molecules around the CCI clusters does not influence the diamond-shaped structures as the corresponding bond lengths and angles are similar to those obtained using only the COSMO solvation model.
The most pronounced changes can be observed in the T-shaped clusters for the Np--Np distances (\textit{vide infra}).
We should note that {theoretical} studies suggest that the impact of the equatorial ligands on the spectroscopic properties of the actinyl unit is small~\cite{uranyl-spectra-1,uranyl-spectra-2,uranyl-spectra-3,uranyl-spectra-4}.

\begin{figure}[htp]
\centering
	\includegraphics[width=0.7\linewidth]{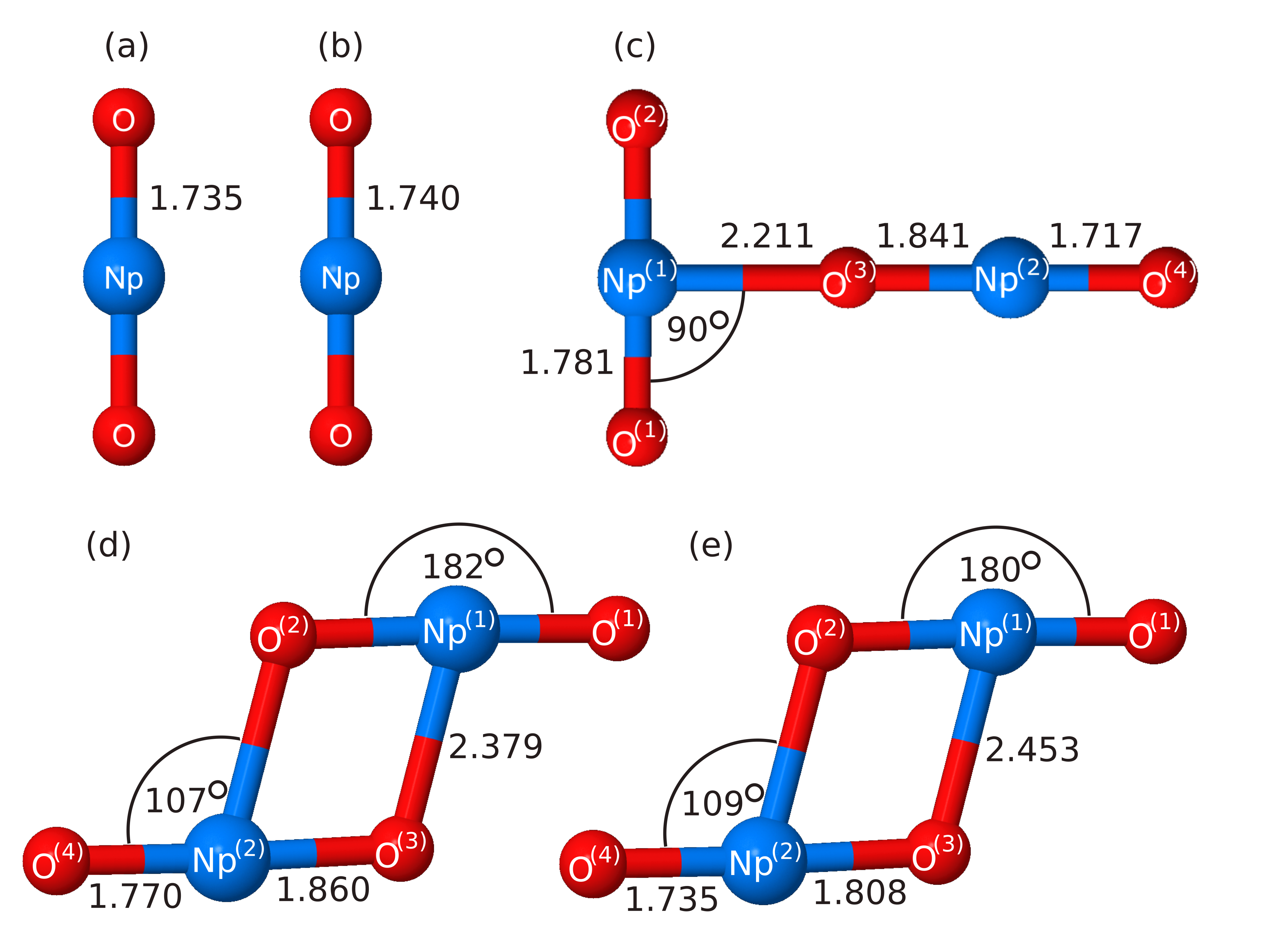}
	\caption{Optimized structural parameters of the ground-state molecules in aqueous solution: (a) NpO$_2^+$ ($N_{\textrm{un}}$=2), (b) NpO$_2^{2+}$ ($N_{\textrm{un}}$=4), (c) the T-shaped CCI of [NpO$_2$]$_2^{2+}$ ($N_{\textrm{un}}$=4), (d) the diamond-shaped CCI of [NpO$_2$]$_2^{2+}$ ($N_{\textrm{un}}$=4), (e) the diamond-shaped CCI of [NpO$_2$]$_2^{3+}$ ($N_{\textrm{un}}$=3).
	All bond lengths are given in \r{A}ngstr\"{o}ms. No stable structure of T-shaped [NpO$_2$]$_2^{3+}$ was found as the neptunyl monomers diverged{, that is, the cluster broke apart}.
    {Note that the valence electrons of NpO$_2^+$ and NpO$_2^{2+}$ occupy non-bonding orbital and hence the Np--O bond distance changes only slightly.}
    }
	\label{fig:geometry}
\end{figure}

\begin{table*}[h]
\caption{BP86 optimized structural parameters (bond lengths [\AA{}] and angles [$^\circ$]) of the T-shaped [NpO$_2$]$_2^{2+}$ in different environments. Calculations for molecules in solution used the COSMO solvation model. $N_{\textrm{un}}$ denotes the number of unpaired electrons. The column ``H$_2$O'' indicates the number of explicit water molecules. $\protect\angle$ symbolizes the angle between Np$^{(1)}$, O$^{(1)}$, and O$^{(3)}$. The bond length \mbox{O$^{(1)}$--Np$^{(1)}$} is equal to \mbox{O$^{(2)}$--Np$^{(1)}$}. Detailed information about the neptunyl--water distance can be found in the ESI. Please, note that both structures with nine explicit water molecules are characterized by $C_s$ symmetry. Their molecular structures ((g) and (h), respectively) are presented in the ESI.}
\begin{tabular*}{0.99\textwidth}{@{\extracolsep{\fill}}llllllllll} \hline
		&&	\scalebox{.75}[1.0]{\textbf{\ce{H2O}}}	&	\scalebox{.75}[1.0]{\textbf{O$^{(1)}$-Np$^{(1)}$}}	&	\scalebox{.75}[1.0]{\textbf{Np$^{(1)}$-O$^{(3)}$}}	&	\scalebox{.75}[1.0]{\textbf{O$^{(3)}$-Np$^{(2)}$}}	&	\scalebox{.75}[1.0]{\textbf{Np$^{(2)}$-O$^{(4)}$}}	&	\textbf{$\angle$}	&	\scalebox{.75}[1.0]{\textbf{Np$^{(1)}$-Np$^{(2)}$}}	\\ \hline
\multirow{3}{*}{\rot{Vacuum~}}	
	&	~[(NpO$_2$)$_2$]$_{[N_{\textrm{un}}=0]}^{2+}$	&	--	&	1.744	&	2.465	&	1.839	&	1.724	&	90	&		4.304	\\
	&	~[(NpO$_2$)$_2$]$_{[N_{\textrm{un}}=2]}^{2+}$	&	--	&	1.741	&	2.465	&	1.862	&	1.742	&	90	&		4.328	\\ \vspace{2mm}
	&	~[(NpO$_2$)$_2$]$_{[N_{\textrm{un}}=4]}^{2+}$	&	--	&	1.747	&	2.500	&	1.851	&	1.741	&	90	&		4.351	\\
\multirow{5}{*}{\rot{Solution~}}	
	&	~[(NpO$_2$)$_2$]$_{[N_{\textrm{un}}=0]}^{2+}$	&	--	&	1.772	&	2.338	&	1.834	&	1.756	&	91	&		4.172	\\
	&	~[(NpO$_2$)$_2$]$_{[N_{\textrm{un}}=2]}^{2+}$	&	--	&	1.778	&	2.344	&	1.856	&	1.775	&	90	&		4.200	\\
	&	~[(NpO$_2$)$_2$]$_{[N_{\textrm{un}}=4]}^{2+}$	&	--	&	1.781	&	2.211	&	1.841	&	1.717	&	90	&		4.052	\\
	&	~[(NpO$_2$)$_2$]$_{[N_{\textrm{un}}=4]}^{2+}$ 	&	9	&	1.827	&	2.392	&	1.868	&	1.813	&	86	&		4.220	\\ 
	&	~[(NpO$_2$)$_2$]$_{[N_{\textrm{un}}=4]}^{2+}$	&	9	&	1.832	&	2.390	&	1.872	&	1.811	&	89	&		4.209	 \\\hline
\end{tabular*}
\end{table*}

\subsection{The T-shaped [NpO$_2$]$_2^{2+}$ molecule}

\subsubsection{Ground-state geometry and spin}

\begin{table} [t] 
\small
\caption{Spin-state energetics and total spin expectation values of the T-shaped [NpO$_2$]$_2^{2+}$ calculated with DFT and COSMO. $N_{\textrm{un}}$ denotes the number of unpaired electrons. The energy differences are calculated with respect to the lowest lying spin state.}
\label{tab:DFTenergyT}
\begin{tabular}{c c c c c }
  \hline
  & \textbf{Compound}  & \textbf{Energy [cm$^{-1}$]} &  \textbf{S$^2$} & \textbf{Ideal S$^2$} \\
  \hline
& [(NpO$_{2})_{2}]_{[N_{\textrm{un}}=0]}^{2+}$       & 16 839 & 0.00 & 0.00 \\
Solution & [(NpO$_{2})_{2}]_{[N_{\textrm{un}}=2]}^{2+}$ & 2 433  & 3.02 &  2.00\\ \vspace{2mm}
& [(NpO$_{2})_{2}]_{[N_{\textrm{un}}=4]}^{2+} $      & 0      & 6.03 & 6.00  \\ 
& [(NpO$_{2})_{2}]_{[N_{\textrm{un}}=0]}^{2+}$       & 17 038 & 0.00 & 0.00 \\
Vacuum & [(NpO$_{2})_{2}]_{[N_{\textrm{un}}=2]}^{2+}$& 3 917  & 3.02 & 2.00\\
& [(NpO$_{2})_{2}]_{[N_{\textrm{un}}=4]}^{2+} $      & 0      & 6.03 & 6.00  \\
  \hline
\end{tabular} 
\end{table}

\begin{table*}[h]
\caption{Mulliken spin-population analysis for the investigated structures (optimized for BP86 and with the COSMO solvation model), where n denotes the difference between alpha- and beta-electrons for a given atomic center and M stands for the multiplicity. The charge and population of the O$^{(2)}$ atom is exactly the same as for O$^{(1)}$ in T-shaped structure or the same as for O$^{(3)}$ in diamond-shaped structure.}
\label{tab:pop}
\centering
\begin{tabular*}{0.9\textwidth}{@{\extracolsep{\fill}}llllllll} \hline
	&		&	\multicolumn{2}{l}{T-shaped cluster}	&	 \multicolumn{4}{l}{Diamond-shaped cluster}		\\ \cline{3-4} \cline{5-8}
	&		&	\multicolumn{2}{l}{[NpO$_2$]$_2^{2+}$}		&	\multicolumn{2}{l}{[NpO$_2$]$_2^{2+}$}		&	\multicolumn{2}{l}{[NpO$_2$]$_2^{3+}$}		\\ \cline{3-4} \cline{5-6} \cline{7-8}
	&		& $N_{\textrm{un}}=3$	&	 $N_{\textrm{un}}=5$	&	 $N_{\textrm{un}}=3$	&	 $N_{\textrm{un}}=5$	&	 $N_{\textrm{un}}=2$	&	 $N_{\textrm{un}}=4$	\\ \hline
Np$^{(1)}$	&	\textbf{charge}	&	~1.64	&	~1.62	&	~1.51	&	~1.75	&	~1.58	&	~1.97	\\
	&	\textbf{n(d)}	&	-0.02	&	~0.05	&	~0.03	&	~0.05	&	~0.01	&	~0.03	\\
	&	\textbf{n(f)}	&	-0.24	&	~2.03	&	~1.09	&	~2.17	&	~0.56	&	~1.65	\\
Np$^{(2)}$	&	\textbf{charge}	&	~1.87	&	~1.89	&	~1.51	&	~1.75	&	~1.58	&	~1.97	\\
	&	\textbf{n(d)}	&	~0.05	&	~0.05	&	~0.03	&	~0.05	&	~0.01	&	~0.03	\\
	&	\textbf{n(f)}	&	~2.41	&	~2.32	&	~1.09	&	~2.17	&	~0.56	&	~1.65	\\
O$^{(1)}$	&	\textbf{charge}	&	-0.35	&	-0.36	&	-0.15	&	-0.33	&	~0.05	&	-0.18	\\
	&	\textbf{n(p)}	&	~0.01	&	-0.10	&	-0.06	&	-0.12	&	-0.03	&	-0.10	\\
O$^{(3)}$	&	\textbf{charge}	&	-0.47	&	-0.45	&	-0.36	&	-0.42	&	-0.13	&	-0.29	\\
	&	\textbf{n(p)}	&	-0.09	&	-0.10	&	-0.05	&	-0.09	&	-0.03	&	-0.08	\\
O$^{(4)}$	&	\textbf{charge}	&	-0.35	&	-0.33	&	-0.15	&	-0.33	&	~0.05	&	-0.18	\\
	&	\textbf{n(p)}	&	-0.13	&	-0.12	&	-0.06	&	-0.12	&	-0.03	&	-0.10	\\ \hline
\end{tabular*}
\end{table*}

The bond lengths and angles of the T-shaped cluster in aqueous solution and in vacuo are summarized in Figure~\ref{fig:geometry}.
In the following, we will refer to the (O$^{(1)}$--Np$^{(1)}$--O$^{(2)}$) monomer as the vertical unit, while (O$^{(3)}$--Np$^{(2)}$--O$^{(4)}$) will be termed the horizontal unit. The Np$^{(1)}$--O$^{(2)}$ bond length of the vertical subunit differs only slightly from the corresponding equilibrium distance of the neptunyl(V) monomer. Solvation effects generally increase equilibrium bond lengths compared to calculations performed in vacuo. \cite{swart2005}
In the horizontal unit, the internal Np$^{(2)}$--O$^{(3)}$ bond is longer than the distant (external) Np$^{(2)}$--O$^{(4)}$ one, which also elongates in solution.
Only the internal bond length decreases in solution.
If the number of unpaired electrons increases, the Np$^{(1)}$--O$^{(1)}$ bond elongates.
The Np$^{(2)}$--O$^{(3)}$ and Np$^{(2)}$--O$^{(4)}$ bond lengths reach their maximum value for the state with $N_{\textrm{un}}$=2. The inter-monomeric distance between the vertical and horizontal unit is similar for the states with $N_{\textrm{un}}$=0 and $N_{\textrm{un}}$=2, while for the state with $N_{\textrm{un}}$=4, this distance is significantly shorter (about 0.1 \AA{}) compared to the other spin states and increases by approximately 0.3 \AA{} to 2.5 \AA{} in vacuo.
Furthermore, the CCI bond distances are considerably shorter in solution than in vacuo.
Finally, slight deformations from linearity are observed for the vertical subunit of the states with $N_{\textrm{un}}$=0 and $N_{\textrm{un}}$=2 in vacuo and amount to $4^\circ$.
Furthermore, Krot and Grigoriev~\cite{krot2004} who investigated the Np--O distances in various actinide compounds report that the Np--O$_{yl}$ bonds (which corresponds to Np$^{(1)}$--O$^{(1)}$, Np$^{(1)}$--O$^{(2)}$, Np$^{(2)}$--O$^{(3)}$, and Np$^{(2)}$--O$^{(4)}$) are between 1.800 and 1.915 \AA{}, while the distances between the neptunium center of one neptunyl unit and the oxygen atom from the other neptunyl unit are in {the range of 2.334--2.540 \AA{}. The DFT-optimized bond lengths tend to be slightly shorter.}
We should emphasize that our optimized Np--O bond lengths are very similar to the bond distances deduced from various neptunium-containing crystalline solids (cf. Table S3 in the ESI for further details). However, these similarities should be considered only qualitatively due to different phases and the varying number and types of ligands surrounding the neptunyl clusters. Most importantly, an explicit solvation model is required for the dications to provide accurate structural parameters.

Spin-state energetics and total spin expectation values are presented in Table \ref{tab:DFTenergyT}.
{We should emphasize that the spin expectation value is calculated for the noninteracting system, where the wave function is a Slater determinant of the Kohn--Sham orbitals and the same expression for \textbf{S$^2$} can be used as in unrestricted Hartree--Fock theory.\mbox{~\cite{Wang1995,Wittbrodt1996,grafenstein2001,Cohen2007}} This approach, however, does not give the same \textbf{S$^2$} expectation value for the real, interacting system, where the two particle density is required.\mbox{~\cite{Wang1995,Wittbrodt1996,grafenstein2001,Cohen2007}}
}
The {state with $N_{\textrm{un}}$=4} is the ground state in both the gas phase and solution, while the {state with $N_{\textrm{un}}$=0} lies highest in energy.
Furthermore, the aqueous environment does not change the relative order of the spin states and energy gaps between different spin states. Note, however, that the {state with $N_{\textrm{un}}$=2} is considerably spin-contaminated (see also Table \ref{tab:DFTenergyT}).
A Mulliken population analysis~\cite{mulliken-pop-analys-1955} of all investigated compounds is presented in Table~\ref{tab:pop}. A positive charge is located on the neptunium atoms, while the oxygen atoms are negatively charged.
{The spin density} is mostly centered on the Np atom and dominated by f-electrons.
The Mulliken population analysis further predicts an excess of $\beta$-electron density on the Np atom in the (spin-contaminated) {state with $N_{\textrm{un}}$=2} of the T-shaped CCI.
Note that this Np atom is coordinated by three oxygen atoms, while the bi-coordinated Np$^{(2)}$ atom bears a large excess of $\alpha$-electron density.
{These differences in spin density} on the Np centers may originate from the observed spin contamination in the {state with $N_{\textrm{un}}$=2}.
{A Bader analysis is presented in the ESI for comparison. In general, it indicates a higher charge concentration on the atomic centers compared to a Mulliken population analysis.
}

Finally, we will compare the molecular structure of the investigated neptunyl clusters to the recently presented uranyl analogues.\mbox{~\cite{tecmer-song2016}}
Both T-shaped actinyl CCI clusters feature a similar distance between the monomers, \textit{i.e.}, similar Ac--O bond lengths (2.3, 2.32, and 2.35 \AA{} for the {states with $N_{\textrm{un}}$=0, $N_{\textrm{un}}$=2, and $N_{\textrm{un}}$=4}, respectively, in the uranyl(V)--uranyl(V) clusters). However, the difference in alpha and beta electrons has a stronger impact on the geometry in neptunyl clusters than in uranyl CCIs, where the CCI bond length is noticeably shorter for the {state with $N_{\textrm{un}}$=4}. Furthermore, the uranyl clusters feature a different energetic order of spin states as the {state with $N_{\textrm{un}}$=2} is the ground state followed by the states with $N_{\textrm{un}}$=0 and $N_{\textrm{un}}$=2.

\subsubsection{Water coordination of T-shaped CCI}

Experimental studies indicate that the first coordination sphere of the bare neptunyl in aqueous solution comprises five water molecules.~\cite{Szabo1996,allen1997,Np-5-waters} Thus, the complete first coordination sphere of the neptunyl--neptunyl T-shaped dimer should contain nine water molecules. Theoretical studies on the monomeric building blocks, however, highlight the difficulty of quantum-chemistry methods to reliably predict the first solvation sphere of neptunyls.~\cite{schreckenbach-2000,vallet2004,neptunyl-solution} In this work, we therefore modelled the first solvation sphere using a different number of water molecules{, starting from a rough explicit solvation model containing two water ligands up to the complete first solvation sphere comprising nine water molecules}. Most importantly, the structural parameters of the neptunyl {monomer} are negligibly affected by the number of explicit solvent molecules (see Table S1 of the ESI). Furthermore, our calculations confirm the trends in the Np--O bond lengths obtained using only the COSMO model. The most pronounced qualitative feature is the small deformation of the neptunyl dimers as Np$^{(1)}$, O$^{(3)}$, and Np$^{(2)}$ are not lying on a straight line. The distances between two neptunium centers (4.220 and 4.209 \AA{} for the {state with $N_{\textrm{un}}$=4} with a complete first coordination sphere) agree well with the experimental result of 4.200 \AA{} reported by Guillaume~\textit{et al.}~\cite{guillaume1983} {Generally,} the inclusion of water molecules to model the first solvation shell is required to reproduce the experimentally observed Np--Np distance. In this work, this is already achieved by including 2 water molecules.

\begin{table} [t] %
\scriptsize
\caption{Energetics of the spin-free CASSCF wave functions with and without CASPT2 corrections for the T-shaped [NpO$_2$]$_2^{2+}$ molecule. The 0 cm$^{-1}$ value is assigned to the lowest energy in the column.}
\label{tab:compare-t2-energies}
\begin{tabular}{m{0.03\columnwidth} >{\centering}m{0.1\columnwidth} >{\centering}m{0.12\columnwidth} >{\centering}m{0.1\columnwidth} m{0.01\columnwidth} m{0.03\columnwidth} >{\centering}m{0.1\columnwidth} c }
\hline
	&	\multicolumn{3}{c}{$\Delta$E [cm$^{-1}$] for NpO$_2$]$_2^{2+}$}					&	&	\multicolumn{3}{c}{{$\Delta$E [cm$^{-1}$] for NpO$_2$]$_2^{2+}$+9H$_2$O}}			\\	\cline{2-4}	\cline{6-8}	\\[-0.9em]
	&	CAS(4,8)SCF 	&	CAS(4,8)SCF	&	CAS(4,8)PT2 	&	&		&	CASSCF(4,8)	&	CAS(4,8)PT2	\\	
	&	ECP	&	DKH2	&	DKH2	&	&		&	DKH2	&	DKH2	\\ \hline	
1$^5$A$_1$	&	0	&	160	&	58	&	&	1$^5$A	&	0	&	0	\\	
2$^5$A$_1$	&	3476	&	3241	&	2546	&	&	2$^5$A	&	5	&	14	\\	
3$^5$A$_1$	&		&	3527	&	2632	&	&	3$^5$A	&	9	&	14	\\	
4$^5$A$_1$	&		&	6470	&	4297	&	&	4$^5$A	&	14	&	27	\\	
1$^5$B$_2$	&	1	&	281	&	0	&	&	5$^5$A	&	2400	&	1599	\\	
2$^5$B$_2$	&	3645	&	3517	&	2555	&	&	6$^5$A	&	2405	&	1611	\\	
3$^5$B$_2$	&		&	6785	&	4373	&	&	7$^5$A	&	2785	&	2080	\\	
4$^5$B$_2$	&		&	7190	&	4579	&	&	8$^5$A	&	2794	&	2091	\\	
1$^5$B$_1$	&	1	&	26	&	128	&	&	1$^3$A	&	6	&	6	\\	
2$^5$B$_1$	&	7119	&	501	&	5215	&	&	2$^3$A	&	11	&	19	\\	
1$^5$A$_2$	&	0	&	0	&	51	&	&	3$^3$A	&	15	&	20	\\	
2$^5$A$_2$	&	3476	&	3161	&	2564	&	&	4$^3$A	&	20	&	33	\\	
1$^3$A$_1$	&		&	7	&	63	&	&	5$^3$A	&	2407	&	1598	\\	
2$^3$A$_1$	&		&	3167	&	2573	&	&	6$^3$A	&	2412	&	1609	\\	
1$^3$B$_1$	&		&	4	&	54	&	&	7$^3$A	&	2790	&	2080	\\	
1$^3$B$_1$	&		&	3319	&	2629	&	&	8$^3$A	&	2798	&	2091	\\	
1$^3$B$_2$	&		&	33	&	141	&	&		&		&		\\	
2$^3$B$_2$	&		&	6507	&	5224	&	&		&		&		\\	
1$^3$A$_2$	&		&	7	&	61	&	&		&		&		\\	
2$^3$A$_2$	&		&	3167	&	2571	&	&		&		&		\\	
 \hline
\end{tabular}
\end{table}

\begin{figure*}[htp]
\centering
  \includegraphics[width=0.6\linewidth]{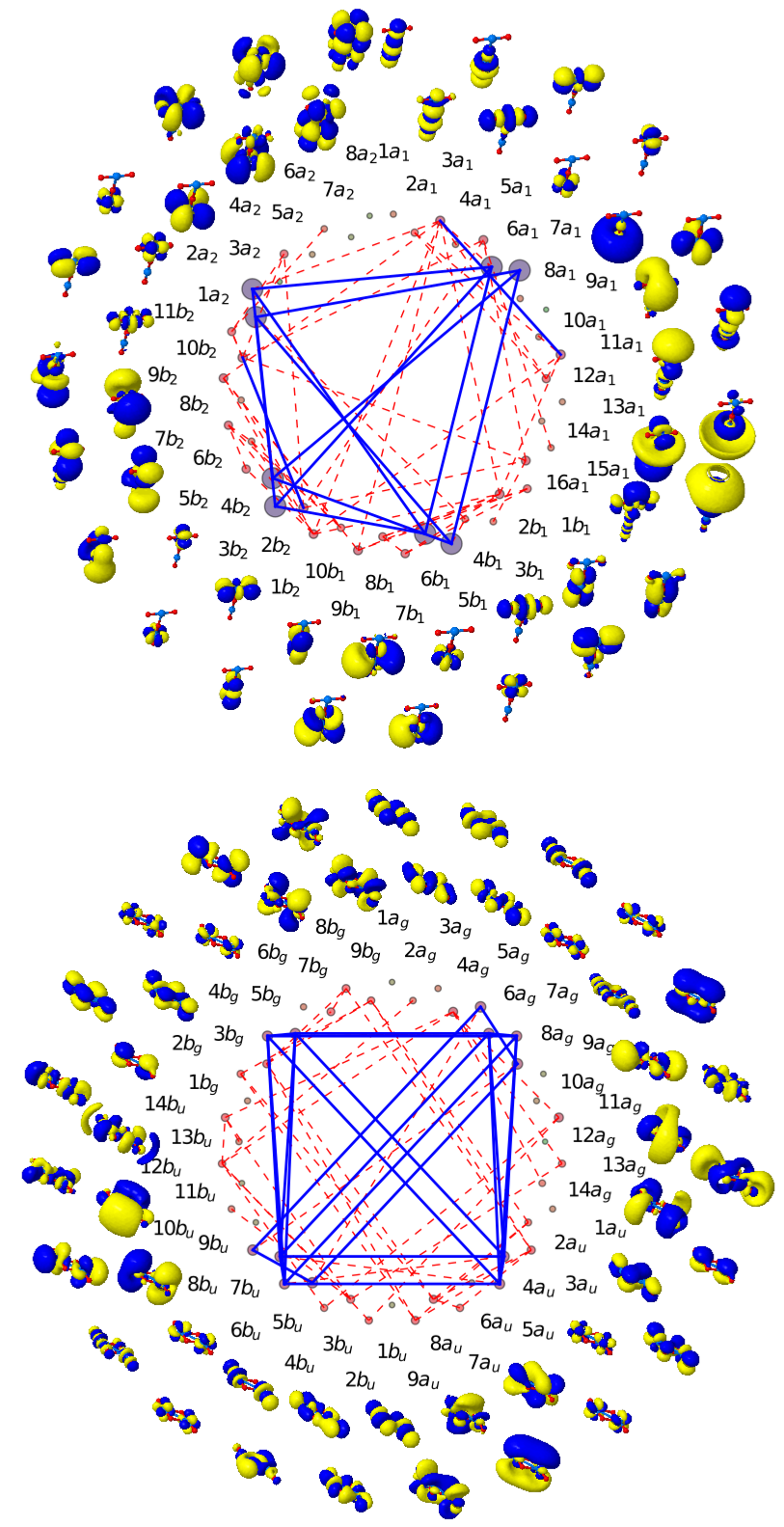}
  \caption{Orbital-pair correlations of the T-shaped (top) and diamond-shaped (bottom) [NpO$_2$]$_2^{2+}$. The values of the single-orbital entropy are coded by the size of the dots corresponding to each orbital. The strongest correlated orbitals are connected by blue lines ($I_{i|j}>10^{-1}$), followed by orbital-pair correlations marked by red lines ($10^{-1}>I_{i|j}>10^{-2}$). Note that the irreducible representations of the $C_{2h}$ and $C_{2v}$ point group have been used to label the orbitals to facilitate the identification of active space orbitals. A translation to the irreducible representations of $D_{\infty h}$ can be found in the ESI.}
  \label{fig:i12-t2-d2}
\end{figure*}

\subsubsection{Orbital-pair correlation measures in the CASSCF natural orbital space}

\begin{figure}[htp]
\centering
  \includegraphics[width=0.7\linewidth]{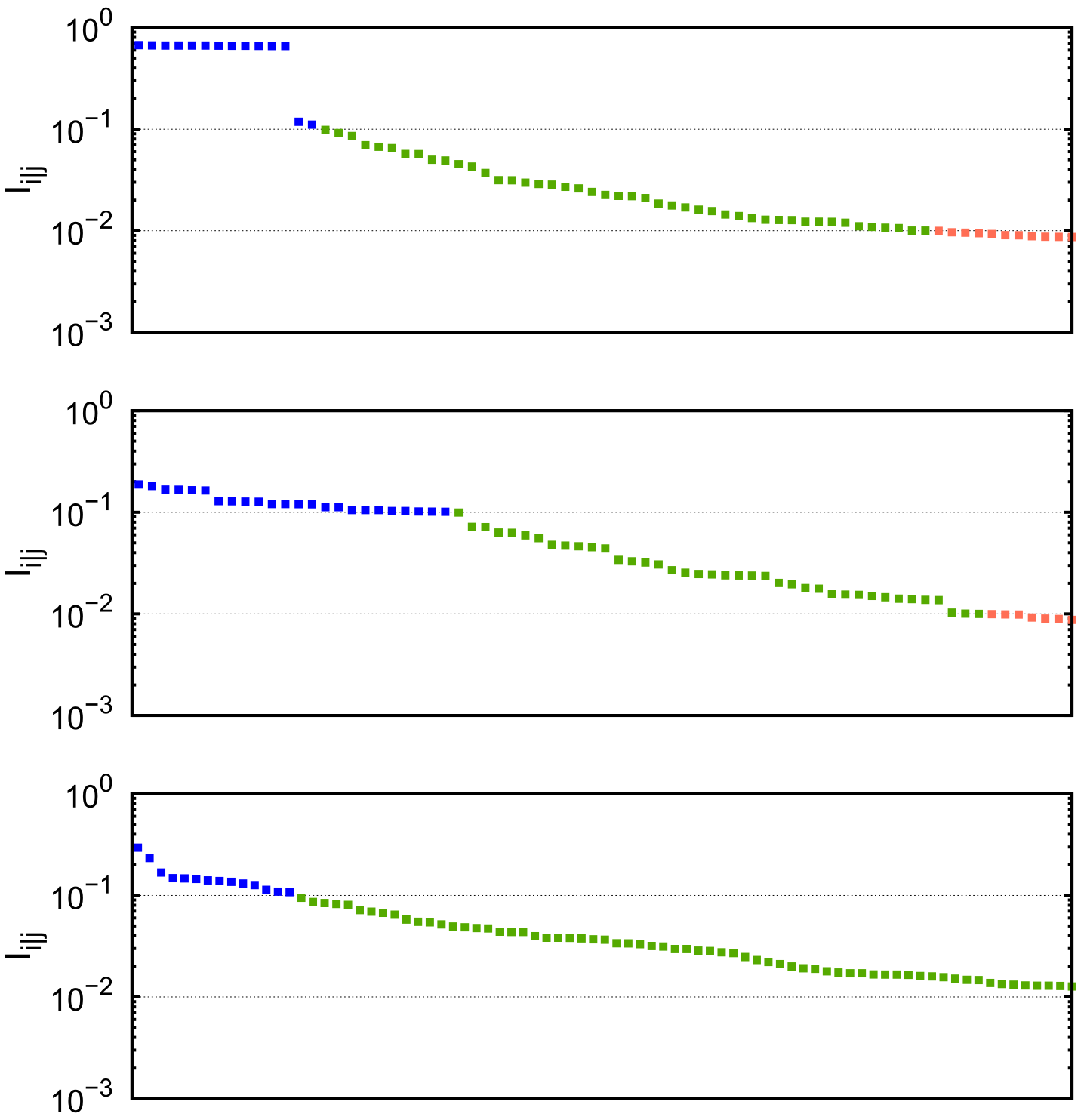}
  \caption{Decaying values of the mutual information 
  (a) for the T-shaped [NpO$_2$]$_2^{2+}$ (first twelve blue points correspond to the strong correlation between $\delta_u$- and $\phi_u$-type orbitals),
  (b) for the diamond-shaped [NpO$_2$]$_2^{2+}$ (the first two blue points correspond to the strong correlation between $\sigma_g$- and $\sigma_g^*$-type orbitals, while the orbital-pair correlation between $\phi_u$- and $\delta_u$-type orbitals are marked by next thirteen points), 
  (c) for the diamond-shaped [NpO$_2$]$_2^{3+}$ (the first blue point corresponds to the strong correlation between {$\pi$-type 4a$_g$ and $\pi^*$-type 11a$_g$ orbitals}).}
  \label{fig:decay}
\end{figure}

The multi-reference nature of cation--cation interactions in the T-shaped ion can be further elucidated using orbital-pair correlation diagrams (see Figure \ref{fig:i12-t2-d2}), which were obtained using the DMRG algorithm.
The correlation between orbital pairs $i,j$ is quantified by the orbital-pair mutual information $I_{i|j}$. Strongly-correlated orbitals correspond to $I_{i|j}>10^{-1}$, while for weakly-correlated orbitals $I_{i|j}<10^{-2}$.
For the four-dimensional basis ($\vert$\zero$\rangle$, $\vert$\up$\rangle$, $\vert$\down$\rangle$, $\vert$\double$\rangle$), the orbital-pair mutual information is bounded between 0 and $\ln16\approx2.7$.
A detailed discussion on the orbital-pair mutual information as a correlation measure can be found in, for instances, refs.~\citenum{mutual_information_1,dmrg-11,dmrg-13,boguslawski2017,dmrg-19,CUO_DMRG,Corinne_2015,barcza2014entanglement,boguslawski2015,ijqc-eratum}.
The decaying values of the mutual information are presented in Figure \ref{fig:decay} (a plot with a linear scale is shown in the ESI). We observe that the CAS(4,8)SCF active space orbitals (6a$_1$, 7a$_1$, 5b$_1$, 6b$_1$, 3b$_2$, 4b$_2$,1a$_2$, and 2a$_2$) are highly correlated ($I_{i|j} > 0.65$) and their contribution to the CCI bonding mechanism is most significant. 
These orbitals are primarily built of 5f atomic orbitals of the neptunium atom and the 2p orbitals of the oxygen atom. This orbital-pair correlations suggest the important role of 5f orbitals in modeling cation--cation interactions.
However, we observe other strongly ($I_{i|j} > 10^{-1}$) correlated bonding and antibonding orbital pairs: orbitals whose character corresponds to $\sigma_u$- (3a$_1$), $\sigma_u^*$- (11a$_1$), $\pi_g$- (2b$_2$), and $\pi_g^*$-type (10b$_2$) orbitals of the linear moieties. Their values of the mutual information are close to other moderately correlated orbital pairs.
Specifically, we have other $\sigma_u$/$\sigma_u^*$-, $\pi_u$/$\pi_u^*$, and $\pi_g$/$\pi_g^*$-type orbitals that belong to the group of moderately correlated orbital pairs ($I_{i|j} \approx 10^{-2}$).
Finally, we observe additional orbital correlations between $\sigma_u$/$\pi_u$-, \mbox{$\sigma_u^*$/$\pi_u$-,} and $\sigma_u$/$\delta_u$-type orbitals.
In certain cases, the decay of $I_{i|j}$ shows a jump, \textit{i.e.}, $I_{i|j}$ has a gap in its spectrum, that can be used to set an \textit{a priori} defined threshold value to identify large and weak values corresponding to static (strong) and dynamic (weak) correlations, respectively.
In more general cases, when $I_{i|j}$ decays smoothly, such a separation of scales is not well defined and we use $I_{i|j}>0.1${ to identify the large contributions}.~\cite{dmrg-13}
We conclude that the $\delta_u$- and $\phi_u$-type orbitals are thus the most important building blocks of a minimal active space for the T-shaped molecule to properly describe static/nondynamic electron correlation.
In order to describe moderate orbital correlations, the active space should also include $\sigma_u$- and $\sigma_u^*$-, $\pi_u^*$-, $\pi_u$-, $\pi_g$-, and $\pi_g^*$-type orbitals whose mutual correlations indicate that these orbitals should not be separated into an active and inactive part of the orbital space.
In the following, we will focus on modelling static electron correlation within the CASSCF approach exploiting the minimal active space only.

\subsubsection{Electronic structures from CASSCF and CASPT2}

The relative energies of all states calculated with the CASSCF+ECP method are shown in Table \ref{tab:compare-t2-energies}. 
The CI vectors of 1$^5$A$_1$, 2$^5$A$_1$, 1$^5$B$_1$, 2$^5$B$_1$, 1$^5$B$_2$, 2$^5$B$_2$, 1$^5$A$_2$, and 2$^5$A$_2$, which are included in the state-averaged wave function, and their most significant determinants are presented in the ESI. 
We observe that the lowest-lying 1$^5$A$_1$, 1$^5$A$_2$, 1$^5$B$_1$ and 1$^5$B$_2$ states are quasi-degenerate.
The corresponding natural occupation numbers of the ground-state wave function are presented in the ESI. 
The active-space orbitals are approximately half-occupied, where the average number of electrons in $\delta_u$-type orbitals is slightly larger than in $\phi_u$-type orbitals. 
Table~\ref{tab:compare-t2-energies} {also summarizes the twelve lowest-lying quintet ($N_{\textrm{un}}$=4) and eight lowest-lying triplet ($N_{\textrm{un}}$=2) states calculated with the CASSCF+DKH2 method.}
Most importantly, the DKH2 Hamiltonian lifts the quasi-degeneracies of the 1$^5$A$_1$, 1$^5$B$_1$, and 1$^5$B$_2$ states as predicted by CASSCF+ECP.
In contrast to our DFT results, where the {states with $N_{\textrm{un}}$=2} were predicted to be significantly higher in energy than the {state with $N_{\textrm{un}}$=4}, CASSCF results in quasi-degenerate triplet ({$N_{\textrm{un}}$=2}) states (1$^3$A$_1$, 1$^3$B$_1$, and 1$^3$A$_2$) and quintet ({$N_{\textrm{un}}$=4}) states (1$^5$A$_2$). 
For both CASSCF+ECP and CASSCF+DKH2, $| \delta_u^{(2)} \phi_u^{(2)} \rangle$-type determinants dominate in the wave functions.
Including dynamical correlation on top of the CASSCF+DKH2 wave function using second-order perturbation theory (indicated as CAS(4,8)PT2 in Table~\ref{tab:compare-t2-energies}) shifts the relative energies and yields a 1$^5$B$_1$ ground state. 
Furthermore, CASPT2 lifts the quasi-degeneracies of states lying energetically close to the ground state and simultaneously decreases the energy gap between higher-lying states.
Since the energy differences between the lowest-lying states are much smaller than 400 cm$^{-1}$, the ground state cannot be unambiguously described by spin-free methods.

\subsubsection{Accurate spectra of the T-shaped molecule in water}

The relative energies of some excited states with respect to the ground state of the T-shaped cluster are listed in Table \ref{tab:so-energies-t}. {The molecular model includes a complete explicit first coordination shell (nine water molecules).}
The vertical excitation energies were calculated using the {CASPT2/SO-RASSI} approach including spin-orbit coupling effects.
The energetics of {all states of some selected model systems (calculated using $C_{2v}$ and $C_1$ point group symmetry) are shown in the ESI, including both the gas-phase model and the hydrated cluster.
Note that the differences in electronic spectra between the gas-phase and the hydrated CCIs are negligible for the lower part of the spectrum (typically much smaller than 1000 cm$^{-1}$/0.1 eV), which is to be expected as those excitations feature atomic f--f transitions.}
Our results indicate mixing between the CASSCF reference states of different irreducible representations and the {triplet ($N_{\textrm{un}}$=2) and quintet ($N_{\textrm{un}}$=4}) states in the low-energy region of the spectrum (see Table \ref{tab:so-energies-t} for more details).
{The lowest-lying states are characterized by a dominant contribution of {quintet states ($N_{\textrm{un}}$=4)}.}
We should note that the excitation spectra of the CCI clusters have many similarities to the spectra of the bare neptunyl(V) \cite{matsika2000, infante2006, neptunyl-ion-spectra-exp-1, neptunyl-ion-spectra-exp-2} and neptunyl(VI) \cite{gomes2008} subunits. The transitions around 9 800 cm$^{-1}$, 10 500 cm$^{-1}$, 13 000--14 500 cm$^{-1}$, 16 000 cm$^{-1}$ and 21 000 cm$^{-1}$ were observed in neptunyl(V), while the calculated excitation energies close to 7 000 cm$^{-1}$ and 8 500--9 700 cm$^{-1}$ are present in neptunyl(VI).
The transitions from the ground state have an oscillator strength below $10^{-5}$ and the spectra are dominated by transitions between excited states. All transitions are listed in the ESI.

\begin{table*} [t]%
\small
\caption{Energetics of the T-shaped {[NpO$_2$]$_2^{2+}$ + 9H$_2$O molecule} calculated with CASPT2/SO-RASSI. The energy differences are calculated with respect to the lowest lying state. All calculated states are presented in the ESI.}
\label{tab:so-energies-t}
\begin{tabular*}{1\textwidth}{@{\extracolsep{\fill}}lll lll lll lll} \hline
& & \multicolumn{10}{c}{\textbf{Composition in terms of the spin-free wave functions}}\\
State no.& \textbf{$\Delta$E [cm$^{-1}$]} & \multicolumn{10}{c}{\textbf{(weight in \%)}}\\ \hline
1  &  0  &  2$^5$A  &  (22)  &  4$^5$A  &  (17)  &  1$^5$A  &  (17)  &  3$^5$A  &  (13)  &  3$^3$A  &  (13) \\  
 5  &  5007  &  1$^5$A  &  (49)  &  3$^3$A  &  (12)  &  3$^5$A  &  (12)  &  2$^5$A  &  (12)  &  2$^3$A  &  (12) \\  
 9  &  6607  &  5$^5$A  &  (49)  &  6$^5$A  &  (49)  &    &    &    &    &    &   \\  
 13  &  7067  &  7$^5$A  &  (49)  &  8$^5$A  &  (49)  &    &    &    &    &    &   \\ 
 17  &  8282  &  3$^5$A  &  (24)  &  1$^3$A  &  (15)  &  4$^3$A  &  (15)  &  2$^3$A  &  (11)  &  2$^5$A  &  (10) \\  
 21  &  9052  &  5$^5$A  &  (42)  &  6$^3$A  &  (41)  &  5$^3$A  &  (8)  &  6$^5$A  &  (7)  &    &   \\  
 23  &  9528  &  7$^5$A  &  (50)  &  8$^3$A  &  (49)  &    &    &    &    &    &   \\  
 25  &  10801  &  1$^5$A  &  (49)  &  1$^3$A  &  (49)  &    &    &    &    &    &   \\  
 26  &  10815  &  2$^3$A  &  (49)  &  2$^5$A  &  (49)  &    &    &    &    &    &   \\  
 35  &  12402  &  5$^5$A  &  (49)  &  5$^3$A  &  (49)  &    &    &    &    &    &   \\  
 37  &  12881  &  7$^3$A  &  (49)  &  7$^5$A  &  (49)  &    &    &    &    &    &   \\  
 \hline
 \end{tabular*}
 \end{table*}

\begin{table*}
\caption{BP86 optimized structural parameters (bond lengths [\AA{}] and angles [$^\circ$]) of the diamond-shaped [NpO$_2$]$_2^{2+}$ and [NpO$_2$]$_2^{3+}$  in different environments. Calculations for molecules in solution used the COSMO solvation model. $N_{\textrm{un}}$ denotes the number of unpaired electrons. The column ``H$_2$O'' indicates the number of explicit water molecules. $\protect\angle_1$ denotes the angle between O$^{(1)}$, Np$^{(1)}$, and O$^{(3)}$, while $\protect\angle_2$ stands for the angle between O$^{(1)}$, Np$^{(1)}$ and O$^{(2)}$. The bond lengths O$^{(1)}$--Np$^{(1)}$ and Np$^{(1)}$--O$^{(2)}$ are symmetric to O$^{(4)}$--Np$^{(2)}$ and Np$^{(2)}$--O$^{(3)}$, respectively.}
\label{tab:geod}
\begin{tabular*}{0.95\textwidth}{@{\extracolsep{\fill}}lllllllll}
	&		&	\scalebox{.75}[1.0]{\textbf{\ce{H2O}}}	&	\scalebox{.75}[1.0]{\textbf{O$^{(1)}$-Np$^{(1)}$}}	&	\scalebox{.75}[1.0]{\textbf{Np$^{(1)}$-O$^{(2)}$}}	&	\scalebox{.75}[1.0]{\textbf{Np$^{(1)}$-O$^{(3)}$}}	&	\scalebox{.75}[1.0]{\textbf{$\angle_1$}}	&	\scalebox{.75}[1.0]{\textbf{$\angle_2$}}	&	\scalebox{.75}[1.0]{\textbf{Np$^{(1)}$-Np$^{(2)}$}}	\\ \hline	
\multirow{3}{*}{\rot{~Vacuum~}} 	
	&	~[(NpO$_2$)$_2$]$_{[N_{\textrm{un}}=0]}^{2+}$	&	--	&	1.738	&	1.851	&	2.386	&	107	&	179	&	3.429	\\
	&	~[(NpO$_2$)$_2$]$_{[N_{\textrm{un}}=2]}^{2+}$	&	--	&	1.737	&	1.854	&	2.390	&	108	&	180	&	3.449	\\ \vspace{2mm}
	&	~[(NpO$_2$)$_2$]$_{[N_{\textrm{un}}=4]}^{2+}$	&	--	&	1.739	&	1.852	&	2.433	&	109	&	183	&	3.509	\\
\multirow{9}{*}{\rot{~Solution~}} 	
	&	~[(NpO$_2$)$_2$]$_{[N_{\textrm{un}}=0]}^{2+}$	&	--	&	1.768	&	1.858	&	2.346	&	105	&	179	&	3.357	\\
	&	~[(NpO$_2$)$_2$]$_{[N_{\textrm{un}}=2]}^{2+}$	&	--	&	1.780	&	1.818	&	2.357	&	106	&	182	&	3.450	\\
	&	~[(NpO$_2$)$_2$]$_{[N_{\textrm{un}}=4]}^{2+}$	&	--	&	1.770	&	1.860	&	2.379	&	107	&	182	&	3.425	\\
	&	~[(NpO$_2$)$_2$]$_{[N_{\textrm{un}}=4]}^{2+}$	&	8	&	1.813	&	1.883	&	2.497	&	108	&	180	&	3.551	\\
	&	~[(NpO$_2$)$_2$]$_{[C_i, N_{\textrm{un}}=4]}^{2+}$	&	8	&	1.811	&	1.880	&	2.476	&	107	&	179	&	3.519	\\
	&	~[(NpO$_2$)$_2$]$_{[N_{\textrm{un}}=1]}^{3+}$	&	--	&	1.734	&	1.815	&	2.406	&	109	&	180	&	3.452	\\
	&	~[(NpO$_2$)$_2$]$_{[N_{\textrm{un}}=3]}^{3+}$	&	--	&	1.735	&	1.808	&	2.453	&	109	&	180	&	3.489	\\
	&	~[(NpO$_2$)$_2$]$_{[N_{\textrm{un}}=3]}^{3+}$	&	8	&	1.775	&	1.840	&	2.503	&	108	&	177	&	3.542	\\ 
	&	~[(NpO$_2$)$_2$]$_{[C_i, N_{\textrm{un}}=3]}^{3+}$	&	8	&	1.772	&	1.828	&	2.609	&	111	&	178	&	3.637	\\ \hline
\end{tabular*}
\end{table*}

\subsection{The diamond-shaped [NpO$_2$]$_2^{2+}$ and [NpO$_2$]$_2^{3+}$ ions}

\subsubsection{Structural properties and spin state energetics}
The impact of the water environment on the molecular structures of the [NpO$_2$]$_2^{2+}$ and [NpO$_2$]$_2^{3+}$ moieties is summarized in Table \ref{tab:geod} (see also ESI for more details). We should emphasize that the differences in bond lengths and angles introduced by the explicit first solvation sphere are minor (up to 0.05 \AA{}) and much smaller than observed in the T-shaped cluster (approximately 0.2 \AA{}). In all clusters, the geometry of the neptunyl subunits changes when the CCI is formed. Most importantly, the Np--O bond length becomes asymmetric: the internal Np$^{(1)}$--O$^{(2)}$ bond length significantly increases compared to the Np--O distance in the monomer, while the terminal Np$^{(1)}$--O$^{(1)}$ bond length remains almost unchanged. For [NpO$_2$]$_2^{2+}$, the CCIs result in increased external Np$^{(1)}$--O$^{(1)}$ bonds when the neptunyl clusters are dissolved in aqueous solution. The opposite behaviour can be observed for the monomeric building blocks.~\cite{swart2005} The presence of solution not only shortens the distance between the oxygen atom of one unit and the neptunium atom of the second unit, but also reduces the angle between this intermolecular bond and the internal arm of the second unit ($\angle\mathrm{Np}^{(1)}\mathrm{-O}^{(2)}-\mathrm{Np}^{(2)}$). The structural properties of the diamond-shaped clusters are only weakly affected by the spin state. When the multiplicity rises, the Np$^{(2)}$--O$^{(2)}$ bond length as well as the O$^{(1)}$--Np$^{(1)}$--O$^{(2)}$ and Np$^{(1)}$--O$^{(2)}$--Np$^{(2)}$ bond angles increase. Specifically, the differences in interatomic distances amount to approximately 0.05 \AA{}, while the differences in bond angles are only $2^\circ$. The distance between the cationic building blocks in the CCI, Np$^{(1)}$--O$^{(3)}$ and Np$^{(2)}$--O$^{(2)}$, is similar to the CCI distance in the T-shaped structures and in agreement with the experimentally measured Np--O bond length in crystalline structures. The diamond-shaped cluster, however, features a much shorter Np--Np bond, which ranges from 3.357 to 3.637 \AA{} in solution compared to the experimentally observed value of 4.2 \AA{}.

Furthermore, a diamond-shaped cluster has been reported in crystalline K[(NpO$_2$)(OH)$_2$]$\dot{}$H$_2$O.~\cite{laura2013} The Np--O bond lengths in monomeric NpO$_2$ are 1.86 and 1.91 \AA{}, while the distance between two Np atoms is 3.5398 \AA{}. Both distances are larger than the theoretically predicted bond lengths. The bond angle (179.2$^\circ$) shows a small deflection from linearity. The interaction between the nearest monomers is characterized by a Np--O distance of 2.382 \AA{}, which is similar to the bond lengths optimized in vacuo and using an implicit solvation model. Note, however, the crystalline {structure is measured} in the solid state, while our study focuses on gas phase and solvated clusters.

\begin{table} [ht] 
\small
\caption{Spin-state energetics and total spin expectation values of the diamond-shaped [NpO$_2$]$_2^{2+}$ and [NpO$_2$]$_2^{3+}$ calculated with DFT and COSMO. $N_{\textrm{un}}$ denotes the number of unpaired electrons. The energy differences are calculated with respect to the lowest lying spin state of [NpO$_2$]$_2^{2+}$ and [NpO$_2$]$_2^{3+}$, respectively.}
\label{tab:DFTEnergiesDiamondCosmo}
\begin{tabular}{ccccc }
\hline
  & \textbf{Compound} & \textbf{Energy [cm$^{-1}$]} &  \textbf{S$^2$} & \textbf{Ideal S$^2$}  \\
\hline
& [(NpO$_{2})_{2}$]$_{[N_{\textrm{un}}=0]}^{2+}$        & 14 655 & 0.00 & 0.00 \\
&[(NpO$_{2})_{2}$]$_{[N_{\textrm{un}}=2]}^{2+}$         & 6 191 & 2.99  & 2.00 \\ \vspace{1mm}
Solution &[(NpO$_{2})_{2}$]$_{[N_{\textrm{un}}=4]}^{2+}    $ & 0      & 6.03  &  6.00 \\
&[(NpO$_{2})_{2}$]$_{[N_{\textrm{un}}=1]}^{3+}$         & 70 300 & 1.75 &  0.75 \\\vspace{3mm}
&[(NpO$_{2})_{2}$]$_{[N_{\textrm{un}}=3]}^{3+}$         & 0 & 3.77 & 3.75 \\
& [(NpO$_{2})_{2}$]$_{[N_{\textrm{un}}=0]}^{2+}$        & 14 375 & 0.00 & 0.00 \\
Vacuum &[(NpO$_{2})_{2}$]$_{[N_{\textrm{un}}=2]}^{2+}$  & 5 981 & 3.00 & 2.00 \\ 
&[(NpO$_{2})_{2}$]$_{[N_{\textrm{un}}=4]}^{2+} $        & 0 & 6.03 & 6.00 \\
 \hline
\end{tabular} 
\end{table}

\begin{table*}[h] %
\scriptsize
\caption{Energetics of the spin-free CASSCF wave functions with and without CASPT2 corrections for the diamond-shaped [NpO$_2$]$_2^{2+}$ and [NpO$_2$]$_2^{3+}$ molecules. PT2 corrections were performed on top of the CASSCF wave function optimized for the DKH Hamiltonian. The value of 0 cm$^{-1}$ is assigned to the lowest energy in each column.}
\label{tab:cas-energies-d}
\begin{tabular*}{\textwidth}{@{\extracolsep{\fill}} cccccc c ccc c cc}
\hline
 \linespread{1.5} 
	&	\multicolumn{5}{c}{[$\Delta$E [cm$^{-1}$] for NpO$_2$]$_2^{2+}$}									& & 	\multicolumn{3}{c}{{$\Delta$E [cm$^{-1}$] for [NpO$_2$]$_2^{2+}$+8H$_2$O}}					& & 	\multicolumn{2}{c}{$\Delta$E [cm$^{-1}$] for [NpO$_2$]$_2^{3+}$}			\\	\cline{2-6}	\cline{8-10}	\cline{12-13}	\\[-0.9em]
	&	CAS(8,12)SCF	&	CAS(8,12)SCF	&	CAS(8,12)PT2	&	CAS(4,8)SCF	&	CAS(4,8)PT2	& & 		&	CAS(8,12)SCF	&	CAS(8,12)SCF	& & 			CAS(8,12)SCF	\\				
	&	ECP	&	DKH2	&	DKH2	&	DKH2	&	DKH2	& & 		&	DKH2	&	DKH2	& & 			ECP	\\	\hline			
1$^5$A$_g$	&	0	&	0	&	12	&	0	&	11	& & 	1$^5$A	&	0	&	15	& & 	1$^4$A$_g$	&	1627	\\				
2$^5$A$_g$	&	5	&	10	&	30	&	13	&	31	& & 	2$^5$A	&	19	&	134	& & 	2$^4$A$_g$	&	4218	\\				
1$^5$A$_u$	&	1	&	20	&	63	&	23	&	116	& & 	3$^5$A	&	19	&	162	& & 	1$^4$A$_u$	&	2220	\\				
2$^5$A$_u$	&	3482	&	3238	&	2459	&	3220	&	2519	& & 	4$^5$A	&	38	&	169	& & 	2$^4$A$_u$	&	4216	\\				
1$^5$B$_u$	&	3478	&	3467	&	2480	&	3455	&	2564	& & 	5$^5$A	&	2478	&	1765	& & 	1$^4$B$_u$	&	0	\\				
2$^5$B$_u$	&	7536	&	6569	&	4258	&	6588	&	4338	& & 	6$^5$A	&	2481	&	1827	& & 	2$^4$B$_u$	&	2230	\\				
1$^5$B$_g$	&	4	&	22	&	53	&	25	&	107	& & 	7$^5$A	&	2499	&	1830	& & 	1$^4$B$_g$	&	1	\\				
2$^5$B$_g$	&	3480	&	3234	&	2459	&	3216	&	2513	& & 	8$^5$A	&	2500	&	1875	& & 	2$^4$B$_g$	&	1635	\\				
1$^3$A$_g$	&		&	3474	&	2462	&	3462	&	2561	& & 	1$^3$A	&	4	&	0	& & 		&		\\				
2$^3$A$_g$	&		&	6579	&	4408	&	6599	&	4482	& & 	2$^3$A	&	24	&	124	& & 		&		\\				
1$^3$A$_u$	&		&	28	&	38	&	32	&	99	& & 	3$^3$A	&	24	&	145	& & 		&		\\				
2$^3$A$_u$	&		&	3242	&	2452	&	3226	&	2514	& & 	4$^3$A	&	43	&	160	& & 		&		\\				
1$^3$B$_u$	&		&	7	&	0	&	8	&	0	& & 	5$^3$A	&	2482	&	1761	& & 		&		\\				
2$^3$B$_u$	&		&	16	&	13	&	21	&	11	& & 	6$^3$A	&	2484	&	1810	& & 		&		\\				
1$^3$B$_g$	&		&	28	&	40	&	32	&	102	& & 	7$^3$A	&	2503	&	1816	& & 		&		\\				
2$^3$B$_g$	&		&	3243	&	2447	&	3226	&	2517	& & 	8$^3$A	&	2503	&	1862	& & 		&		\\				
\hline
\end{tabular*}
\end{table*}

In contrast to diamond-shaped [NpO$_2$]$_2^{2+}$, the equilibrium structure of the [NpO$_2$]$_2^{3+}$ cluster could be optimized only if the solvent is considered in the theoretical model. This can be done either implicitly using COSMO or explicitly by considering the first solvation shell. This observation suggest that the [NpO$_2$]$_2^{3+}$ dication is formed in solution, but not in vacuo.
The interior bond length of the neptunyl subunits decreases compared to the 2+ charged system. This compound features a slightly longer Np$^{(2)}$--O$^{(2)}$ bond length. The qualitative impact of the number of unpaired electrons on the CCIs is similar as in [NpO$_2$]$_2^{2+}$. For low-spin states, the Np$^{(2)}$ and O$^{(2)}$ atoms come closer to each other, while the O$^{(1)}$--Np$^{(2)}$--O$^{(2)}$ and O$^{(1)}$--Np$^{(1)}$--O$^{(2)}$ angles decrease.
The equilibrium Np--O distance between the neptunyl units is 2.349--2.406 \AA{} for the {state with $N_{\textrm{un}}$=1} state and 2.453 \AA{} for the {state with $N_{\textrm{un}}$=3}. Our results {are similar to} the Np(V)--O$_{\mathrm{Np(VI)}}$ distance observed~\cite{exp_cci_np5_np6} in the mixed valence (Np(VI)O$_2$Cl$_2$)(Np(V)O$_2$Cl(thf)$_3$)$_2$ cluster, where the interaction distances are 2.317 and 2.303 \AA{}. The observed differences might be due to the presence of different solvents and counter ions used in experiments.

Spin-state energetics and total spin expectation values of [NpO$_2$]$_2^{2+}$ predicted by DFT/COSMO are shown in Table \ref{tab:DFTEnergiesDiamondCosmo}, together with the ideal \textbf{S}$^2$ expectation values.
Most importantly, the environment does not affect spin state energetics considerably (differences in relative energies amount to 280 cm$^{-1}$). 
For both gas phase and solution, the {state with $N_{\textrm{un}}$=4} of diamond-shaped [NpO$_2$]$_2^{2+}$ is the ground state, while the energy differences with respect to the {state with $N_{\textrm{un}}$=4} increase for a decreasing number of unpaired electrons (see Table \ref{tab:DFTEnergiesDiamondCosmo}).
For [NpO$_2$]$_2^{3+}$, the {state with $N_{\textrm{un}}$=1} lies about 70 300 cm$^{-1}$ higher in energy than the {state with $N_{\textrm{un}}$=3}. We should note that the {state with $N_{\textrm{un}}$=1} of [NpO$_2$]$_2^{3+}$ and the {state with $N_{\textrm{un}}$=2} of [NpO$_2$]$_2^{2+}$ are significantly spin-contaminated. Since, however, both states lie above each corresponding ground state, spin-contamination does not pose a problem in the determination of the ground state in [NpO$_2$]$_2^{2+}$ and [NpO$_2$]$_2^{3+}$, respectively.
Thus, our analysis of spin-state energetics suggest that the {state with $N_{\textrm{un}}$=4} [NpO$_2$]$_2^{2+}$ and {$N_{\textrm{un}}$=3} [NpO$_2$]$_2^{3+}$ may be considered as ground states.

A Mulliken population analysis can be found in Table~\ref{tab:pop}. 
Similar to the T-shaped cluster, a positive charge is located on the neptunium atoms, while the oxygen atoms are negatively charged.
An exception is the {state with $N_{\textrm{un}}$=1} of [NpO$_2$]$_2^{2+}$, where the terminal oxygen atom O$^{(1)}$/O$^{(4)}$ bears a slight positive charge.
As observed for the T-shaped isomer, the difference between the $\alpha$- and $\beta$-electron density is mostly centered on the Np atom and dominated by f-electrons. Thus, the unpaired $\alpha$-electrons are located mainly on the Np atoms.
Note that for the {state with $N_{\textrm{un}}$=2}[NpO$_2$]$_2^{2+}$ and {$N_{\textrm{un}}$=1} [NpO$_2$]$_2^{3+}$ the predicted spin population on the Np atom is much smaller than for the spin state of higher multiplicity.
These differences in {spin magnetization} and charge density on the neptunium centers and oxygen atoms, respectively, may originate from the spin contamination observed for the {states with $N_{\textrm{un}}$=1 and $N_{\textrm{un}}$=2}.
{A Bader analysis is presented in the ESI for comparison. As observed for the T-shaped cluster, it yields a higher charge concentration on the atomic centers compared to a Mulliken population analysis.}

The relative energies of {[NpO$_2$]$_2^{2+}$} (with and without explicit solvent) calculated using different exchange--correlation functionals (PBE0, PBE0-D3, B3LYP, and B3LYP-D3) are presented in Table S4 in the ESI.
If the solvent is modeled using COSMO, the diamond-shaped cluster is lower in energy than the T-shaped cluster of the same charge for all exchange--correlation functionals investigated (by approximately 6 to 9 kcal/mol).
If the solvent is included explicitly, hybrid functionals predict the T-shaped CCI to be energetically lower than the diamond-shaped isomer (about 15 to 20 kcal/mol), while for BP86 the diamond-shaped cluster is lower in energy by approximately 9 kcal/mol.
We should emphasize that a D3 correction does not significantly affect the relative energies of the CCI isomers (differences are between 1 to 4 kcal/mol).
Experimental studies~\cite{guillaume1983} suggest that the T-shaped dimer dominates in solution.
Thus, hybrid functionals as well as an explicit treatment of the first solvation shell have to be considered in order to accurately predict the relative stability of the CCI structural isomers.

Finally, the inter-atomic distances in the diamond-shaped clusters are very similar to the corresponding uranyl(V)--uranyl(V) and uranyl(V)--uranyl(VI) structures optimized using DFT/COSMO.\mbox{~\cite{tecmer-song2016}} All internal Np--O bonds are slightly shorter (up to 0.06 \AA), while the Np--O distance between two monomers are longer (up to 0.12 \AA). The cation--cation interactions preserve the linearity in both the neptunyl monomer and the uranyl cations. For [NpO$_2$]$_2^{2+}$, we observe a different ordering of spin states as the {state with $N_{\textrm{un}}$=2} has the lowest energy in uranyl CCIs, followed by the corresponding states with $N_{\textrm{un}}$=0 and $N_{\textrm{un}}$=2. Both clusters with charge $3+$ have the same order of states.

\subsubsection{Molecular structures with an explicit first coordination sphere}

Similar to the T-shaped CCIs, we included a different number of water molecules (four to eight) to model the first solvation shell (see Table S2 in the ESI for a complete picture). In general, the structural parameters {of the monomers} are only slightly affected by the number of water molecules.
Specifically, the neptunyl dimers with a complete first coordination sphere (\textit{e.g.}, eight water molecules) were investigated in two configurations, one in $C_{2\mathrm{h}}$ symmetry and one in $C_\mathrm{i}$. In [(NpO$_2$)$_2$]$_{[N_{\textrm{un}}=5]}^{2+}$, all non-CCI Np--O bonds are almost unaffected by explicitly including the first coordination sphere as the differences in bond lengths amount to 0.05 \AA{}, which is less than 3$\%$ of a total bond length. We further observe that the ligated water molecules slightly reduce the bending in the monomers. The Np$^{(1)}$--O$^{(3)}$ distance elongates if the complete explicit first coordination sphere is included into the model. This bond length exceeds the values obtained in vacuo.
The [(NpO$_2$)$_2$]$_{[N_{\textrm{un}}=4]}^{3+}$ molecule is more sensitive to solvation effects and hence changes in bond lengths are more pronounced. However, the qualitative trends are the same as for [(NpO$_2$)$_2$]$_{[N_{\textrm{un}}=5]}^{2+}$. The ``yl'' bond distances are elongated up to 0.09 \AA{}. The linearity of the monomers is slightly distorted (up to 5$^\circ$) in the compound with eight surrounding water molecules in the first coordination sphere.

\subsubsection{Correlation in neptunyl(V)--neptunyl(V)}

Figure \ref{fig:decay}(b) shows the decaying values of the orbital-pair mutual information (obtained using the DMRG algorithm) for the diamond-shaped neptunyl(V)--neptunyl(V) cluster, while the orbital-pair correlation diagram is presented in Figure \ref{fig:i12-t2-d2}. The two largest values of the mutual information for the diamond-shaped species ($I_{i|j} = 0.18 $) are noticeably smaller than observed in the T-shaped molecule and show up for the bonding and antibonding $\sigma_g$-type orbitals (5a$_g$/8a$_g$ and 5b$_u$/8b$_u$) of the same irreducible representation.
Furthermore, we do not observe a gap in the mutual information as in the T-shaped cluster and $I_{i|j}$ decays rather gradually.
The orbital-pair correlations in decreasing order correspond to (a) the strong correlation between the $\phi_u$-type orbitals (6a$_g$, 4a$_u$, 6b$_u$, and 4b$_g$) and between the $\delta_u$-type orbitals (7a$_g$, 5a$_u$, 7b$_u$, and 5b$_g$), (b) correlations between $\delta_u$- and $\phi_u$-type orbitals of the same irreducible representation, (c) correlations between $\delta_u$/$\delta_u$-, $\phi_u$/$\phi_u$-, $\delta_u$/$\phi_u$-, and $\sigma_g$/$\sigma_g^*$-type orbital-pairs of different irreducible representations, and (d) moderately correlated $\pi_u$- and $\pi_u^*$-type orbitals ($I_{i|j} < 10^{-1}$).
A balanced active space that describes mostly nondynamic correlation should include primarily the most relevant $\delta_u$-, $\phi_u$-, $\sigma_g$-, and $\sigma_g^*$-type orbitals, and preferably the $\pi_u$- and $\pi_u^*$-type orbitals.

\subsubsection{Orbital-pair correlations in neptunyl(V)--neptunyl(VI)}

The decaying values of the orbital-pair mutual information for [NpO$_2$]$_2^{3+}$ are presented in Figure \ref{fig:decay}(c), while the orbital-pair correlation diagram is presented in the ESI. 
In contrast to the neptunyl(V)--neptunyl(V) cluster, the number of orbital-pairs that are strongly correlated ($I_{i|j} > 10^{-1}$) decreases, while the number of moderately correlated orbital pairs increases. 
Furthermore, the strong correlation between the $\sigma_g$/$\sigma_g^*$-type (5a$_g$/8a$_g$ and 5b$_u$/8b$_u$) orbitals is present in both diamond-shaped CCIs. {Their importance for static correlation has also been emphasized in transition state calculations.\mbox{~\cite{laura2013}}}
Although $\delta_u$- (6a$_g$, 4a$_u$, 6b$_u$, and 4b$_g$) and $\phi_u$-type (7a$_g$, 5a$_u$, 7b$_u$, and 5b$_g$) orbitals are moderately correlated with each other, the corresponding values of the mutual information mostly do not pass the threshold of $I_{i|j} = 10^{-1}$. 
Among the strongest orbital-pair correlations, we encounter correlations between $\pi_u$-type orbitals and $\pi_u^*$-type orbitals as well as other diffused orbitals (see the ESI for more details). 
If we consider only orbitals that are important for nondynamic correlation ($I_{i|j} > 10^{-1}$), the corresponding active space should contain $\delta_u$-, $\phi_u$-, $\sigma_g$-, $\sigma_g^*$-, $\pi_u$-, and $\pi_u^*$-type orbitals. 
However, due to the large number of moderately correlated orbitals, a balanced active space for [NpO$_2$]$_2^{3+}$ requires more than 30 orbitals that is computationally infeasible within the CASSCF approach.
Due to convergence difficulties, we only included $\delta_u$-, $\phi_u$-, $\sigma_g$-, and $\sigma_g^*$-type orbitals in our CASSCF calculations.

\subsubsection{Electronic structures of the [NpO$_2$]$_2^{2+}$ and [NpO$_2$]$_2^{3+}$ from CASSCF calculations}

For [NpO$_2$]$_2^{2+}$, the CASSCF+ECP wave functions {are optimized to minimize the average energy of the} 1$^5$A$_\mathrm{g}$, 2$^5$A$_\mathrm{g}$, 1$^5$A$_\mathrm{u}$, 2$^5$A$_\mathrm{u}$, 1$^5$B$_\mathrm{u}$, 2$^5$B$_\mathrm{u}$, 1$^5$B$_\mathrm{g}$, and 2$^5$B$_\mathrm{g}$ states.
The relative energies of the corresponding states are presented in Table \ref{tab:cas-energies-d}. 
As the 1$^5$A$_\mathrm{g}$ is lowest in energy, both DFT and CASSCF predict the  $^5$A$_\mathrm{g}$ state to be the ground state. 
Note, however, that the energy difference between the 2$^5$A$_\mathrm{g}$, 1$^5$A$_\mathrm{u}$, and 1$^5$B$_\mathrm{g}$ states is lower than 5 cm$^{-1}$ and thus these states are quasi-degenerate with the ground state.
The dominant ground-state electronic configuration is $| \sigma_g^{2} \delta_u^2 \phi_u^2 \rangle $, where electrons are equally distributed into $\delta_u$- and $\phi_u$-type orbitals.
The $\sigma_g$ orbitals are almost doubly occupied, while the average occupation of the $\sigma_g^*$  orbitals is close to zero. 
The occupation numbers of the $\delta_u$-type orbitals are slightly above 0.5, while those of $\phi_u$-type orbitals are slightly below 0.5. 
{The active space orbitals are non-bonding. Within a single-determinant approximation, these orbitals are the highest occupied and the lowest unoccupied orbitals, which explains the fact that the {molecular geometries} of the [NpO$_2$]$_2^{2+}$ and [NpO$_2$]$_2^{3+}$ clusters are rather similar in DFT/GGA calculations.}

The energies of the states calculated using the CAS(8,12)SCF+DKH2 and CAS(4,8)SCF+DKH2 methods are also presented in Table \ref{tab:cas-energies-d}, while the most important composite determinants are collected in the ESI.
Our CAS(8,12)SCF+DKH2 and CAS(4,8)SCF+DKH2 calculations are characterized by consistent ordering of the states, however, they differ from the CAS(8,12)SCF+ECP ordering. 
For all these methods, the 1$^5$A$_\mathrm{g}$ is the ground state.
The relative energies of the lowest-lying {triplet states ($N_{\textrm{un}}$=2)} are close to the ground state energy, which contradicts our DFT results where the energy difference between the {states with $N_{\textrm{un}}$=2 and $N_{\textrm{un}}$=4} is approximately three orders of magnitude larger.
Furthermore, the CI expansion of the {triplet states ($N_{\textrm{un}}$=2)} features more Slater determinants with smaller weights ($|\mathrm{CI}| \approx 0.2 $), while the {quintet states ($N_{\textrm{un}}$=4)} comprise Slater determinants with intermediate to large weights ($|\mathrm{CI}| > 0.35 $).
After inclusion of dynamical correlation with CASPT2, the relative energies and the ordering of the states changes for both the minimal CAS(4,8) and the extended CAS(8,12) active spaces (see also Table \ref{tab:cas-energies-d}). Both active spaces indicate that the ground state is a $^3$B$_\mathrm{u}$ ({$N_{\textrm{un}}$=2}) state and is composed of many small-weighted determinants (cf. ESI).

The relative energies of different CI vectors of the diamond-shaped [NpO$_2$]$_2^{3+}$ calculated using CAS(8,12)+ECP is summarized in Table \ref{tab:cas-energies-d}. {We should stress that this CCI cluster is stable only in solution, where the cluster does not break apart. However, in our CAS(8,12)+ECP calculations, we only consider the gas-phase model. The corresponding CASSCF and CASPT2/SO-RASSI calculations for the hydrated cluster resulted in broken-symmetry solutions.}
Most importantly, the 1$^4$B$_\mathrm{u}$ state is lowest in energy, followed by a quasi-degenerate 1$^4$B$_\mathrm{g}$ state.
The most {significant} configurations include $| \sigma_g^{2} \phi_u^2  \delta_u^1 \rangle $-type determinants. The occupation numbers of the state-averaged ground-state wave function are presented in the ESI.

\begin{table*}[h] %
\small
\caption{Energetics of the diamond-shaped [NpO$_2$]$_2^{2+}$ + 8H$_2$O molecule calculated with CASPT2/SO-RASSI. For each molecule, the energy differences are calculated with respect to the lowest lying state. The first columns contains labels assigned to the states. All calculated states are presented in the ESI.}
\label{tab:so-energies-d2}
\begin{tabular*}{1\textwidth}{@{\extracolsep{\fill}}lll lll lll lll} \hline
& & \multicolumn{10}{c}{\textbf{Composition in terms of the spin-free wave functions}}\\
State no.& \textbf{$\Delta$E [cm$^{-1}$]} & \multicolumn{10}{c}{\textbf{(weight in \%)}}\\ \hline
 1  &  0  &  2$^5$A  &  (37)  &  4$^5$A  &  (36)  &  1$^5$A  &  (23)  &  8$^5$A  &  (1)  &    &   \\  
 3  &  4856  &  3$^5$A  &  (44)  &  1$^3$A  &  (22)  &  2$^3$A  &  (10)  &  4$^3$A  &  (10)  &  3$^3$A  &  (4) \\  
 5  &  5695  &  1$^3$A  &  (46)  &  3$^5$A  &  (20)  &  2$^3$A  &  (19)  &  4$^3$A  &  (9)  &  8$^3$A  &  (1) \\  
 9  &  5747  &  4$^3$A  &  (39)  &  2$^3$A  &  (28)  &  3$^5$A  &  (27)  &  1$^3$A  &  (1)  &    &   \\  
 13  &  7331  &  8$^5$A  &  (42)  &  7$^5$A  &  (40)  &  1$^5$A  &  (7)  &  6$^5$A  &  (6)  &  2$^5$A  &  (1) \\  
 15  &  7353  &  5$^5$A  &  (48)  &  6$^5$A  &  (40)  &  7$^5$A  &  (6)  &  3$^5$A  &  (2)  &  7$^3$A  &  (1) \\  
 17  &  7418  &  5$^5$A  &  (25)  &  8$^3$A  &  (24)  &  6$^5$A  &  (23)  &  7$^3$A  &  (23)  &  7$^5$A  &  (2) \\  
 21  &  9718  &  8$^3$A  &  (33)  &  6$^5$A  &  (32)  &  5$^5$A  &  (10)  &  1$^3$A  &  (9)  &  7$^3$A  &  (6) \\  
 23  &  9764  &  7$^3$A  &  (37)  &  5$^5$A  &  (37)  &  6$^5$A  &  (6)  &  8$^3$A  &  (6)  &  7$^5$A  &  (5) \\  
 25  &  11537  &  4$^5$A  &  (32)  &  4$^3$A  &  (30)  &  2$^3$A  &  (16)  &  2$^5$A  &  (11)  &  1$^5$A  &  (2) \\  
 29  &  11551  &  1$^5$A  &  (41)  &  4$^5$A  &  (27)  &  2$^5$A  &  (22)  &  8$^5$A  &  (7)  &    &   \\  
 \hline
\end{tabular*}
\end{table*}

\subsubsection{Calculated excited states including spin-orbit corrections}

Including spin-orbit coupling using the configuration interaction approach with CASSCF wave functions results in 64 excited states that are presented in the ESI. Table \ref{tab:so-energies-d2} contains some of those states (the quasi-degenerate states have been omitted). {We should emphasize that the model includes static correlation effects (CASSCF), a dynamic correlation correction (CASPT2), spin-orbit effects (SO-RASSI), and solvation effects (eight explicit water molecules surrounding the neptunyl--neptunyl cluster). The corresponding spectra for the bare dications can be found in the ESI. Note that the differences in electronic spectra between the gas-phase model and the hydrated cluster are negligible for the lower part of the spectrum (typically much smaller than 1000 cm$^{-1}$/0.1 eV), which is to be expected as those excitations feature atomic f--f transitions.}
All calculated states consist of more than one component and are degenerate or nearly-degenerate.
The doubly-degenerate ground states contains only quintet ($N_{\textrm{un}}$=4) components, while the excited states include both {triplet ($N_{\textrm{un}}$=2) and quintet ($N_{\textrm{un}}$=4) states}.
One similar energy, 9 700 cm$^{-1}$, was calculated for bare neptunyl(VI).
The transitions between excited states feature larger intensities than transitions from the ground states.

\begin{table}[h] %
\small
\caption{{Binding energies of neptunyl clusters with respect to the monomers. $\Delta$E$_u$ denotes the difference between the energy of the dimer and monomers for the unrelaxed geometry (unrelaxed binding energy). $\Delta$E$_p$ stands for the difference between the energies of the unrelaxed monomers with respect to the energies of the relaxed monomers (preparation energy). $\Delta$E is the binding energy calculated as the sum of $\Delta$E$_u$ and $\Delta$E$_p$ (relaxed binding energy). All energies are given in kcal/mol.}}
\label{tab:be}
\begin{tabular*}{0.48\textwidth}{@{\extracolsep{\fill}}lllll} \hline
	&		&	$\Delta$E$_u$	&	$\Delta$E$_p$	&	$\Delta$E\\\hline
~T-shaped \\
~(NpO$_2$)$_2^{2+}$ + 9 H$_2$O	&	PBE0	&	-11.0	&	~27.4	&	~16.4	\\
&	PBE0-D3	&	-14.3	&	~27.0	&	~12.7	\\
~(NpO$_2$)$_2^{2+}$ &	CASSCF	&	~42.3	&	~7.1	&	~49.4	\\
&	CASPT2	&	~45.1	&	~4.4	&	~49.6	\\
&	CASPT2/SO-RASSI	&	~49.6	&	~4.2	&	~53.8	\\
~(NpO$_2$)$_2^{2+}$ + 9 H$_2$O	&	CASSCF	&	~9.2	&	~31.5	&	~40.6	\\
&	CASPT2	&	~9.9	&	~28.9	&	~38.8	\\
&	CASPT2/SO-RASSI	&	~15.0	&	~28.8	&	~43.9	\\\hline
~diamond-shaped \\
~(NpO$_2$)$_2^{2+}$ + 8 H$_2$O	&	PBE0	&	-27.1	&	~19.2	&	-7.9	\\
&	PBE0-D3	&	-32.1	&	~18.2	&	-13.9	\\
~(NpO$_2$)$_2^{2+}$ &	CASSCF	&	~27.2	&	~20.2	&	~47.4	\\
&	CASPT2	&	~34.8	&	~9.3	&	~44.1	\\
&	CASPT2/SO-RASSI	&	~36.7	&	~9.3	&	~46.0	\\
~(NpO$_2$)$_2^{2+}$ + 8 H$_2$O	&	CASSCF	&	-1.6	&	~40.6	&	~38.9	\\
&	CASPT2	&	-10.9	&	~29.0	&	~18.1	\\
&	CASPT2/SO-RASSI	&	-7.5	&	~28.8	&	~21.4	\\
 \hline
\end{tabular*}
\end{table}

\subsection{Stability of clusters with respect to the monomers}

Table\mbox{~\ref{tab:be}} presents the relaxed and unrelaxed binding energies as well as preparation energies of [NpO$_2$]$_2^{2+}$ (both diamond and T-shaped). The unrelaxed binding energies $\Delta E_u$ suggest that the diamond-shaped cluster is more stable than the T-shaped structure as the corresponding energy differences are (more) negative for the diamond-shaped conformation. 
Furthermore, DFT calculations provide negative (relaxed) binding energies $\Delta E$ for the diamond-shaped and positive binding energies for the T-shaped structure suggesting that the latter is unstable in water. The inclusion of dispersion forces through a D3 correction slightly strengthen the CCI interaction (3--6 kcal/mol), but its influence is rather small with respect to the energy difference between the T-shaped and diamond-shaped structures (24--31 kcal/mol).
Wave-function-based methods (CASSCF, CASPT2, and CASPT2/SO-RASSI) yield positive (relaxed) binding energies $\Delta E$ for all structures indicating that our dication models might be too simplistic to capture all important chemical properties of the solvated CCIs. The inclusion of explicit water molecules into the CCI model lowers the binding energies significantly (8--26 kcal/mol), while spin-orbit corrections do not significantly affect the binding energies (2--5 kcal/mol).
This suggests that a second coordination sphere or an implicit solvation model might be crucial to obtain proper binding energies.
{Furthermore}, our CCI models might be further improved by enlarging the active orbital space with molecular orbitals centered on the water ligands, which would allow us to explicitly treat the correlation between the neptunyl and water molecules.
A different factor might arise from an insufficient treatment of dynamical correlation as the difference in binding energies in CASSCF and CASPT2 amounts up to 10 kcal/mol.
To conclude, large active space calculations for the hydrated clusters combined with a dynamic energy correction, like the DMRG-tailored coupled cluster method,\mbox{~\cite{dmrg-tcc}} might improve binding energies and provide a better description of the stability of CCI clusters.

Finally, an energy decomposition analysis (EDA) for both the hydrated and gas-phase CCIs calculated within DFT/COSMO is presented in the ESI.
Most importantly, both the EDA and the binding energies presented in Table\mbox{~\ref{tab:be}} show similar trends indicating that the diamond-shaped [NpO$_2$]$_2^{2+}$ cluster has a negative interaction energy, while the T-shaped [NpO$_2$]$_2^{2+}$ compound features a positive interaction energy.

\section{Conclusions}

The CCIs are an important structure-forming factor for many actinide compounds containing uranyl and neptunyl as simple building blocks.  
The presence of CCIs affects the characteristic UV-Vis and IR spectra as well as other properties of bare actinyl ions.  
Specifically for neptunyls, the CCIs are responsible for the formation of the diamond- and T-shaped [NpO$_2$]$_2^{2+}$ and [NpO$_2$]$_2^{3+}$ clusters.
In this work, we analyzed the properties of neptunyl(V)--neptunyl(V), neptunyl(V)--neptunyl(VI), and neptunyl(VI)--neptunyl(VI) clusters in different geometrical arrangements, spin-states, and environments.
{Although we found molecular geometries that minimize the energy of the neptunyl clusters, our energetic analysis of the complexes and monomeric building blocks does not confirm the stability of the dimers.}

{The theoretically predicted Np--Np distance of the T-shaped compound agrees well with experimental results if the water molecules are included in model. Our calculations suggest that the CCI are sensitive towards the surrounding, which agrees well with the observations made by Madic \textit{et al}.}~\cite{CCI_79} 
{The calculated f--f transitions in the electronic spectra of neptunyl clusters are similar to the calculated spectra of the bare neptunyl(V). This explains why only small changes in absorption spectra were observed by Guillaume \textit{et al}.}~\cite{guillaume1981}
{Although experiments focus on the T-shaped interactions, our study agrees with the conclusions by Vlaisavljevich \textit{et al.} that the diamond-shaped [NpO$_2$]$_2^{2+}$ displays stronger binding interactions in solution than its T-shaped structural isomer.}~\cite{laura2013}

The electronic structures of the stable neptunyl CCI clusters were studied within the CASSCF approach with two different scalar-relativistic Hamiltonians: (a) effective core potentials (ECP) and (b) the second order Douglas-Kroll-Hess Hamiltonian (DKH2).
We found that the order of the states and the symmetry of the ground state depends on the choice of the (scalar) relativistic Hamiltonian.
The ground-state structure of the electronic wave function highlights the strong multi-reference nature of the diamond-shaped [NpO$_2$]$_2^{2+}$ cluster, where the $\delta_u$-type and $\phi_u$-type orbitals are quasi-degenerate. 
The \mbox{Np--O} bond in the NpO$_2^{+}$ monomer has a completely different nature than the CCIs. The main contribution to the chemical bond in the bare complex is made by 6p$_{\sigma}$/5f$_{\sigma}$ hybrid orbitals as well as 6d$_{\pi}$, while the correlation measurements show that 5f$_{\phi}$ orbitals play an important role in CCIs of actinide compounds.~\cite{clark2004}
The lowest-lying excited states in all studied clusters are mostly degenerate or quasi-degenerate.
Furthermore, our study suggests that the ground-state of the investigated clusters is strongly affected by both electron correlation effects and spin-orbit coupling. 
Specifically, accounting for dynamical correlation using a CASPT2 correction changes the ground-state from a {quintet state ($N_{\textrm{un}}$=4)} to a {triplet state ($N_{\textrm{un}}$=2)}. Inclusion of spin-orbit coupling entails mixing between {triplet ($N_{\textrm{un}}$=2) and quintet ($N_{\textrm{un}}$=4) states }for the T-shaped CCI, while the diamond-shaped CCI features only {quintet ($N_{\textrm{un}}$=4) states}.

Although the active spaces of all investigated molecules contained similar orbitals, an orbital correlation analysis highlights different correlations between those active space orbitals in different CCI clusters. 
In the T-shaped compound, the $\delta_u$/$\phi_u$-type orbitals dominate the static/nondynamic correlation picture.
In the diamond-shaped [NpO$_2$]$_2^{2+}$, the $\sigma_g$- and $\sigma_g^*$-type orbitals are as important as $\delta_u$- and $\phi_u$-type orbitals.
Reducing the number of electrons increases the significance of diffused and $\pi_u$-type orbitals and simultaneously decreases the correlation between $\delta_u$- and $\phi_u$-type orbitals.
Finally, our orbital-correlation analysis suggests that a balanced active space for neptunyl-containing CCIs that allows us to describe both nondynamic and static correlation should contain approximately 30 orbitals ($\delta_u$-, $\phi_u$-, bonding and antibonding combinations of $\sigma_g$-, $\sigma_u$-, $\pi_u$-, and $\pi_g$-type orbitals of each monomer).
However, such large active spaces are computationally infeasible for conventional electronic structure methods like CASSCF.
Furthermore, the need to include additional water molecules in the structural model impedes the routine application of CASSCF as some of the oxygen orbitals of the water molecule have to be considered in active space calculations.
A four-component DMRG \cite{dmrg_relativistic_1, dmrg_relativistic_2} study of all presented CCIs that includes all strongly and moderately correlated orbitals as well as explicit water molecules is currently under investigation in our laboratory.

\section{Acknowledgment}
A.~\L.~and K.~B.~acknowledge financial support from the National Science Centre, Poland (SONATA BIS 5 Grant No.~2015/18/E/ST4/00584).
K.~B.~gratefully acknowledges funding from a Marie-Sk\l{}odowska-Curie Individual Fellowship project no.~702635--PCCDX and a scholarship for outstanding young scientists from the Ministry of Science and Higher Education.
P.~T.~thanks the POLONEZ fellowship program of the National Science Center (Poland), No. 2015/19/P/ST4/02480.
This project had received funding from the European Union's Horizon 2020 research and innovation programme under the Marie Sk{\l}odowska--Curie grant agreement No. 665778.
\"{O}.~L.~thanks the Hungarian National Research, Development and Innovation Office (grant no.  K120569), and the Hungarian Quantum Technology National Excellence Program (project no. 2017-1.2.1-NKP-2017-00001). 
Calculations have been carried out using resources provided by Wroclaw Centre for Networking and Supercomputing (http://wcss.pl), Grant No.~412.


\normalem
\bibliography{rsc}

\begin{thebibliography}{142}%
\makeatletter
\providecommand \@ifxundefined [1]{%
 \@ifx{#1\undefined}
}%
\providecommand \@ifnum [1]{%
 \ifnum #1\expandafter \@firstoftwo
 \else \expandafter \@secondoftwo
 \fi
}%
\providecommand \@ifx [1]{%
 \ifx #1\expandafter \@firstoftwo
 \else \expandafter \@secondoftwo
 \fi
}%
\providecommand \natexlab [1]{#1}%
\providecommand \enquote  [1]{``#1''}%
\providecommand \bibnamefont  [1]{#1}%
\providecommand \bibfnamefont [1]{#1}%
\providecommand \citenamefont [1]{#1}%
\providecommand \href@noop [0]{\@secondoftwo}%
\providecommand \href [0]{\begingroup \@sanitize@url \@href}%
\providecommand \@href[1]{\@@startlink{#1}\@@href}%
\providecommand \@@href[1]{\endgroup#1\@@endlink}%
\providecommand \@sanitize@url [0]{\catcode `\\12\catcode `\$12\catcode
  `\&12\catcode `\#12\catcode `\^12\catcode `\_12\catcode `\%12\relax}%
\providecommand \@@startlink[1]{}%
\providecommand \@@endlink[0]{}%
\providecommand \url  [0]{\begingroup\@sanitize@url \@url }%
\providecommand \@url [1]{\endgroup\@href {#1}{\urlprefix }}%
\providecommand \urlprefix  [0]{URL }%
\providecommand \Eprint [0]{\href }%
\providecommand \doibase [0]{http://dx.doi.org/}%
\providecommand \selectlanguage [0]{\@gobble}%
\providecommand \bibinfo  [0]{\@secondoftwo}%
\providecommand \bibfield  [0]{\@secondoftwo}%
\providecommand \translation [1]{[#1]}%
\providecommand \BibitemOpen [0]{}%
\providecommand \bibitemStop [0]{}%
\providecommand \bibitemNoStop [0]{.\EOS\space}%
\providecommand \EOS [0]{\spacefactor3000\relax}%
\providecommand \BibitemShut  [1]{\csname bibitem#1\endcsname}%
\let\auto@bib@innerbib\@empty
\bibitem [{\citenamefont {Sessler}\ \emph {et~al.}(2001)\citenamefont
  {Sessler}, \citenamefont {Seidel}, \citenamefont {Vivian}, \citenamefont
  {Lynch}, \citenamefont {Scott},\ and\ \citenamefont {Keogh}}]{sessler2001}%
  \BibitemOpen
  \bibfield  {author} {\bibinfo {author} {\bibfnamefont {J.~L.}\ \bibnamefont
  {Sessler}}, \bibinfo {author} {\bibfnamefont {D.}~\bibnamefont {Seidel}},
  \bibinfo {author} {\bibfnamefont {A.~E.}\ \bibnamefont {Vivian}}, \bibinfo
  {author} {\bibfnamefont {V.}~\bibnamefont {Lynch}}, \bibinfo {author}
  {\bibfnamefont {B.~L.}\ \bibnamefont {Scott}}, \ and\ \bibinfo {author}
  {\bibfnamefont {D.~W.}\ \bibnamefont {Keogh}},\ }\href@noop {} {\bibfield
  {journal} {\bibinfo  {journal} {Angew. Chem., Int. Ed.}\ }\textbf {\bibinfo
  {volume} {40}},\ \bibinfo {pages} {591} (\bibinfo {year} {2001})}\BibitemShut
  {NoStop}%
\bibitem [{\citenamefont {Tian}\ \emph {et~al.}(2005)\citenamefont {Tian},
  \citenamefont {Xu},\ and\ \citenamefont {Rao}}]{tian2005}%
  \BibitemOpen
  \bibfield  {author} {\bibinfo {author} {\bibfnamefont {G.~X.}\ \bibnamefont
  {Tian}}, \bibinfo {author} {\bibfnamefont {J.~D.}\ \bibnamefont {Xu}}, \ and\
  \bibinfo {author} {\bibfnamefont {L.~F.}\ \bibnamefont {Rao}},\ }\href@noop
  {} {\bibfield  {journal} {\bibinfo  {journal} {Angew. Chem., Int. Ed.}\ ,\
  \bibinfo {pages} {6200}} (\bibinfo {year} {2005})}\BibitemShut {NoStop}%
\bibitem [{\citenamefont {Tian}\ \emph {et~al.}(2007)\citenamefont {Tian},
  \citenamefont {Rao},\ and\ \citenamefont {Oliver}}]{tian2007}%
  \BibitemOpen
  \bibfield  {author} {\bibinfo {author} {\bibfnamefont {G.~X.}\ \bibnamefont
  {Tian}}, \bibinfo {author} {\bibfnamefont {L.~F.}\ \bibnamefont {Rao}}, \
  and\ \bibinfo {author} {\bibfnamefont {A.}~\bibnamefont {Oliver}},\
  }\href@noop {} {\bibfield  {journal} {\bibinfo  {journal} {Chem. Commun.}\
  }\textbf {\bibinfo {volume} {0}},\ \bibinfo {pages} {4119} (\bibinfo {year}
  {2007})}\BibitemShut {NoStop}%
\bibitem [{\citenamefont {Sarsfield}\ \emph {et~al.}(2007)\citenamefont
  {Sarsfield}, \citenamefont {Taylor},\ and\ \citenamefont
  {Maher}}]{neptunium-important-1}%
  \BibitemOpen
  \bibfield  {author} {\bibinfo {author} {\bibfnamefont {M.~J.}\ \bibnamefont
  {Sarsfield}}, \bibinfo {author} {\bibfnamefont {R.~J.}\ \bibnamefont
  {Taylor}}, \ and\ \bibinfo {author} {\bibfnamefont {C.~J.}\ \bibnamefont
  {Maher}},\ }\href@noop {} {\bibfield  {journal} {\bibinfo  {journal}
  {Radiochim. Acta}\ }\textbf {\bibinfo {volume} {95}},\ \bibinfo {pages} {677}
  (\bibinfo {year} {2007})}\BibitemShut {NoStop}%
\bibitem [{\citenamefont {Clark}\ \emph {et~al.}(1996)\citenamefont {Clark},
  \citenamefont {Conradson}, \citenamefont {Ekberg}, \citenamefont {Hess},
  \citenamefont {Neu}, \citenamefont {Palmer}, \citenamefont {Runde},\ and\
  \citenamefont {Tait}}]{neptunium-important-2}%
  \BibitemOpen
  \bibfield  {author} {\bibinfo {author} {\bibfnamefont {D.~L.}\ \bibnamefont
  {Clark}}, \bibinfo {author} {\bibfnamefont {S.~D.}\ \bibnamefont
  {Conradson}}, \bibinfo {author} {\bibfnamefont {S.~A.}\ \bibnamefont
  {Ekberg}}, \bibinfo {author} {\bibfnamefont {N.~J.}\ \bibnamefont {Hess}},
  \bibinfo {author} {\bibfnamefont {M.~P.}\ \bibnamefont {Neu}}, \bibinfo
  {author} {\bibfnamefont {P.~D.}\ \bibnamefont {Palmer}}, \bibinfo {author}
  {\bibfnamefont {W.}~\bibnamefont {Runde}}, \ and\ \bibinfo {author}
  {\bibfnamefont {C.~D.}\ \bibnamefont {Tait}},\ }\href@noop {} {\bibfield
  {journal} {\bibinfo  {journal} {J. Am. Chem. Soc.}\ }\textbf {\bibinfo
  {volume} {113}},\ \bibinfo {pages} {2089} (\bibinfo {year}
  {1996})}\BibitemShut {NoStop}%
\bibitem [{\citenamefont {Law}\ \emph {et~al.}(2010)\citenamefont {Law},
  \citenamefont {Geissler}, \citenamefont {Lloyd}, \citenamefont {Livens},
  \citenamefont {Boothman}, \citenamefont {Begg}, \citenamefont {Denecke},
  \citenamefont {Rothe}, \citenamefont {Dardenne}, \citenamefont {Burke},
  \citenamefont {Charnock},\ and\ \citenamefont
  {Morris}}]{neptunium-important-3}%
  \BibitemOpen
  \bibfield  {author} {\bibinfo {author} {\bibfnamefont {G.~T.~W.}\
  \bibnamefont {Law}}, \bibinfo {author} {\bibfnamefont {A.}~\bibnamefont
  {Geissler}}, \bibinfo {author} {\bibfnamefont {J.~R.}\ \bibnamefont {Lloyd}},
  \bibinfo {author} {\bibfnamefont {F.~R.}\ \bibnamefont {Livens}}, \bibinfo
  {author} {\bibfnamefont {C.}~\bibnamefont {Boothman}}, \bibinfo {author}
  {\bibfnamefont {J.~D.~C.}\ \bibnamefont {Begg}}, \bibinfo {author}
  {\bibfnamefont {M.~A.}\ \bibnamefont {Denecke}}, \bibinfo {author}
  {\bibfnamefont {J.}~\bibnamefont {Rothe}}, \bibinfo {author} {\bibfnamefont
  {K.}~\bibnamefont {Dardenne}}, \bibinfo {author} {\bibfnamefont {I.~T.}\
  \bibnamefont {Burke}}, \bibinfo {author} {\bibfnamefont {J.~M.}\ \bibnamefont
  {Charnock}}, \ and\ \bibinfo {author} {\bibfnamefont {K.}~\bibnamefont
  {Morris}},\ }\href@noop {} {\bibfield  {journal} {\bibinfo  {journal}
  {Environ. Sci. Technol.}\ }\textbf {\bibinfo {volume} {44}},\ \bibinfo
  {pages} {8924} (\bibinfo {year} {2010})}\BibitemShut {NoStop}%
\bibitem [{\citenamefont {Mathur}\ \emph {et~al.}(1996)\citenamefont {Mathur},
  \citenamefont {Murali}, \citenamefont {Krishna}, \citenamefont {Iyer},
  \citenamefont {Chitnis}, \citenamefont {Wattal}, \citenamefont {Theyyunni},
  \citenamefont {Ramanujam}, \citenamefont {Dhami},\ and\ \citenamefont
  {Gopalakrishnan}}]{purex01}%
  \BibitemOpen
  \bibfield  {author} {\bibinfo {author} {\bibfnamefont {J.~N.}\ \bibnamefont
  {Mathur}}, \bibinfo {author} {\bibfnamefont {M.~S.}\ \bibnamefont {Murali}},
  \bibinfo {author} {\bibfnamefont {M.~V.~B.}\ \bibnamefont {Krishna}},
  \bibinfo {author} {\bibfnamefont {R.~H.}\ \bibnamefont {Iyer}}, \bibinfo
  {author} {\bibfnamefont {R.~R.}\ \bibnamefont {Chitnis}}, \bibinfo {author}
  {\bibfnamefont {P.~K.}\ \bibnamefont {Wattal}}, \bibinfo {author}
  {\bibfnamefont {T.~K.}\ \bibnamefont {Theyyunni}}, \bibinfo {author}
  {\bibfnamefont {A.}~\bibnamefont {Ramanujam}}, \bibinfo {author}
  {\bibfnamefont {P.~S.}\ \bibnamefont {Dhami}}, \ and\ \bibinfo {author}
  {\bibfnamefont {V.}~\bibnamefont {Gopalakrishnan}},\ }\href@noop {}
  {\bibfield  {journal} {\bibinfo  {journal} {Sep. Sci. Technol.}\ }\textbf
  {\bibinfo {volume} {31}},\ \bibinfo {pages} {2045} (\bibinfo {year}
  {1996})}\BibitemShut {NoStop}%
\bibitem [{\citenamefont {Taylor}\ \emph {et~al.}(2016)\citenamefont {Taylor},
  \citenamefont {Carrott}, \citenamefont {Galan}, \citenamefont {Geist},
  \citenamefont {H\'{e}res}, \citenamefont {Maher}, \citenamefont {Mason},
  \citenamefont {Malmbeck}, \citenamefont {Miguirditchian}, \citenamefont
  {Modolo}, \citenamefont {Rhodes}, \citenamefont {Sarsfield},\ and\
  \citenamefont {Wilden}}]{ganex01}%
  \BibitemOpen
  \bibfield  {author} {\bibinfo {author} {\bibfnamefont {R.}~\bibnamefont
  {Taylor}}, \bibinfo {author} {\bibfnamefont {M.}~\bibnamefont {Carrott}},
  \bibinfo {author} {\bibfnamefont {H.}~\bibnamefont {Galan}}, \bibinfo
  {author} {\bibfnamefont {A.}~\bibnamefont {Geist}}, \bibinfo {author}
  {\bibfnamefont {X.}~\bibnamefont {H\'{e}res}}, \bibinfo {author}
  {\bibfnamefont {C.}~\bibnamefont {Maher}}, \bibinfo {author} {\bibfnamefont
  {C.}~\bibnamefont {Mason}}, \bibinfo {author} {\bibfnamefont
  {R.}~\bibnamefont {Malmbeck}}, \bibinfo {author} {\bibfnamefont
  {M.}~\bibnamefont {Miguirditchian}}, \bibinfo {author} {\bibfnamefont
  {G.}~\bibnamefont {Modolo}}, \bibinfo {author} {\bibfnamefont
  {C.}~\bibnamefont {Rhodes}}, \bibinfo {author} {\bibfnamefont
  {M.}~\bibnamefont {Sarsfield}}, \ and\ \bibinfo {author} {\bibfnamefont
  {A.}~\bibnamefont {Wilden}},\ }\href@noop {} {\bibfield  {journal} {\bibinfo
  {journal} {Procedia Chem.}\ }\textbf {\bibinfo {volume} {21}},\ \bibinfo
  {pages} {524} (\bibinfo {year} {2016})}\BibitemShut {NoStop}%
\bibitem [{\citenamefont {Carrott}\ \emph {et~al.}(2015)\citenamefont
  {Carrott}, \citenamefont {Gregson},\ and\ \citenamefont {Taylor}}]{ganex02}%
  \BibitemOpen
  \bibfield  {author} {\bibinfo {author} {\bibfnamefont {M.}~\bibnamefont
  {Carrott}}, \bibinfo {author} {\bibfnamefont {C.}~\bibnamefont {Gregson}}, \
  and\ \bibinfo {author} {\bibfnamefont {R.}~\bibnamefont {Taylor}},\
  }\href@noop {} {\bibfield  {journal} {\bibinfo  {journal} {NNL Science}\
  }\textbf {\bibinfo {volume} {3}},\ \bibinfo {pages} {14} (\bibinfo {year}
  {2015})}\BibitemShut {NoStop}%
\bibitem [{\citenamefont {Burn}\ \emph {et~al.}(2017)\citenamefont {Burn},
  \citenamefont {Martin},\ and\ \citenamefont {Nash}}]{ccis-in-solution-4}%
  \BibitemOpen
  \bibfield  {author} {\bibinfo {author} {\bibfnamefont {A.~G.}\ \bibnamefont
  {Burn}}, \bibinfo {author} {\bibfnamefont {L.~R.}\ \bibnamefont {Martin}}, \
  and\ \bibinfo {author} {\bibfnamefont {K.~L.}\ \bibnamefont {Nash}},\
  }\href@noop {} {\bibfield  {journal} {\bibinfo  {journal} {J. Solution
  Chem.}\ }\textbf {\bibinfo {volume} {46}},\ \bibinfo {pages} {1299} (\bibinfo
  {year} {2017})}\BibitemShut {NoStop}%
\bibitem [{\citenamefont {Cotton}(2005)}]{Cotton_book}%
  \BibitemOpen
  \bibfield  {author} {\bibinfo {author} {\bibfnamefont {S.}~\bibnamefont
  {Cotton}},\ }\href@noop {} {\emph {\bibinfo {title} {Lanthanide and actinde
  chemistry}}}\ (\bibinfo  {publisher} {Wiley},\ \bibinfo {address}
  {Chichester},\ \bibinfo {year} {2005})\BibitemShut {NoStop}%
\bibitem [{\citenamefont {Morss}\ \emph {et~al.}(2011)\citenamefont {Morss},
  \citenamefont {Edelstein}, \citenamefont {Fuger},\ and\ \citenamefont
  {Katz}}]{Actinides_bible}%
  \BibitemOpen
  \bibinfo {editor} {\bibfnamefont {L.~R.}\ \bibnamefont {Morss}}, \bibinfo
  {editor} {\bibfnamefont {N.}~\bibnamefont {Edelstein}}, \bibinfo {editor}
  {\bibfnamefont {J.}~\bibnamefont {Fuger}}, \ and\ \bibinfo {editor}
  {\bibfnamefont {J.~J.}\ \bibnamefont {Katz}},\ eds.,\ \href@noop {} {\emph
  {\bibinfo {title} {The chemistry of the actinide and transactinide
  elements}}},\ \bibinfo {edition} {4th}\ ed.\ (\bibinfo  {publisher}
  {Springer},\ \bibinfo {address} {Dordrecht, The Netherlands},\ \bibinfo
  {year} {2011})\BibitemShut {NoStop}%
\bibitem [{\citenamefont {Pepper}\ and\ \citenamefont
  {Bursten}(1991)}]{bursten_91}%
  \BibitemOpen
  \bibfield  {author} {\bibinfo {author} {\bibfnamefont {M.}~\bibnamefont
  {Pepper}}\ and\ \bibinfo {author} {\bibfnamefont {B.~E.}\ \bibnamefont
  {Bursten}},\ }\href@noop {} {\bibfield  {journal} {\bibinfo  {journal} {Chem.
  Rev.}\ }\textbf {\bibinfo {volume} {91}},\ \bibinfo {pages} {719} (\bibinfo
  {year} {1991})}\BibitemShut {NoStop}%
\bibitem [{\citenamefont {Krot}\ and\ \citenamefont
  {Grigoriev}(2004)}]{krot2004}%
  \BibitemOpen
  \bibfield  {author} {\bibinfo {author} {\bibfnamefont {N.~N.}\ \bibnamefont
  {Krot}}\ and\ \bibinfo {author} {\bibfnamefont {M.~S.}\ \bibnamefont
  {Grigoriev}},\ }\href@noop {} {\bibfield  {journal} {\bibinfo  {journal}
  {Russ. Chem. Rev.}\ }\textbf {\bibinfo {volume} {73}},\ \bibinfo {pages} {89}
  (\bibinfo {year} {2004})}\BibitemShut {NoStop}%
\bibitem [{\citenamefont {Denning}(2007)}]{denning2007}%
  \BibitemOpen
  \bibfield  {author} {\bibinfo {author} {\bibfnamefont {R.~G.}\ \bibnamefont
  {Denning}},\ }\href@noop {} {\bibfield  {journal} {\bibinfo  {journal}
  {J.~Phys.~Chem.~A}\ }\textbf {\bibinfo {volume} {111}},\ \bibinfo {pages}
  {4125} (\bibinfo {year} {2007})}\BibitemShut {NoStop}%
\bibitem [{\citenamefont {Wang}\ \emph
  {et~al.}(2012{\natexlab{a}})\citenamefont {Wang}, \citenamefont {{van
  Gunsteren}},\ and\ \citenamefont {Chai}}]{actinoid_rev_2012}%
  \BibitemOpen
  \bibfield  {author} {\bibinfo {author} {\bibfnamefont {D.}~\bibnamefont
  {Wang}}, \bibinfo {author} {\bibfnamefont {W.~F.}\ \bibnamefont {{van
  Gunsteren}}}, \ and\ \bibinfo {author} {\bibfnamefont {Z.}~\bibnamefont
  {Chai}},\ }\href@noop {} {\bibfield  {journal} {\bibinfo  {journal} {Chem.
  Soc. Rev.}\ }\textbf {\bibinfo {volume} {41}},\ \bibinfo {pages} {5836}
  (\bibinfo {year} {2012}{\natexlab{a}})}\BibitemShut {NoStop}%
\bibitem [{\citenamefont {Gomes}\ \emph {et~al.}(2015)\citenamefont {Gomes},
  \citenamefont {R\'{e}al}, \citenamefont {Schimmelpfennig}, \citenamefont
  {Wahlgren},\ and\ \citenamefont {Vallet}}]{gomes2015applied}%
  \BibitemOpen
  \bibfield  {author} {\bibinfo {author} {\bibfnamefont {A.~S.~P.}\
  \bibnamefont {Gomes}}, \bibinfo {author} {\bibfnamefont {F.}~\bibnamefont
  {R\'{e}al}}, \bibinfo {author} {\bibfnamefont {B.}~\bibnamefont
  {Schimmelpfennig}}, \bibinfo {author} {\bibfnamefont {U.}~\bibnamefont
  {Wahlgren}}, \ and\ \bibinfo {author} {\bibfnamefont {V.}~\bibnamefont
  {Vallet}},\ }\enquote {\bibinfo {title} {Applied computational actinide
  chemistry},}\ in\ \href@noop {} {\emph {\bibinfo {booktitle} {Computational
  Methods in Lanthanide and Actinide Chemistry}}}\ (\bibinfo  {publisher}
  {Wiley-Blackwell},\ \bibinfo {year} {2015})\ Chap.~\bibinfo {chapter} {11},
  pp.\ \bibinfo {pages} {269--298}\BibitemShut {NoStop}%
\bibitem [{\citenamefont {McKee}\ and\ \citenamefont
  {Swart}(2005)}]{swart2005}%
  \BibitemOpen
  \bibfield  {author} {\bibinfo {author} {\bibfnamefont {M.~L.}\ \bibnamefont
  {McKee}}\ and\ \bibinfo {author} {\bibfnamefont {M.}~\bibnamefont {Swart}},\
  }\href@noop {} {\bibfield  {journal} {\bibinfo  {journal} {Inorg. Chem.}\
  }\textbf {\bibinfo {volume} {44}},\ \bibinfo {pages} {6975} (\bibinfo {year}
  {2005})}\BibitemShut {NoStop}%
\bibitem [{\citenamefont {Fromager}\ \emph {et~al.}(2005)\citenamefont
  {Fromager}, \citenamefont {Vallet}, \citenamefont {Schimmelpfennig},
  \citenamefont {Macak}, \citenamefont {Privalov},\ and\ \citenamefont
  {Wahlgren}}]{fromager2005}%
  \BibitemOpen
  \bibfield  {author} {\bibinfo {author} {\bibfnamefont {E.}~\bibnamefont
  {Fromager}}, \bibinfo {author} {\bibfnamefont {V.}~\bibnamefont {Vallet}},
  \bibinfo {author} {\bibfnamefont {B.}~\bibnamefont {Schimmelpfennig}},
  \bibinfo {author} {\bibfnamefont {P.}~\bibnamefont {Macak}}, \bibinfo
  {author} {\bibfnamefont {T.}~\bibnamefont {Privalov}}, \ and\ \bibinfo
  {author} {\bibfnamefont {U.}~\bibnamefont {Wahlgren}},\ }\href@noop {}
  {\bibfield  {journal} {\bibinfo  {journal} {J. Phys. Chem}\ }\textbf
  {\bibinfo {volume} {109}},\ \bibinfo {pages} {4957} (\bibinfo {year}
  {2005})}\BibitemShut {NoStop}%
\bibitem [{\citenamefont {Kov\'{a}cs}\ and\ \citenamefont
  {Konings}(2011)}]{kovacs2011}%
  \BibitemOpen
  \bibfield  {author} {\bibinfo {author} {\bibfnamefont {A.}~\bibnamefont
  {Kov\'{a}cs}}\ and\ \bibinfo {author} {\bibfnamefont {R.~J.~M.}\ \bibnamefont
  {Konings}},\ }\href@noop {} {\bibfield  {journal} {\bibinfo  {journal} {J.
  Phys. Chem. A}\ }\textbf {\bibinfo {volume} {115}},\ \bibinfo {pages} {6646}
  (\bibinfo {year} {2011})}\BibitemShut {NoStop}%
\bibitem [{\citenamefont {Vlaisavljevich}\ \emph {et~al.}(2013)\citenamefont
  {Vlaisavljevich}, \citenamefont {Mir\'{o}}, \citenamefont {Ma}, \citenamefont
  {Sigmon}, \citenamefont {Burns}, \citenamefont {Cramer},\ and\ \citenamefont
  {Gagliardi}}]{laura2013}%
  \BibitemOpen
  \bibfield  {author} {\bibinfo {author} {\bibfnamefont {B.}~\bibnamefont
  {Vlaisavljevich}}, \bibinfo {author} {\bibfnamefont {P.}~\bibnamefont
  {Mir\'{o}}}, \bibinfo {author} {\bibfnamefont {D.}~\bibnamefont {Ma}},
  \bibinfo {author} {\bibfnamefont {G.~E.}\ \bibnamefont {Sigmon}}, \bibinfo
  {author} {\bibfnamefont {P.~C.}\ \bibnamefont {Burns}}, \bibinfo {author}
  {\bibfnamefont {C.~J.}\ \bibnamefont {Cramer}}, \ and\ \bibinfo {author}
  {\bibfnamefont {L.}~\bibnamefont {Gagliardi}},\ }\href@noop {} {\bibfield
  {journal} {\bibinfo  {journal} {Chem. Eur. J.}\ }\textbf {\bibinfo {volume}
  {19}},\ \bibinfo {pages} {2937} (\bibinfo {year} {2013})}\BibitemShut
  {NoStop}%
\bibitem [{\citenamefont {Gendron}\ \emph {et~al.}(2014)\citenamefont
  {Gendron}, \citenamefont {P\'{a}ez-Hern\'{a}ndez}, \citenamefont {Notter},
  \citenamefont {Pritchard}, \citenamefont {Bolvin},\ and\ \citenamefont
  {Autschbach}}]{gendron2014}%
  \BibitemOpen
  \bibfield  {author} {\bibinfo {author} {\bibfnamefont {F.}~\bibnamefont
  {Gendron}}, \bibinfo {author} {\bibfnamefont {D.}~\bibnamefont
  {P\'{a}ez-Hern\'{a}ndez}}, \bibinfo {author} {\bibfnamefont {F.-P.}\
  \bibnamefont {Notter}}, \bibinfo {author} {\bibfnamefont {B.}~\bibnamefont
  {Pritchard}}, \bibinfo {author} {\bibfnamefont {H.}~\bibnamefont {Bolvin}}, \
  and\ \bibinfo {author} {\bibfnamefont {J.}~\bibnamefont {Autschbach}},\
  }\href {\doibase 10.1002/chem.201305039} {\bibfield  {journal} {\bibinfo
  {journal} {Chem. Eur. J.}\ }\textbf {\bibinfo {volume} {20}},\ \bibinfo
  {pages} {7994} (\bibinfo {year} {2014})}\BibitemShut {NoStop}%
\bibitem [{\citenamefont {Infante}\ \emph {et~al.}(2006)\citenamefont
  {Infante}, \citenamefont {Gomes},\ and\ \citenamefont
  {Visscher}}]{infante2006}%
  \BibitemOpen
  \bibfield  {author} {\bibinfo {author} {\bibfnamefont {I.}~\bibnamefont
  {Infante}}, \bibinfo {author} {\bibfnamefont {A.~S.~P.}\ \bibnamefont
  {Gomes}}, \ and\ \bibinfo {author} {\bibfnamefont {L.}~\bibnamefont
  {Visscher}},\ }\href@noop {} {\bibfield  {journal} {\bibinfo  {journal} {J.
  Chem. Phys.}\ }\textbf {\bibinfo {volume} {125}},\ \bibinfo {pages} {074301}
  (\bibinfo {year} {2006})}\BibitemShut {NoStop}%
\bibitem [{\citenamefont {Matsika}\ and\ \citenamefont
  {Pitzer}(2000)}]{matsika2000}%
  \BibitemOpen
  \bibfield  {author} {\bibinfo {author} {\bibfnamefont {S.}~\bibnamefont
  {Matsika}}\ and\ \bibinfo {author} {\bibfnamefont {R.~M.}\ \bibnamefont
  {Pitzer}},\ }\href@noop {} {\bibfield  {journal} {\bibinfo  {journal} {J.
  Phys. Chem. A}\ }\textbf {\bibinfo {volume} {104}},\ \bibinfo {pages} {4064}
  (\bibinfo {year} {2000})}\BibitemShut {NoStop}%
\bibitem [{\citenamefont {Kov\'{a}cs}\ and\ \citenamefont
  {Infante}(2015)}]{neptunyl-spectra-theory-1}%
  \BibitemOpen
  \bibfield  {author} {\bibinfo {author} {\bibfnamefont {A.}~\bibnamefont
  {Kov\'{a}cs}}\ and\ \bibinfo {author} {\bibfnamefont {I.}~\bibnamefont
  {Infante}},\ }\href@noop {} {\bibfield  {journal} {\bibinfo  {journal} {J.
  Chem. Phys.}\ }\textbf {\bibinfo {volume} {143}},\ \bibinfo {pages} {074305}
  (\bibinfo {year} {2015})}\BibitemShut {NoStop}%
\bibitem [{\citenamefont {Eisenstein}\ and\ \citenamefont
  {Pryce}(1966)}]{neptunyl-ion-spectra-exp-1}%
  \BibitemOpen
  \bibfield  {author} {\bibinfo {author} {\bibfnamefont {J.~C.}\ \bibnamefont
  {Eisenstein}}\ and\ \bibinfo {author} {\bibfnamefont {M.~H.~L.}\ \bibnamefont
  {Pryce}},\ }\href@noop {} {\bibfield  {journal} {\bibinfo  {journal} {J. Res.
  Natl. Bur. Stand.}\ }\textbf {\bibinfo {volume} {70A}},\ \bibinfo {pages}
  {165} (\bibinfo {year} {1966})}\BibitemShut {NoStop}%
\bibitem [{\citenamefont {Matsika}\ \emph {et~al.}(2001)\citenamefont
  {Matsika}, \citenamefont {Zhang}, \citenamefont {Brozell}, \citenamefont
  {Bladeau},\ and\ \citenamefont {Pitzer}}]{neptunyl-ion-spectra-exp-2}%
  \BibitemOpen
  \bibfield  {author} {\bibinfo {author} {\bibfnamefont {S.}~\bibnamefont
  {Matsika}}, \bibinfo {author} {\bibfnamefont {Z.}~\bibnamefont {Zhang}},
  \bibinfo {author} {\bibfnamefont {S.~R.}\ \bibnamefont {Brozell}}, \bibinfo
  {author} {\bibfnamefont {J.~P.}\ \bibnamefont {Bladeau}}, \ and\ \bibinfo
  {author} {\bibfnamefont {R.~M.}\ \bibnamefont {Pitzer}},\ }\href@noop {}
  {\bibfield  {journal} {\bibinfo  {journal} {J. Phys. Chem. A}\ }\textbf
  {\bibinfo {volume} {105}},\ \bibinfo {pages} {3825} (\bibinfo {year}
  {2001})}\BibitemShut {NoStop}%
\bibitem [{\citenamefont {Eisenstein}\ and\ \citenamefont
  {Pryce}(1965)}]{neptunyl-ion-spectra-exp-3}%
  \BibitemOpen
  \bibfield  {author} {\bibinfo {author} {\bibfnamefont {J.~C.}\ \bibnamefont
  {Eisenstein}}\ and\ \bibinfo {author} {\bibfnamefont {M.~H.~L.}\ \bibnamefont
  {Pryce}},\ }\href@noop {} {\bibfield  {journal} {\bibinfo  {journal} {J. Res.
  Nat. Bur. Stand.}\ }\textbf {\bibinfo {volume} {69A}},\ \bibinfo {pages}
  {217} (\bibinfo {year} {1965})}\BibitemShut {NoStop}%
\bibitem [{\citenamefont {Madic}\ \emph {et~al.}(1979)\citenamefont {Madic},
  \citenamefont {Guillaume}, \citenamefont {Morriseau},\ and\ \citenamefont
  {Moulin}}]{CCI_79}%
  \BibitemOpen
  \bibfield  {author} {\bibinfo {author} {\bibfnamefont {C.}~\bibnamefont
  {Madic}}, \bibinfo {author} {\bibfnamefont {B.}~\bibnamefont {Guillaume}},
  \bibinfo {author} {\bibfnamefont {J.~C.}\ \bibnamefont {Morriseau}}, \ and\
  \bibinfo {author} {\bibfnamefont {J.~P.}\ \bibnamefont {Moulin}},\
  }\href@noop {} {\bibfield  {journal} {\bibinfo  {journal}
  {J.~Inorg.~Nucl.~Chem.}\ }\textbf {\bibinfo {volume} {41}},\ \bibinfo {pages}
  {1027} (\bibinfo {year} {1979})}\BibitemShut {NoStop}%
\bibitem [{\citenamefont {Tecmer}\ \emph {et~al.}(2016)\citenamefont {Tecmer},
  \citenamefont {Hong},\ and\ \citenamefont {Boguslawski}}]{tecmer-song2016}%
  \BibitemOpen
  \bibfield  {author} {\bibinfo {author} {\bibfnamefont {P.}~\bibnamefont
  {Tecmer}}, \bibinfo {author} {\bibfnamefont {S.~W.}\ \bibnamefont {Hong}}, \
  and\ \bibinfo {author} {\bibfnamefont {K.}~\bibnamefont {Boguslawski}},\
  }\href@noop {} {\bibfield  {journal} {\bibinfo  {journal} {Phys. Chem. Chem.
  Phys.}\ }\textbf {\bibinfo {volume} {18}},\ \bibinfo {pages} {18305}
  (\bibinfo {year} {2016})}\BibitemShut {NoStop}%
\bibitem [{\citenamefont {Baker}(2012)}]{baker2012}%
  \BibitemOpen
  \bibfield  {author} {\bibinfo {author} {\bibfnamefont {R.~J.}\ \bibnamefont
  {Baker}},\ }\href@noop {} {\bibfield  {journal} {\bibinfo  {journal} {Chem.
  Eur. J.}\ }\textbf {\bibinfo {volume} {18}},\ \bibinfo {pages} {16258}
  (\bibinfo {year} {2012})}\BibitemShut {NoStop}%
\bibitem [{\citenamefont {Copping}\ \emph {et~al.}(2012)\citenamefont
  {Copping}, \citenamefont {Mougel}, \citenamefont {Auwer}, \citenamefont
  {Berthon}, \citenamefont {Moisy},\ and\ \citenamefont
  {Mazzanti}}]{tetrameric-cci}%
  \BibitemOpen
  \bibfield  {author} {\bibinfo {author} {\bibfnamefont {R.}~\bibnamefont
  {Copping}}, \bibinfo {author} {\bibfnamefont {V.}~\bibnamefont {Mougel}},
  \bibinfo {author} {\bibfnamefont {C.~D.}\ \bibnamefont {Auwer}}, \bibinfo
  {author} {\bibfnamefont {C.}~\bibnamefont {Berthon}}, \bibinfo {author}
  {\bibfnamefont {P.}~\bibnamefont {Moisy}}, \ and\ \bibinfo {author}
  {\bibfnamefont {M.}~\bibnamefont {Mazzanti}},\ }\href@noop {} {\bibfield
  {journal} {\bibinfo  {journal} {Dalton Trans.}\ }\textbf {\bibinfo {volume}
  {41}},\ \bibinfo {pages} {10900} (\bibinfo {year} {2012})}\BibitemShut
  {NoStop}%
\bibitem [{\citenamefont {Edelstein}(2015)}]{neptunyl-ion-spectra-exp-4}%
  \BibitemOpen
  \bibfield  {author} {\bibinfo {author} {\bibfnamefont {N.~M.}\ \bibnamefont
  {Edelstein}},\ }\href@noop {} {\bibfield  {journal} {\bibinfo  {journal} {J.
  Phys. Chem. A}\ }\textbf {\bibinfo {volume} {119}},\ \bibinfo {pages} {11146}
  (\bibinfo {year} {2015})}\BibitemShut {NoStop}%
\bibitem [{\citenamefont {Sullivan}\ \emph {et~al.}(1960)\citenamefont
  {Sullivan}, \citenamefont {Hindman},\ and\ \citenamefont
  {Zielen}}]{sullivan1960}%
  \BibitemOpen
  \bibfield  {author} {\bibinfo {author} {\bibfnamefont {J.~C.}\ \bibnamefont
  {Sullivan}}, \bibinfo {author} {\bibfnamefont {J.~C.}\ \bibnamefont
  {Hindman}}, \ and\ \bibinfo {author} {\bibfnamefont {A.~J.}\ \bibnamefont
  {Zielen}},\ }\href@noop {} {\bibfield  {journal} {\bibinfo  {journal} {J. Am.
  Chem. Soc.}\ }\textbf {\bibinfo {volume} {82}},\ \bibinfo {pages} {5288}
  (\bibinfo {year} {1960})}\BibitemShut {NoStop}%
\bibitem [{\citenamefont {Sullivan}\ \emph {et~al.}(1961)\citenamefont
  {Sullivan}, \citenamefont {Hindman},\ and\ \citenamefont
  {Zielen}}]{sullivan1961}%
  \BibitemOpen
  \bibfield  {author} {\bibinfo {author} {\bibfnamefont {J.~C.}\ \bibnamefont
  {Sullivan}}, \bibinfo {author} {\bibfnamefont {J.~C.}\ \bibnamefont
  {Hindman}}, \ and\ \bibinfo {author} {\bibfnamefont {A.~J.}\ \bibnamefont
  {Zielen}},\ }\href@noop {} {\bibfield  {journal} {\bibinfo  {journal} {J. Am.
  Chem. Soc.}\ }\textbf {\bibinfo {volume} {83}},\ \bibinfo {pages} {3373}
  (\bibinfo {year} {1961})}\BibitemShut {NoStop}%
\bibitem [{\citenamefont {Sullivan}(1962)}]{sullivan1962}%
  \BibitemOpen
  \bibfield  {author} {\bibinfo {author} {\bibfnamefont {J.~C.}\ \bibnamefont
  {Sullivan}},\ }\href@noop {} {\bibfield  {journal} {\bibinfo  {journal} {J.
  Am. Chem. Soc.}\ }\textbf {\bibinfo {volume} {84}},\ \bibinfo {pages} {4256}
  (\bibinfo {year} {1962})}\BibitemShut {NoStop}%
\bibitem [{\citenamefont {Guillaume}\ \emph {et~al.}(1981)\citenamefont
  {Guillaume}, \citenamefont {Hobart},\ and\ \citenamefont
  {Bourges}}]{guillaume1981}%
  \BibitemOpen
  \bibfield  {author} {\bibinfo {author} {\bibfnamefont {B.}~\bibnamefont
  {Guillaume}}, \bibinfo {author} {\bibfnamefont {D.~E.}\ \bibnamefont
  {Hobart}}, \ and\ \bibinfo {author} {\bibfnamefont {J.~Y.}\ \bibnamefont
  {Bourges}},\ }\href@noop {} {\bibfield  {journal} {\bibinfo  {journal} {J.
  Inorg. Nucl. Chem.}\ }\textbf {\bibinfo {volume} {43}},\ \bibinfo {pages}
  {3295} (\bibinfo {year} {1981})}\BibitemShut {NoStop}%
\bibitem [{\citenamefont {Guillaume}\ \emph {et~al.}(1983)\citenamefont
  {Guillaume}, \citenamefont {Hahn},\ and\ \citenamefont
  {Narten}}]{guillaume1983}%
  \BibitemOpen
  \bibfield  {author} {\bibinfo {author} {\bibfnamefont {B.}~\bibnamefont
  {Guillaume}}, \bibinfo {author} {\bibfnamefont {R.~L.}\ \bibnamefont {Hahn}},
  \ and\ \bibinfo {author} {\bibfnamefont {A.~H.}\ \bibnamefont {Narten}},\
  }\href@noop {} {\bibfield  {journal} {\bibinfo  {journal} {Inorg. Chem.}\
  }\textbf {\bibinfo {volume} {22}},\ \bibinfo {pages} {109} (\bibinfo {year}
  {1983})}\BibitemShut {NoStop}%
\bibitem [{\citenamefont {Guillaume}\ \emph {et~al.}(1982)\citenamefont
  {Guillaume}, \citenamefont {Begun},\ and\ \citenamefont
  {Hahn}}]{ccis-in-solution-2}%
  \BibitemOpen
  \bibfield  {author} {\bibinfo {author} {\bibfnamefont {B.}~\bibnamefont
  {Guillaume}}, \bibinfo {author} {\bibfnamefont {G.~M.}\ \bibnamefont
  {Begun}}, \ and\ \bibinfo {author} {\bibfnamefont {R.~L.}\ \bibnamefont
  {Hahn}},\ }\href@noop {} {\bibfield  {journal} {\bibinfo  {journal} {Inorg.
  Chem.}\ }\textbf {\bibinfo {volume} {21}},\ \bibinfo {pages} {1159} (\bibinfo
  {year} {1982})}\BibitemShut {NoStop}%
\bibitem [{\citenamefont {Charushnikova}\ \emph {et~al.}(2010)\citenamefont
  {Charushnikova}, \citenamefont {Bosse}, \citenamefont {Guillaumont},\ and\
  \citenamefont {Moisy}}]{ccis-in-solution-3}%
  \BibitemOpen
  \bibfield  {author} {\bibinfo {author} {\bibfnamefont {I.}~\bibnamefont
  {Charushnikova}}, \bibinfo {author} {\bibfnamefont {E.}~\bibnamefont
  {Bosse}}, \bibinfo {author} {\bibfnamefont {D.}~\bibnamefont {Guillaumont}},
  \ and\ \bibinfo {author} {\bibfnamefont {P.}~\bibnamefont {Moisy}},\
  }\href@noop {} {\bibfield  {journal} {\bibinfo  {journal} {Inorg. Chem.}\
  }\textbf {\bibinfo {volume} {49}},\ \bibinfo {pages} {2077} (\bibinfo {year}
  {2010})}\BibitemShut {NoStop}%
\bibitem [{\citenamefont {Arnold}\ \emph {et~al.}(2009)\citenamefont {Arnold},
  \citenamefont {Love},\ and\ \citenamefont {Patel}}]{Arnold2009}%
  \BibitemOpen
  \bibfield  {author} {\bibinfo {author} {\bibfnamefont {P.~L.}\ \bibnamefont
  {Arnold}}, \bibinfo {author} {\bibfnamefont {J.~B.}\ \bibnamefont {Love}}, \
  and\ \bibinfo {author} {\bibfnamefont {D.}~\bibnamefont {Patel}},\
  }\href@noop {} {\bibfield  {journal} {\bibinfo  {journal} {Coord. Chem.
  Rev.}\ }\textbf {\bibinfo {volume} {253}},\ \bibinfo {pages} {1973} (\bibinfo
  {year} {2009})}\BibitemShut {NoStop}%
\bibitem [{\citenamefont {Jin}\ \emph {et~al.}(2011{\natexlab{a}})\citenamefont
  {Jin}, \citenamefont {Skanthakumar},\ and\ \citenamefont
  {Soderholm}}]{Jin2011}%
  \BibitemOpen
  \bibfield  {author} {\bibinfo {author} {\bibfnamefont {G.~B.}\ \bibnamefont
  {Jin}}, \bibinfo {author} {\bibfnamefont {S.}~\bibnamefont {Skanthakumar}}, \
  and\ \bibinfo {author} {\bibfnamefont {L.}~\bibnamefont {Soderholm}},\
  }\href@noop {} {\bibfield  {journal} {\bibinfo  {journal} {Inorg. Chem.}\
  }\textbf {\bibinfo {volume} {50}},\ \bibinfo {pages} {52035} (\bibinfo {year}
  {2011}{\natexlab{a}})}\BibitemShut {NoStop}%
\bibitem [{\citenamefont {Jin}\ \emph {et~al.}(2011{\natexlab{b}})\citenamefont
  {Jin}, \citenamefont {Skanthakumar},\ and\ \citenamefont
  {Soderholm}}]{Jin2011-2}%
  \BibitemOpen
  \bibfield  {author} {\bibinfo {author} {\bibfnamefont {G.~B.}\ \bibnamefont
  {Jin}}, \bibinfo {author} {\bibfnamefont {S.}~\bibnamefont {Skanthakumar}}, \
  and\ \bibinfo {author} {\bibfnamefont {L.}~\bibnamefont {Soderholm}},\
  }\href@noop {} {\bibfield  {journal} {\bibinfo  {journal} {Inorg. Chem.}\
  }\textbf {\bibinfo {volume} {50}},\ \bibinfo {pages} {6297} (\bibinfo {year}
  {2011}{\natexlab{b}})}\BibitemShut {NoStop}%
\bibitem [{\citenamefont {Wang}\ \emph {et~al.}(2011)\citenamefont {Wang},
  \citenamefont {Alekseev}, \citenamefont {Depmeier},\ and\ \citenamefont
  {Albrecht-Schmitt}}]{wang2011}%
  \BibitemOpen
  \bibfield  {author} {\bibinfo {author} {\bibfnamefont {S.}~\bibnamefont
  {Wang}}, \bibinfo {author} {\bibfnamefont {E.~V.}\ \bibnamefont {Alekseev}},
  \bibinfo {author} {\bibfnamefont {W.}~\bibnamefont {Depmeier}}, \ and\
  \bibinfo {author} {\bibfnamefont {T.~E.}\ \bibnamefont {Albrecht-Schmitt}},\
  }\href@noop {} {\bibfield  {journal} {\bibinfo  {journal} {Inorg. Chem.}\
  }\textbf {\bibinfo {volume} {50}},\ \bibinfo {pages} {4692} (\bibinfo {year}
  {2011})}\BibitemShut {NoStop}%
\bibitem [{\citenamefont {Wang}\ \emph
  {et~al.}(2012{\natexlab{b}})\citenamefont {Wang}, \citenamefont {Diwu},
  \citenamefont {Alekseev}, \citenamefont {Jouffret}, \citenamefont
  {Depmeier},\ and\ \citenamefont {Albrecht-Schmitt}}]{wang2012}%
  \BibitemOpen
  \bibfield  {author} {\bibinfo {author} {\bibfnamefont {S.}~\bibnamefont
  {Wang}}, \bibinfo {author} {\bibfnamefont {J.}~\bibnamefont {Diwu}}, \bibinfo
  {author} {\bibfnamefont {E.~V.}\ \bibnamefont {Alekseev}}, \bibinfo {author}
  {\bibfnamefont {L.~J.}\ \bibnamefont {Jouffret}}, \bibinfo {author}
  {\bibfnamefont {W.}~\bibnamefont {Depmeier}}, \ and\ \bibinfo {author}
  {\bibfnamefont {T.~E.}\ \bibnamefont {Albrecht-Schmitt}},\ }\href@noop {}
  {\bibfield  {journal} {\bibinfo  {journal} {Inorg. Chem.}\ }\textbf {\bibinfo
  {volume} {51}},\ \bibinfo {pages} {7016} (\bibinfo {year}
  {2012}{\natexlab{b}})}\BibitemShut {NoStop}%
\bibitem [{\citenamefont {Diwu}\ \emph {et~al.}(2012)\citenamefont {Diwu},
  \citenamefont {Wang},\ and\ \citenamefont {Albrecht-Schmitt}}]{diwu2012}%
  \BibitemOpen
  \bibfield  {author} {\bibinfo {author} {\bibfnamefont {J.}~\bibnamefont
  {Diwu}}, \bibinfo {author} {\bibfnamefont {S.}~\bibnamefont {Wang}}, \ and\
  \bibinfo {author} {\bibfnamefont {T.~E.}\ \bibnamefont {Albrecht-Schmitt}},\
  }\href@noop {} {\bibfield  {journal} {\bibinfo  {journal} {Inorg. Chem.}\
  }\textbf {\bibinfo {volume} {51}},\ \bibinfo {pages} {4088} (\bibinfo {year}
  {2012})}\BibitemShut {NoStop}%
\bibitem [{\citenamefont {Forbes}\ \emph {et~al.}(2006)\citenamefont {Forbes},
  \citenamefont {Burns}, \citenamefont {Soderholm},\ and\ \citenamefont
  {Skanthakumar}}]{ccis-in-solids-1}%
  \BibitemOpen
  \bibfield  {author} {\bibinfo {author} {\bibfnamefont {T.~Z.}\ \bibnamefont
  {Forbes}}, \bibinfo {author} {\bibfnamefont {P.~C.}\ \bibnamefont {Burns}},
  \bibinfo {author} {\bibfnamefont {L.}~\bibnamefont {Soderholm}}, \ and\
  \bibinfo {author} {\bibfnamefont {S.}~\bibnamefont {Skanthakumar}},\
  }\href@noop {} {\bibfield  {journal} {\bibinfo  {journal} {Chem. Mater.}\
  }\textbf {\bibinfo {volume} {18}},\ \bibinfo {pages} {1643} (\bibinfo {year}
  {2006})}\BibitemShut {NoStop}%
\bibitem [{\citenamefont {Almond}\ \emph {et~al.}(2007)\citenamefont {Almond},
  \citenamefont {Skanthakumar}, \citenamefont {Soderholm},\ and\ \citenamefont
  {Burns}}]{ccis-in-solids-2}%
  \BibitemOpen
  \bibfield  {author} {\bibinfo {author} {\bibfnamefont {P.~M.}\ \bibnamefont
  {Almond}}, \bibinfo {author} {\bibfnamefont {S.}~\bibnamefont
  {Skanthakumar}}, \bibinfo {author} {\bibfnamefont {L.}~\bibnamefont
  {Soderholm}}, \ and\ \bibinfo {author} {\bibfnamefont {P.~C.}\ \bibnamefont
  {Burns}},\ }\href@noop {} {\bibfield  {journal} {\bibinfo  {journal} {Chem.
  Mater.}\ }\textbf {\bibinfo {volume} {19}},\ \bibinfo {pages} {280} (\bibinfo
  {year} {2007})}\BibitemShut {NoStop}%
\bibitem [{\citenamefont {Skanthakumar}\ \emph {et~al.}(2008)\citenamefont
  {Skanthakumar}, \citenamefont {Antonio},\ and\ \citenamefont
  {Soderholm}}]{ccis-in-solids-3}%
  \BibitemOpen
  \bibfield  {author} {\bibinfo {author} {\bibfnamefont {S.}~\bibnamefont
  {Skanthakumar}}, \bibinfo {author} {\bibfnamefont {M.~R.}\ \bibnamefont
  {Antonio}}, \ and\ \bibinfo {author} {\bibfnamefont {L.}~\bibnamefont
  {Soderholm}},\ }\href@noop {} {\bibfield  {journal} {\bibinfo  {journal}
  {Inorg. Chem.}\ }\textbf {\bibinfo {volume} {47}},\ \bibinfo {pages} {4591}
  (\bibinfo {year} {2008})}\BibitemShut {NoStop}%
\bibitem [{\citenamefont {Basile}\ \emph {et~al.}(2018)\citenamefont {Basile},
  \citenamefont {Cole},\ and\ \citenamefont {Forbes}}]{ic-cci-neptunyl-2018}%
  \BibitemOpen
  \bibfield  {author} {\bibinfo {author} {\bibfnamefont {M.}~\bibnamefont
  {Basile}}, \bibinfo {author} {\bibfnamefont {E.}~\bibnamefont {Cole}}, \ and\
  \bibinfo {author} {\bibfnamefont {T.~Z.}\ \bibnamefont {Forbes}},\
  }\href@noop {} {\bibfield  {journal} {\bibinfo  {journal} {Inorg.~Chem.}\
  }\textbf {\bibinfo {volume} {57}},\ \bibinfo {pages} {6016} (\bibinfo {year}
  {2018})}\BibitemShut {NoStop}%
\bibitem [{\citenamefont {Vallet}\ \emph {et~al.}(2004)\citenamefont {Vallet},
  \citenamefont {Privalov}, \citenamefont {Wahlgren},\ and\ \citenamefont
  {Grenthe}}]{vallet2004}%
  \BibitemOpen
  \bibfield  {author} {\bibinfo {author} {\bibfnamefont {V.}~\bibnamefont
  {Vallet}}, \bibinfo {author} {\bibfnamefont {T.}~\bibnamefont {Privalov}},
  \bibinfo {author} {\bibfnamefont {U.}~\bibnamefont {Wahlgren}}, \ and\
  \bibinfo {author} {\bibfnamefont {I.}~\bibnamefont {Grenthe}},\ }\href@noop
  {} {\bibfield  {journal} {\bibinfo  {journal} {J. Am. Chem. Soc.}\ }\textbf
  {\bibinfo {volume} {126}},\ \bibinfo {pages} {7766} (\bibinfo {year}
  {2004})}\BibitemShut {NoStop}%
\bibitem [{\citenamefont {Choppin}\ and\ \citenamefont
  {Rao}(1984)}]{choppin1984}%
  \BibitemOpen
  \bibfield  {author} {\bibinfo {author} {\bibfnamefont {G.~R.}\ \bibnamefont
  {Choppin}}\ and\ \bibinfo {author} {\bibfnamefont {L.~F.}\ \bibnamefont
  {Rao}},\ }\href@noop {} {\bibfield  {journal} {\bibinfo  {journal}
  {Radiochim. Acta}\ }\textbf {\bibinfo {volume} {37}},\ \bibinfo {pages} {143}
  (\bibinfo {year} {1984})}\BibitemShut {NoStop}%
\bibitem [{\citenamefont {Rao}\ \emph {et~al.}(1979)\citenamefont {Rao},
  \citenamefont {Gudi}, \citenamefont {Bagawde},\ and\ \citenamefont
  {Patil}}]{rao1979}%
  \BibitemOpen
  \bibfield  {author} {\bibinfo {author} {\bibfnamefont {P.~R.~V.}\
  \bibnamefont {Rao}}, \bibinfo {author} {\bibfnamefont {N.~M.}\ \bibnamefont
  {Gudi}}, \bibinfo {author} {\bibfnamefont {S.~V.}\ \bibnamefont {Bagawde}}, \
  and\ \bibinfo {author} {\bibfnamefont {S.~K.}\ \bibnamefont {Patil}},\
  }\href@noop {} {\bibfield  {journal} {\bibinfo  {journal} {J. Inorg. Nucl.
  Chem.}\ }\textbf {\bibinfo {volume} {41}},\ \bibinfo {pages} {235} (\bibinfo
  {year} {1979})}\BibitemShut {NoStop}%
\bibitem [{\citenamefont {Gainar}\ and\ \citenamefont
  {Sykes}(1964)}]{gainar1983}%
  \BibitemOpen
  \bibfield  {author} {\bibinfo {author} {\bibfnamefont {I.}~\bibnamefont
  {Gainar}}\ and\ \bibinfo {author} {\bibfnamefont {K.~W.}\ \bibnamefont
  {Sykes}},\ }\href@noop {} {\bibfield  {journal} {\bibinfo  {journal} {J.
  Chem. Soc.}\ ,\ \bibinfo {pages} {4452}} (\bibinfo {year}
  {1964})}\BibitemShut {NoStop}%
\bibitem [{\citenamefont {Halperin}\ and\ \citenamefont
  {Oliver}(1983)}]{halperin1983}%
  \BibitemOpen
  \bibfield  {author} {\bibinfo {author} {\bibfnamefont {J.}~\bibnamefont
  {Halperin}}\ and\ \bibinfo {author} {\bibfnamefont {J.~H.}\ \bibnamefont
  {Oliver}},\ }\href@noop {} {\bibfield  {journal} {\bibinfo  {journal}
  {Radiochim. Acta}\ }\textbf {\bibinfo {volume} {33}},\ \bibinfo {pages} {29}
  (\bibinfo {year} {1983})}\BibitemShut {NoStop}%
\bibitem [{\citenamefont {Roesch}\ \emph {et~al.}(1990)\citenamefont {Roesch},
  \citenamefont {Dittrich}, \citenamefont {Buklanov}, \citenamefont {Milanov},
  \citenamefont {Khalkin},\ and\ \citenamefont {Dreyer}}]{roesch1990}%
  \BibitemOpen
  \bibfield  {author} {\bibinfo {author} {\bibfnamefont {F.}~\bibnamefont
  {Roesch}}, \bibinfo {author} {\bibfnamefont {S.}~\bibnamefont {Dittrich}},
  \bibinfo {author} {\bibfnamefont {G.~V.}\ \bibnamefont {Buklanov}}, \bibinfo
  {author} {\bibfnamefont {M.}~\bibnamefont {Milanov}}, \bibinfo {author}
  {\bibfnamefont {V.~A.}\ \bibnamefont {Khalkin}}, \ and\ \bibinfo {author}
  {\bibfnamefont {R.}~\bibnamefont {Dreyer}},\ }\href@noop {} {\bibfield
  {journal} {\bibinfo  {journal} {Radiochim. Acta}\ }\textbf {\bibinfo {volume}
  {49}},\ \bibinfo {pages} {29} (\bibinfo {year} {1990})}\BibitemShut {NoStop}%
\bibitem [{\citenamefont {Reiher}\ and\ \citenamefont
  {Wolf}(2009)}]{reiher-wolf}%
  \BibitemOpen
  \bibfield  {author} {\bibinfo {author} {\bibfnamefont {M.}~\bibnamefont
  {Reiher}}\ and\ \bibinfo {author} {\bibfnamefont {A.}~\bibnamefont {Wolf}},\
  }\href@noop {} {\emph {\bibinfo {title} {Relativistic quantum chemistry. The
  fundamental theory of molecular science}}}\ (\bibinfo  {publisher} {Wiley},\
  \bibinfo {address} {Dordrecht},\ \bibinfo {year} {2009})\BibitemShut
  {NoStop}%
\bibitem [{\citenamefont {Tecmer}\ \emph {et~al.}(2011)\citenamefont {Tecmer},
  \citenamefont {Gomes}, \citenamefont {Ekstr\"om},\ and\ \citenamefont
  {Visscher}}]{pawel1}%
  \BibitemOpen
  \bibfield  {author} {\bibinfo {author} {\bibfnamefont {P.}~\bibnamefont
  {Tecmer}}, \bibinfo {author} {\bibfnamefont {A.~S.~P.}\ \bibnamefont
  {Gomes}}, \bibinfo {author} {\bibfnamefont {U.}~\bibnamefont {Ekstr\"om}}, \
  and\ \bibinfo {author} {\bibfnamefont {L.}~\bibnamefont {Visscher}},\
  }\href@noop {} {\bibfield  {journal} {\bibinfo  {journal}
  {Phys.~Chem.~Chem.~Phys.}\ }\textbf {\bibinfo {volume} {13}},\ \bibinfo
  {pages} {6249} (\bibinfo {year} {2011})}\BibitemShut {NoStop}%
\bibitem [{\citenamefont {Dolg}\ and\ \citenamefont
  {Cao}(2012)}]{Dolg_rev_2012}%
  \BibitemOpen
  \bibfield  {author} {\bibinfo {author} {\bibfnamefont {M.}~\bibnamefont
  {Dolg}}\ and\ \bibinfo {author} {\bibfnamefont {X.}~\bibnamefont {Cao}},\
  }\href@noop {} {\bibfield  {journal} {\bibinfo  {journal} {Chem. Rev.}\
  }\textbf {\bibinfo {volume} {112}},\ \bibinfo {pages} {403} (\bibinfo {year}
  {2012})}\BibitemShut {NoStop}%
\bibitem [{\citenamefont {Tecmer}\ \emph {et~al.}(2014)\citenamefont {Tecmer},
  \citenamefont {Boguslawski}, \citenamefont {Legeza},\ and\ \citenamefont
  {Reiher}}]{CUO_DMRG}%
  \BibitemOpen
  \bibfield  {author} {\bibinfo {author} {\bibfnamefont {P.}~\bibnamefont
  {Tecmer}}, \bibinfo {author} {\bibfnamefont {K.}~\bibnamefont {Boguslawski}},
  \bibinfo {author} {\bibfnamefont {{\"O}.}~\bibnamefont {Legeza}}, \ and\
  \bibinfo {author} {\bibfnamefont {M.}~\bibnamefont {Reiher}},\ }\href@noop {}
  {\bibfield  {journal} {\bibinfo  {journal} {Phys. Chem. Chem. Phys}\ }\textbf
  {\bibinfo {volume} {16}},\ \bibinfo {pages} {719} (\bibinfo {year}
  {2014})}\BibitemShut {NoStop}%
\bibitem [{\citenamefont {Bagus}\ \emph {et~al.}(2012)\citenamefont {Bagus},
  \citenamefont {Ilton}, \citenamefont {Martin}, \citenamefont {Jensen},\ and\
  \citenamefont {Knecht}}]{SO_in_actinides}%
  \BibitemOpen
  \bibfield  {author} {\bibinfo {author} {\bibfnamefont {P.~S.}\ \bibnamefont
  {Bagus}}, \bibinfo {author} {\bibfnamefont {E.~S.}\ \bibnamefont {Ilton}},
  \bibinfo {author} {\bibfnamefont {R.~L.}\ \bibnamefont {Martin}}, \bibinfo
  {author} {\bibfnamefont {H.~J.~A.}\ \bibnamefont {Jensen}}, \ and\ \bibinfo
  {author} {\bibfnamefont {S.}~\bibnamefont {Knecht}},\ }\href@noop {}
  {\bibfield  {journal} {\bibinfo  {journal} {Chem. Phys. Lett.}\ }\textbf
  {\bibinfo {volume} {546}},\ \bibinfo {pages} {58} (\bibinfo {year}
  {2012})}\BibitemShut {NoStop}%
\bibitem [{\citenamefont {Tecmer}\ \emph {et~al.}(2013)\citenamefont {Tecmer},
  \citenamefont {Govind}, \citenamefont {Kowalski}, \citenamefont {{De Jong}},\
  and\ \citenamefont {Visscher}}]{pawel_saldien}%
  \BibitemOpen
  \bibfield  {author} {\bibinfo {author} {\bibfnamefont {P.}~\bibnamefont
  {Tecmer}}, \bibinfo {author} {\bibfnamefont {N.}~\bibnamefont {Govind}},
  \bibinfo {author} {\bibfnamefont {K.}~\bibnamefont {Kowalski}}, \bibinfo
  {author} {\bibfnamefont {W.~A.}\ \bibnamefont {{De Jong}}}, \ and\ \bibinfo
  {author} {\bibfnamefont {L.}~\bibnamefont {Visscher}},\ }\href@noop {}
  {\bibfield  {journal} {\bibinfo  {journal} {J. Chem. Phys}\ }\textbf
  {\bibinfo {volume} {139}},\ \bibinfo {pages} {034301} (\bibinfo {year}
  {2013})}\BibitemShut {NoStop}%
\bibitem [{\citenamefont {Tecmer}\ \emph {et~al.}(2017)\citenamefont {Tecmer},
  \citenamefont {Boguslawski},\ and\ \citenamefont
  {K{\c{e}}dziera}}]{Tecmer2016}%
  \BibitemOpen
  \bibfield  {author} {\bibinfo {author} {\bibfnamefont {P.}~\bibnamefont
  {Tecmer}}, \bibinfo {author} {\bibfnamefont {K.}~\bibnamefont {Boguslawski}},
  \ and\ \bibinfo {author} {\bibfnamefont {D.}~\bibnamefont {K{\c{e}}dziera}},\
  }in\ \href@noop {} {\emph {\bibinfo {booktitle} {Handbook of computational
  chemistry}}},\ Vol.~\bibinfo {volume} {2},\ \bibinfo {editor} {edited by\
  \bibinfo {editor} {\bibfnamefont {J.}~\bibnamefont {Leszczy{\'{n}}ski}}}\
  (\bibinfo  {publisher} {Springer Netherlands},\ \bibinfo {address}
  {Dordrecht},\ \bibinfo {year} {2017})\ pp.\ \bibinfo {pages}
  {885--926}\BibitemShut {NoStop}%
\bibitem [{\citenamefont {Boguslawski}\ \emph {et~al.}(2017)\citenamefont
  {Boguslawski}, \citenamefont {R\'{e}al}, \citenamefont {Tecmer},
  \citenamefont {Duperrouzel}, \citenamefont {Gomes}, \citenamefont {Legeza},
  \citenamefont {Ayers},\ and\ \citenamefont {Vallet}}]{boguslawski2017}%
  \BibitemOpen
  \bibfield  {author} {\bibinfo {author} {\bibfnamefont {K.}~\bibnamefont
  {Boguslawski}}, \bibinfo {author} {\bibfnamefont {F.}~\bibnamefont
  {R\'{e}al}}, \bibinfo {author} {\bibfnamefont {P.}~\bibnamefont {Tecmer}},
  \bibinfo {author} {\bibfnamefont {C.}~\bibnamefont {Duperrouzel}}, \bibinfo
  {author} {\bibfnamefont {A.~S.~P.}\ \bibnamefont {Gomes}}, \bibinfo {author}
  {\bibfnamefont {{\"O}.}~\bibnamefont {Legeza}}, \bibinfo {author}
  {\bibfnamefont {P.~W.}\ \bibnamefont {Ayers}}, \ and\ \bibinfo {author}
  {\bibfnamefont {V.}~\bibnamefont {Vallet}},\ }\href@noop {} {\bibfield
  {journal} {\bibinfo  {journal} {Phys. Chem. Chem. Phys.}\ }\textbf {\bibinfo
  {volume} {19}},\ \bibinfo {pages} {4317} (\bibinfo {year}
  {2017})}\BibitemShut {NoStop}%
\bibitem [{\citenamefont {Kov\'{a}cs}\ \emph {et~al.}(2015)\citenamefont
  {Kov\'{a}cs}, \citenamefont {Konings}, \citenamefont {Gibson}, \citenamefont
  {Infante},\ and\ \citenamefont {Gagliardi}}]{actinide_oxides_1}%
  \BibitemOpen
  \bibfield  {author} {\bibinfo {author} {\bibfnamefont {A.}~\bibnamefont
  {Kov\'{a}cs}}, \bibinfo {author} {\bibfnamefont {R.~J.~M.}\ \bibnamefont
  {Konings}}, \bibinfo {author} {\bibfnamefont {J.~K.}\ \bibnamefont {Gibson}},
  \bibinfo {author} {\bibfnamefont {I.}~\bibnamefont {Infante}}, \ and\
  \bibinfo {author} {\bibfnamefont {L.}~\bibnamefont {Gagliardi}},\ }\href@noop
  {} {\bibfield  {journal} {\bibinfo  {journal} {Chem. Rev.}\ }\textbf
  {\bibinfo {volume} {115}},\ \bibinfo {pages} {1725} (\bibinfo {year}
  {2015})}\BibitemShut {NoStop}%
\bibitem [{\citenamefont {Maron}\ \emph {et~al.}(1999)\citenamefont {Maron},
  \citenamefont {Leininger}, \citenamefont {Schimmelpfennig}, \citenamefont
  {V.~Vallet}, \citenamefont {Teichteil}, \citenamefont {Gropen},\ and\
  \citenamefont {Wahlgren}}]{computational-difficulties-casscf}%
  \BibitemOpen
  \bibfield  {author} {\bibinfo {author} {\bibfnamefont {L.}~\bibnamefont
  {Maron}}, \bibinfo {author} {\bibfnamefont {T.}~\bibnamefont {Leininger}},
  \bibinfo {author} {\bibfnamefont {B.}~\bibnamefont {Schimmelpfennig}},
  \bibinfo {author} {\bibfnamefont {J.~L.~H.}\ \bibnamefont {V.~Vallet}},
  \bibinfo {author} {\bibfnamefont {C.}~\bibnamefont {Teichteil}}, \bibinfo
  {author} {\bibfnamefont {O.}~\bibnamefont {Gropen}}, \ and\ \bibinfo {author}
  {\bibfnamefont {U.}~\bibnamefont {Wahlgren}},\ }\href@noop {} {\bibfield
  {journal} {\bibinfo  {journal} {Chem. Phys.}\ }\textbf {\bibinfo {volume}
  {244}},\ \bibinfo {pages} {195} (\bibinfo {year} {1999})}\BibitemShut
  {NoStop}%
\bibitem [{\citenamefont {Clavagu\'{e}ra-Sarrio}\ \emph
  {et~al.}(2004)\citenamefont {Clavagu\'{e}ra-Sarrio}, \citenamefont {Vallet},
  \citenamefont {Maynau},\ and\ \citenamefont
  {Marsden}}]{computational-difficulties-casscf-2}%
  \BibitemOpen
  \bibfield  {author} {\bibinfo {author} {\bibfnamefont {C.}~\bibnamefont
  {Clavagu\'{e}ra-Sarrio}}, \bibinfo {author} {\bibfnamefont {V.}~\bibnamefont
  {Vallet}}, \bibinfo {author} {\bibfnamefont {D.}~\bibnamefont {Maynau}}, \
  and\ \bibinfo {author} {\bibfnamefont {C.~J.}\ \bibnamefont {Marsden}},\
  }\href@noop {} {\bibfield  {journal} {\bibinfo  {journal} {J. Chem. Phys.}\
  }\textbf {\bibinfo {volume} {121}},\ \bibinfo {pages} {5312} (\bibinfo {year}
  {2004})}\BibitemShut {NoStop}%
\bibitem [{\citenamefont {Ahlrichs}\ \emph {et~al.}(1989)\citenamefont
  {Ahlrichs}, \citenamefont {B\"{a}r}, \citenamefont {H\"{a}ser}, \citenamefont
  {Horn},\ and\ \citenamefont {K\"{o}lmel}}]{ahlrichs_89}%
  \BibitemOpen
  \bibfield  {author} {\bibinfo {author} {\bibfnamefont {R.}~\bibnamefont
  {Ahlrichs}}, \bibinfo {author} {\bibfnamefont {M.}~\bibnamefont {B\"{a}r}},
  \bibinfo {author} {\bibfnamefont {M.}~\bibnamefont {H\"{a}ser}}, \bibinfo
  {author} {\bibfnamefont {H.}~\bibnamefont {Horn}}, \ and\ \bibinfo {author}
  {\bibfnamefont {C.}~\bibnamefont {K\"{o}lmel}},\ }\href@noop {} {\bibfield
  {journal} {\bibinfo  {journal} {Chem.~Phys.~Lett.}\ }\textbf {\bibinfo
  {volume} {162}},\ \bibinfo {pages} {165} (\bibinfo {year}
  {1989})}\BibitemShut {NoStop}%
\bibitem [{tur()}]{turbomole}%
  \BibitemOpen
  \href@noop {} {\enquote {\bibinfo {title} {{TURBOMOLE V7.0 2015}, a
  development of {University of Karlsruhe} and {Forschungszentrum Karlsruhe
  GmbH}, 1989-2007, {TURBOMOLE GmbH}, since 2007; available from \\ {\tt
  http://www.turbomole.com}.}}\ }\BibitemShut {NoStop}%
\bibitem [{tur(2015)}]{turbomole-manual}%
  \BibitemOpen
  \href@noop {} {\enquote {\bibinfo {title} {{TURBOMOLE program package for ab
  initio electronic structure calculations user's manual}},}\ } (\bibinfo
  {year} {2015})\BibitemShut {NoStop}%
\bibitem [{\citenamefont {Weigend}\ \emph {et~al.}(1998)\citenamefont
  {Weigend}, \citenamefont {H\"{a}ser}, \citenamefont {Patzelt},\ and\
  \citenamefont {Ahlrichs}}]{def2-tzvp_o}%
  \BibitemOpen
  \bibfield  {author} {\bibinfo {author} {\bibfnamefont {F.}~\bibnamefont
  {Weigend}}, \bibinfo {author} {\bibfnamefont {M.}~\bibnamefont {H\"{a}ser}},
  \bibinfo {author} {\bibfnamefont {H.}~\bibnamefont {Patzelt}}, \ and\
  \bibinfo {author} {\bibfnamefont {R.}~\bibnamefont {Ahlrichs}},\ }\href@noop
  {} {\bibfield  {journal} {\bibinfo  {journal} {Chem. Phys. Lett.}\ }\textbf
  {\bibinfo {volume} {294}},\ \bibinfo {pages} {143} (\bibinfo {year}
  {1998})}\BibitemShut {NoStop}%
\bibitem [{\citenamefont {K\"{u}chle}\ \emph {et~al.}(1994)\citenamefont
  {K\"{u}chle}, \citenamefont {Dolg}, \citenamefont {Stoll},\ and\
  \citenamefont {Preuss}}]{ECP_Np_1}%
  \BibitemOpen
  \bibfield  {author} {\bibinfo {author} {\bibfnamefont {W.}~\bibnamefont
  {K\"{u}chle}}, \bibinfo {author} {\bibfnamefont {M.}~\bibnamefont {Dolg}},
  \bibinfo {author} {\bibfnamefont {H.}~\bibnamefont {Stoll}}, \ and\ \bibinfo
  {author} {\bibfnamefont {H.}~\bibnamefont {Preuss}},\ }\href@noop {}
  {\bibfield  {journal} {\bibinfo  {journal} {J.~Chem.~Phys.}\ }\textbf
  {\bibinfo {volume} {100}},\ \bibinfo {pages} {7535} (\bibinfo {year}
  {1994})}\BibitemShut {NoStop}%
\bibitem [{\citenamefont {Cao}\ \emph {et~al.}(2003)\citenamefont {Cao},
  \citenamefont {Dolg},\ and\ \citenamefont {Stoll}}]{ECP_Np_2}%
  \BibitemOpen
  \bibfield  {author} {\bibinfo {author} {\bibfnamefont {X.}~\bibnamefont
  {Cao}}, \bibinfo {author} {\bibfnamefont {M.}~\bibnamefont {Dolg}}, \ and\
  \bibinfo {author} {\bibfnamefont {H.}~\bibnamefont {Stoll}},\ }\href@noop {}
  {\bibfield  {journal} {\bibinfo  {journal} {J.~Chem.~Phys.}\ }\textbf
  {\bibinfo {volume} {118}},\ \bibinfo {pages} {487} (\bibinfo {year}
  {2003})}\BibitemShut {NoStop}%
\bibitem [{\citenamefont {Eichkorn}\ \emph {et~al.}(1997)\citenamefont
  {Eichkorn}, \citenamefont {Weigend}, \citenamefont {Treutler},\ and\
  \citenamefont {Ahlrichs}}]{def-tzvp_np}%
  \BibitemOpen
  \bibfield  {author} {\bibinfo {author} {\bibfnamefont {K.}~\bibnamefont
  {Eichkorn}}, \bibinfo {author} {\bibfnamefont {F.}~\bibnamefont {Weigend}},
  \bibinfo {author} {\bibfnamefont {O.}~\bibnamefont {Treutler}}, \ and\
  \bibinfo {author} {\bibfnamefont {R.}~\bibnamefont {Ahlrichs}},\ }\href
  {\doibase 10.1007/s002140050244} {\bibfield  {journal} {\bibinfo  {journal}
  {Theor. Chem. Acc.}\ }\textbf {\bibinfo {volume} {97}},\ \bibinfo {pages}
  {119} (\bibinfo {year} {1997})}\BibitemShut {NoStop}%
\bibitem [{\citenamefont {Becke}(1988)}]{becke88}%
  \BibitemOpen
  \bibfield  {author} {\bibinfo {author} {\bibfnamefont {A.~D.}\ \bibnamefont
  {Becke}},\ }\href@noop {} {\bibfield  {journal} {\bibinfo  {journal} {Phys.
  Rev. A}\ }\textbf {\bibinfo {volume} {38}},\ \bibinfo {pages} {3098}
  (\bibinfo {year} {1988})}\BibitemShut {NoStop}%
\bibitem [{\citenamefont {Perdew}(1986)}]{perdew86}%
  \BibitemOpen
  \bibfield  {author} {\bibinfo {author} {\bibfnamefont {J.~P.}\ \bibnamefont
  {Perdew}},\ }\href@noop {} {\bibfield  {journal} {\bibinfo  {journal} {Phys.
  Rev. B}\ }\textbf {\bibinfo {volume} {33}},\ \bibinfo {pages} {8822}
  (\bibinfo {year} {1986})}\BibitemShut {NoStop}%
\bibitem [{\citenamefont {Schreckenbach}\ and\ \citenamefont
  {Shamov}(2010)}]{schreckenbach_dft}%
  \BibitemOpen
  \bibfield  {author} {\bibinfo {author} {\bibfnamefont {G.}~\bibnamefont
  {Schreckenbach}}\ and\ \bibinfo {author} {\bibfnamefont {G.~A.}\ \bibnamefont
  {Shamov}},\ }\href@noop {} {\bibfield  {journal} {\bibinfo  {journal}
  {Acc.~Chem.~Res.}\ }\textbf {\bibinfo {volume} {43}},\ \bibinfo {pages} {19}
  (\bibinfo {year} {2010})}\BibitemShut {NoStop}%
\bibitem [{\citenamefont {Grimme}\ \emph {et~al.}(2010)\citenamefont {Grimme},
  \citenamefont {Antony}, \citenamefont {Ehrlich},\ and\ \citenamefont
  {Krieg}}]{Grimme-D3-2010}%
  \BibitemOpen
  \bibfield  {author} {\bibinfo {author} {\bibfnamefont {S.}~\bibnamefont
  {Grimme}}, \bibinfo {author} {\bibfnamefont {J.}~\bibnamefont {Antony}},
  \bibinfo {author} {\bibfnamefont {S.}~\bibnamefont {Ehrlich}}, \ and\
  \bibinfo {author} {\bibfnamefont {H.}~\bibnamefont {Krieg}},\ }\href@noop {}
  {\bibfield  {journal} {\bibinfo  {journal} {J.~Chem.~Phys.}\ }\textbf
  {\bibinfo {volume} {132}},\ \bibinfo {pages} {154104} (\bibinfo {year}
  {2010})}\BibitemShut {NoStop}%
\bibitem [{\citenamefont {Rotzinger}(2018)}]{neptunyl-solution}%
  \BibitemOpen
  \bibfield  {author} {\bibinfo {author} {\bibfnamefont {F.~P.}\ \bibnamefont
  {Rotzinger}},\ }\href@noop {} {\bibfield  {journal} {\bibinfo  {journal}
  {Inorg. Chem.}\ }\textbf {\bibinfo {volume} {57}},\ \bibinfo {pages} {2425}
  (\bibinfo {year} {2018})}\BibitemShut {NoStop}%
\bibitem [{\citenamefont {Klamt}\ and\ \citenamefont
  {Sch{\"u}rmann}(1993)}]{COSMO}%
  \BibitemOpen
  \bibfield  {author} {\bibinfo {author} {\bibfnamefont {A.}~\bibnamefont
  {Klamt}}\ and\ \bibinfo {author} {\bibfnamefont {G.}~\bibnamefont
  {Sch{\"u}rmann}},\ }\href@noop {} {\bibfield  {journal} {\bibinfo  {journal}
  {J. Chem. Soc. Perkin Trans.}\ }\textbf {\bibinfo {volume} {2}},\ \bibinfo
  {pages} {799} (\bibinfo {year} {1993})}\BibitemShut {NoStop}%
\bibitem [{\citenamefont {Werner}\ \emph
  {et~al.}(2012{\natexlab{a}})\citenamefont {Werner}, \citenamefont {Knowles},\
  and\ \citenamefont {{Knizia \textit{et al.}}}}]{molpro2012}%
  \BibitemOpen
  \bibfield  {author} {\bibinfo {author} {\bibfnamefont {H.}~\bibnamefont
  {Werner}}, \bibinfo {author} {\bibfnamefont {P.~J.}\ \bibnamefont {Knowles}},
  \ and\ \bibinfo {author} {\bibfnamefont {G.}~\bibnamefont {{Knizia \textit{et
  al.}}}},\ }\href@noop {} {\enquote {\bibinfo {title} {Molpro, version 2012.1,
  a package of \emph{ab initio} programs},}\ } (\bibinfo {year}
  {2012}{\natexlab{a}}),\ \bibinfo {note} {see
  http://www.molpro.net}\BibitemShut {NoStop}%
\bibitem [{\citenamefont {Werner}\ \emph
  {et~al.}(2012{\natexlab{b}})\citenamefont {Werner}, \citenamefont {Knowles},
  \citenamefont {Knizia}, \citenamefont {Manby},\ and\ \citenamefont
  {Sch\"{u}tz}}]{molpro2012_2}%
  \BibitemOpen
  \bibfield  {author} {\bibinfo {author} {\bibfnamefont {H.}~\bibnamefont
  {Werner}}, \bibinfo {author} {\bibfnamefont {P.~J.}\ \bibnamefont {Knowles}},
  \bibinfo {author} {\bibfnamefont {G.}~\bibnamefont {Knizia}}, \bibinfo
  {author} {\bibfnamefont {F.~R.}\ \bibnamefont {Manby}}, \ and\ \bibinfo
  {author} {\bibfnamefont {M.}~\bibnamefont {Sch\"{u}tz}},\ }\href@noop {}
  {\bibfield  {journal} {\bibinfo  {journal} {WIREs Comput. Mol. Sci.}\
  }\textbf {\bibinfo {volume} {2}},\ \bibinfo {pages} {242} (\bibinfo {year}
  {2012}{\natexlab{b}})}\BibitemShut {NoStop}%
\bibitem [{\citenamefont {Werner}\ and\ \citenamefont
  {Knowles}(1985)}]{casscf-01}%
  \BibitemOpen
  \bibfield  {author} {\bibinfo {author} {\bibfnamefont {H.}~\bibnamefont
  {Werner}}\ and\ \bibinfo {author} {\bibfnamefont {P.~J.}\ \bibnamefont
  {Knowles}},\ }\href@noop {} {\bibfield  {journal} {\bibinfo  {journal} {J.
  Chem. Phys.}\ }\textbf {\bibinfo {volume} {82}},\ \bibinfo {pages} {5053}
  (\bibinfo {year} {1985})}\BibitemShut {NoStop}%
\bibitem [{\citenamefont {Werner}\ and\ \citenamefont
  {Knowles}(1988)}]{casscf-02}%
  \BibitemOpen
  \bibfield  {author} {\bibinfo {author} {\bibfnamefont {H.}~\bibnamefont
  {Werner}}\ and\ \bibinfo {author} {\bibfnamefont {P.~J.}\ \bibnamefont
  {Knowles}},\ }\href@noop {} {\bibfield  {journal} {\bibinfo  {journal} {J.
  Chem. Phys.}\ }\textbf {\bibinfo {volume} {89}},\ \bibinfo {pages} {5803}
  (\bibinfo {year} {1988})}\BibitemShut {NoStop}%
\bibitem [{\citenamefont {Cao}\ and\ \citenamefont {Dolg}(2004)}]{U_ECP}%
  \BibitemOpen
  \bibfield  {author} {\bibinfo {author} {\bibfnamefont {X.}~\bibnamefont
  {Cao}}\ and\ \bibinfo {author} {\bibfnamefont {M.}~\bibnamefont {Dolg}},\
  }\href@noop {} {\bibfield  {journal} {\bibinfo  {journal}
  {J.~Mol.~Struct.~(Theochem)}\ }\textbf {\bibinfo {volume} {673}},\ \bibinfo
  {pages} {203} (\bibinfo {year} {2004})}\BibitemShut {NoStop}%
\bibitem [{\citenamefont {{Dunning
  Jr.}}(1989)}]{basis-Dunning-JCP1989-90-1007}%
  \BibitemOpen
  \bibfield  {author} {\bibinfo {author} {\bibfnamefont {T.~H.}\ \bibnamefont
  {{Dunning Jr.}}},\ }\href@noop {} {\bibfield  {journal} {\bibinfo  {journal}
  {J. Chem. Phys.}\ }\textbf {\bibinfo {volume} {90}},\ \bibinfo {pages} {1007}
  (\bibinfo {year} {1989})}\BibitemShut {NoStop}%
\bibitem [{jmo()}]{jmol}%
  \BibitemOpen
  \href@noop {} {}\bibinfo {note} {Jmol: an open-source Java viewer for
  chemical structures in 3D. {\tt http://www.jmol.org/}}\BibitemShut {NoStop}%
\bibitem [{\citenamefont {Tecmer}\ \emph {et~al.}(2015)\citenamefont {Tecmer},
  \citenamefont {Boguslawski},\ and\ \citenamefont
  {Ayers}}]{uranyl-dissociation}%
  \BibitemOpen
  \bibfield  {author} {\bibinfo {author} {\bibfnamefont {P.}~\bibnamefont
  {Tecmer}}, \bibinfo {author} {\bibfnamefont {K.}~\bibnamefont {Boguslawski}},
  \ and\ \bibinfo {author} {\bibfnamefont {P.~W.}\ \bibnamefont {Ayers}},\
  }\href@noop {} {\bibfield  {journal} {\bibinfo  {journal} {Phys. Chem. Chem.
  Phys.}\ }\textbf {\bibinfo {volume} {17}},\ \bibinfo {pages} {14427}
  (\bibinfo {year} {2015})}\BibitemShut {NoStop}%
\bibitem [{\citenamefont {Andersson}\ \emph {et~al.}(1990)\citenamefont
  {Andersson}, \citenamefont {Malmqvist}, \citenamefont {Roos}, \citenamefont
  {Sadlej},\ and\ \citenamefont {Woli{\'n}ski}}]{caspt21}%
  \BibitemOpen
  \bibfield  {author} {\bibinfo {author} {\bibfnamefont {K.}~\bibnamefont
  {Andersson}}, \bibinfo {author} {\bibfnamefont {P.-{\AA{}}.}\ \bibnamefont
  {Malmqvist}}, \bibinfo {author} {\bibfnamefont {B.~O.}\ \bibnamefont {Roos}},
  \bibinfo {author} {\bibfnamefont {A.~J.}\ \bibnamefont {Sadlej}}, \ and\
  \bibinfo {author} {\bibfnamefont {K.}~\bibnamefont {Woli{\'n}ski}},\
  }\href@noop {} {\bibfield  {journal} {\bibinfo  {journal} {J.~Phys.~Chem.}\
  }\textbf {\bibinfo {volume} {94}},\ \bibinfo {pages} {5483} (\bibinfo {year}
  {1990})}\BibitemShut {NoStop}%
\bibitem [{\citenamefont {Andersson}\ \emph {et~al.}(1992)\citenamefont
  {Andersson}, \citenamefont {Malmqvist},\ and\ \citenamefont
  {Roos}}]{caspt22}%
  \BibitemOpen
  \bibfield  {author} {\bibinfo {author} {\bibfnamefont {K.}~\bibnamefont
  {Andersson}}, \bibinfo {author} {\bibfnamefont {P.-{\AA{}}.}\ \bibnamefont
  {Malmqvist}}, \ and\ \bibinfo {author} {\bibfnamefont {B.~O.}\ \bibnamefont
  {Roos}},\ }\href@noop {} {\bibfield  {journal} {\bibinfo  {journal}
  {J.~Chem.~Phys.}\ }\textbf {\bibinfo {volume} {96}},\ \bibinfo {pages} {1218}
  (\bibinfo {year} {1992})}\BibitemShut {NoStop}%
\bibitem [{\citenamefont {Finley}\ \emph {et~al.}(1998)\citenamefont {Finley},
  \citenamefont {Malmqvist}, \citenamefont {Roos},\ and\ \citenamefont
  {Serrano-Andr\'es}}]{MS-CASPT2}%
  \BibitemOpen
  \bibfield  {author} {\bibinfo {author} {\bibfnamefont {J.}~\bibnamefont
  {Finley}}, \bibinfo {author} {\bibfnamefont {P.-{\AA{}}.}\ \bibnamefont
  {Malmqvist}}, \bibinfo {author} {\bibfnamefont {B.~O.}\ \bibnamefont {Roos}},
  \ and\ \bibinfo {author} {\bibfnamefont {L.}~\bibnamefont
  {Serrano-Andr\'es}},\ }\href@noop {} {\bibfield  {journal} {\bibinfo
  {journal} {Chem.~Phys.~Lett.}\ }\textbf {\bibinfo {volume} {288}},\ \bibinfo
  {pages} {299} (\bibinfo {year} {1998})}\BibitemShut {NoStop}%
\bibitem [{\citenamefont {{G. Karlstr\"{o}m, R. Lindh, P.-{\AA{}}. Malmqvist,
  B. O. Roos, U. Ryde, V. Veryazov, P.-O. Widmark, M. Cossi, B.
  Schimmelpfennig, P. Neogr\'{a}dy and L. Seijo}}(2003)}]{Molcas6}%
  \BibitemOpen
  \bibfield  {author} {\bibinfo {author} {\bibnamefont {{G. Karlstr\"{o}m, R.
  Lindh, P.-{\AA{}}. Malmqvist, B. O. Roos, U. Ryde, V. Veryazov, P.-O.
  Widmark, M. Cossi, B. Schimmelpfennig, P. Neogr\'{a}dy and L. Seijo}}},\
  }\href@noop {} {\bibfield  {journal} {\bibinfo  {journal} {Comput. Mater.
  Sci.}\ }\textbf {\bibinfo {volume} {28}},\ \bibinfo {pages} {222} (\bibinfo
  {year} {2003})}\BibitemShut {NoStop}%
\bibitem [{\citenamefont {{F. Aquilante, L. De Vico, N. Ferr{\'{e}}, G. Ghigo,
  P.-{\r{A}}. Malmqvist, P. Neogr{\'{a}}dy, T. B. Pedersen, M. Pitonak, M.
  Reiher, B. O. Roos, L. Serrano-Andr{\'{e}}s, M. Urban, V. Veryazov and R.
  Lindh}}(2010)}]{Molcas7}%
  \BibitemOpen
  \bibfield  {author} {\bibinfo {author} {\bibnamefont {{F. Aquilante, L. De
  Vico, N. Ferr{\'{e}}, G. Ghigo, P.-{\r{A}}. Malmqvist, P. Neogr{\'{a}}dy, T.
  B. Pedersen, M. Pitonak, M. Reiher, B. O. Roos, L. Serrano-Andr{\'{e}}s, M.
  Urban, V. Veryazov and R. Lindh}}},\ }\href@noop {} {\bibfield  {journal}
  {\bibinfo  {journal} {J. Comput. Chem.}\ }\textbf {\bibinfo {volume} {31}},\
  \bibinfo {pages} {224} (\bibinfo {year} {2010})}\BibitemShut {NoStop}%
\bibitem [{\citenamefont {{V. Veryazov, P.-O. Widmark, L. Serrano-Andr{\'{e}}s,
  R. Lindh and B. O. Roos}}(2004)}]{Molcas-code}%
  \BibitemOpen
  \bibfield  {author} {\bibinfo {author} {\bibnamefont {{V. Veryazov, P.-O.
  Widmark, L. Serrano-Andr{\'{e}}s, R. Lindh and B. O. Roos}}},\ }\href@noop {}
  {\bibfield  {journal} {\bibinfo  {journal} {Int. J. Quantum Chem.}\ }\textbf
  {\bibinfo {volume} {100}},\ \bibinfo {pages} {626} (\bibinfo {year}
  {2004})}\BibitemShut {NoStop}%
\bibitem [{\citenamefont {Aquilante}\ \emph {et~al.}(2016)\citenamefont
  {Aquilante}, \citenamefont {Autschbach},\ and\ \citenamefont {{Carlson
  \textit{et al.}}}}]{Molcas8}%
  \BibitemOpen
  \bibfield  {author} {\bibinfo {author} {\bibfnamefont {F.}~\bibnamefont
  {Aquilante}}, \bibinfo {author} {\bibfnamefont {J.}~\bibnamefont
  {Autschbach}}, \ and\ \bibinfo {author} {\bibfnamefont {R.~K.}\ \bibnamefont
  {{Carlson \textit{et al.}}}},\ }\href@noop {} {\bibfield  {journal} {\bibinfo
   {journal} {J. Comput. Chem.}\ }\textbf {\bibinfo {volume} {37}},\ \bibinfo
  {pages} {506} (\bibinfo {year} {2016})}\BibitemShut {NoStop}%
\bibitem [{\citenamefont {Douglas}\ and\ \citenamefont {Kroll}(1974)}]{dkh1}%
  \BibitemOpen
  \bibfield  {author} {\bibinfo {author} {\bibfnamefont {N.}~\bibnamefont
  {Douglas}}\ and\ \bibinfo {author} {\bibfnamefont {N.~M.}\ \bibnamefont
  {Kroll}},\ }\href@noop {} {\bibfield  {journal} {\bibinfo  {journal}
  {Ann.~Phys.}\ }\textbf {\bibinfo {volume} {82}},\ \bibinfo {pages} {89}
  (\bibinfo {year} {1974})}\BibitemShut {NoStop}%
\bibitem [{\citenamefont {Hess}(1986)}]{dkh2}%
  \BibitemOpen
  \bibfield  {author} {\bibinfo {author} {\bibfnamefont {B.~A.}\ \bibnamefont
  {Hess}},\ }\href@noop {} {\bibfield  {journal} {\bibinfo  {journal}
  {Phys.~Rev.~A}\ }\textbf {\bibinfo {volume} {33}},\ \bibinfo {pages} {3742}
  (\bibinfo {year} {1986})}\BibitemShut {NoStop}%
\bibitem [{\citenamefont {Roos}\ \emph {et~al.}(2004)\citenamefont {Roos},
  \citenamefont {Lindh}, \citenamefont {Malmqvist}, \citenamefont {Veryazov},\
  and\ \citenamefont {Widmark}}]{basis-ano-rcc-main}%
  \BibitemOpen
  \bibfield  {author} {\bibinfo {author} {\bibfnamefont {B.~O.}\ \bibnamefont
  {Roos}}, \bibinfo {author} {\bibfnamefont {R.}~\bibnamefont {Lindh}},
  \bibinfo {author} {\bibfnamefont {P.-{\AA{}}.}\ \bibnamefont {Malmqvist}},
  \bibinfo {author} {\bibfnamefont {V.}~\bibnamefont {Veryazov}}, \ and\
  \bibinfo {author} {\bibfnamefont {P.-O.}\ \bibnamefont {Widmark}},\
  }\href@noop {} {\bibfield  {journal} {\bibinfo  {journal} {J. Phys. Chem. A}\
  }\textbf {\bibinfo {volume} {108}},\ \bibinfo {pages} {2851} (\bibinfo {year}
  {2004})}\BibitemShut {NoStop}%
\bibitem [{\citenamefont {Roos}\ \emph {et~al.}(2005)\citenamefont {Roos},
  \citenamefont {Lindh}, \citenamefont {Malmqvist}, \citenamefont {Veryazov},\
  and\ \citenamefont {Widmark}}]{basis-ano-rcc-actinides}%
  \BibitemOpen
  \bibfield  {author} {\bibinfo {author} {\bibfnamefont {B.~O.}\ \bibnamefont
  {Roos}}, \bibinfo {author} {\bibfnamefont {R.}~\bibnamefont {Lindh}},
  \bibinfo {author} {\bibfnamefont {P.-{\AA{}}.}\ \bibnamefont {Malmqvist}},
  \bibinfo {author} {\bibfnamefont {V.}~\bibnamefont {Veryazov}}, \ and\
  \bibinfo {author} {\bibfnamefont {P.-O.}\ \bibnamefont {Widmark}},\
  }\href@noop {} {\bibfield  {journal} {\bibinfo  {journal} {Chem. Phys.
  Lett.}\ }\textbf {\bibinfo {volume} {409}},\ \bibinfo {pages} {295} (\bibinfo
  {year} {2005})}\BibitemShut {NoStop}%
\bibitem [{\citenamefont {Ghigo}\ \emph {et~al.}(2004)\citenamefont {Ghigo},
  \citenamefont {Roos},\ and\ \citenamefont {Malmqvist}}]{caspt2-h0}%
  \BibitemOpen
  \bibfield  {author} {\bibinfo {author} {\bibfnamefont {G.}~\bibnamefont
  {Ghigo}}, \bibinfo {author} {\bibfnamefont {B.~O.}\ \bibnamefont {Roos}}, \
  and\ \bibinfo {author} {\bibfnamefont {P.-{\AA{}}.}\ \bibnamefont
  {Malmqvist}},\ }\href@noop {} {\bibfield  {journal} {\bibinfo  {journal}
  {Chem. Phys. Letters}\ }\textbf {\bibinfo {volume} {396}},\ \bibinfo {pages}
  {142} (\bibinfo {year} {2004})}\BibitemShut {NoStop}%
\bibitem [{\citenamefont {Hess}\ \emph {et~al.}(1996)\citenamefont {Hess},
  \citenamefont {Marian}, \citenamefont {Wahlgren},\ and\ \citenamefont
  {Gropen}}]{AMFI_1}%
  \BibitemOpen
  \bibfield  {author} {\bibinfo {author} {\bibfnamefont {B.~A.}\ \bibnamefont
  {Hess}}, \bibinfo {author} {\bibfnamefont {C.~M.}\ \bibnamefont {Marian}},
  \bibinfo {author} {\bibfnamefont {U.}~\bibnamefont {Wahlgren}}, \ and\
  \bibinfo {author} {\bibfnamefont {O.}~\bibnamefont {Gropen}},\ }\href@noop {}
  {\bibfield  {journal} {\bibinfo  {journal} {Chem. Phys. Lett.}\ }\textbf
  {\bibinfo {volume} {251}},\ \bibinfo {pages} {365} (\bibinfo {year}
  {1996})}\BibitemShut {NoStop}%
\bibitem [{\citenamefont {Schimmelpfennig}\ \emph
  {et~al.}(1998{\natexlab{a}})\citenamefont {Schimmelpfennig}, \citenamefont
  {Maron}, \citenamefont {Wahlgren}, \citenamefont {Teichteil}, \citenamefont
  {Fagerli},\ and\ \citenamefont {Gropen}}]{AMFI_2}%
  \BibitemOpen
  \bibfield  {author} {\bibinfo {author} {\bibfnamefont {B.}~\bibnamefont
  {Schimmelpfennig}}, \bibinfo {author} {\bibfnamefont {L.}~\bibnamefont
  {Maron}}, \bibinfo {author} {\bibfnamefont {U.}~\bibnamefont {Wahlgren}},
  \bibinfo {author} {\bibfnamefont {C.}~\bibnamefont {Teichteil}}, \bibinfo
  {author} {\bibfnamefont {H.}~\bibnamefont {Fagerli}}, \ and\ \bibinfo
  {author} {\bibfnamefont {O.}~\bibnamefont {Gropen}},\ }\href@noop {}
  {\bibfield  {journal} {\bibinfo  {journal} {Chem. Phys. Lett.}\ }\textbf
  {\bibinfo {volume} {286}},\ \bibinfo {pages} {261} (\bibinfo {year}
  {1998}{\natexlab{a}})}\BibitemShut {NoStop}%
\bibitem [{\citenamefont {Schimmelpfennig}\ \emph
  {et~al.}(1998{\natexlab{b}})\citenamefont {Schimmelpfennig}, \citenamefont
  {Maron}, \citenamefont {Wahlgren}, \citenamefont {Teichteil}, \citenamefont
  {Fagerli},\ and\ \citenamefont {Gropen}}]{AMFI_3}%
  \BibitemOpen
  \bibfield  {author} {\bibinfo {author} {\bibfnamefont {B.}~\bibnamefont
  {Schimmelpfennig}}, \bibinfo {author} {\bibfnamefont {L.}~\bibnamefont
  {Maron}}, \bibinfo {author} {\bibfnamefont {U.}~\bibnamefont {Wahlgren}},
  \bibinfo {author} {\bibfnamefont {C.}~\bibnamefont {Teichteil}}, \bibinfo
  {author} {\bibfnamefont {H.}~\bibnamefont {Fagerli}}, \ and\ \bibinfo
  {author} {\bibfnamefont {O.}~\bibnamefont {Gropen}},\ }\href@noop {}
  {\bibfield  {journal} {\bibinfo  {journal} {Chem. Phys. Lett.}\ }\textbf
  {\bibinfo {volume} {286}},\ \bibinfo {pages} {267} (\bibinfo {year}
  {1998}{\natexlab{b}})}\BibitemShut {NoStop}%
\bibitem [{\citenamefont {Malmqvist}\ \emph {et~al.}(2002)\citenamefont
  {Malmqvist}, \citenamefont {Roos},\ and\ \citenamefont
  {Schimmelpfennig}}]{rassi}%
  \BibitemOpen
  \bibfield  {author} {\bibinfo {author} {\bibfnamefont {P.-{\AA{}}.}\
  \bibnamefont {Malmqvist}}, \bibinfo {author} {\bibfnamefont {B.~O.}\
  \bibnamefont {Roos}}, \ and\ \bibinfo {author} {\bibfnamefont
  {B.}~\bibnamefont {Schimmelpfennig}},\ }\href@noop {} {\bibfield  {journal}
  {\bibinfo  {journal} {Chem. Phys. Lett.}\ }\textbf {\bibinfo {volume}
  {357}},\ \bibinfo {pages} {230} (\bibinfo {year} {2002})}\BibitemShut
  {NoStop}%
\bibitem [{\citenamefont {White}(1992)}]{white}%
  \BibitemOpen
  \bibfield  {author} {\bibinfo {author} {\bibfnamefont {S.~R.}\ \bibnamefont
  {White}},\ }\href@noop {} {\bibfield  {journal} {\bibinfo  {journal} {Phys.
  Rev. Lett.}\ }\textbf {\bibinfo {volume} {69}},\ \bibinfo {pages} {2863}
  (\bibinfo {year} {1992})}\BibitemShut {NoStop}%
\bibitem [{\citenamefont {White}(1993)}]{white2}%
  \BibitemOpen
  \bibfield  {author} {\bibinfo {author} {\bibfnamefont {S.~R.}\ \bibnamefont
  {White}},\ }\href@noop {} {\bibfield  {journal} {\bibinfo  {journal} {Phys.
  Rev. B}\ }\textbf {\bibinfo {volume} {48}},\ \bibinfo {pages} {10345}
  (\bibinfo {year} {1993})}\BibitemShut {NoStop}%
\bibitem [{\citenamefont {White}\ and\ \citenamefont
  {Martin}(1999)}]{white-qc}%
  \BibitemOpen
  \bibfield  {author} {\bibinfo {author} {\bibfnamefont {S.~R.}\ \bibnamefont
  {White}}\ and\ \bibinfo {author} {\bibfnamefont {R.~L.}\ \bibnamefont
  {Martin}},\ }\href@noop {} {\bibfield  {journal} {\bibinfo  {journal}
  {J.~Chem.~Phys.}\ }\textbf {\bibinfo {volume} {110}},\ \bibinfo {pages}
  {4127} (\bibinfo {year} {1999})}\BibitemShut {NoStop}%
\bibitem [{\citenamefont {Marti}\ and\ \citenamefont
  {Reiher}(2010)}]{marti2010b}%
  \BibitemOpen
  \bibfield  {author} {\bibinfo {author} {\bibfnamefont {K.~H.}\ \bibnamefont
  {Marti}}\ and\ \bibinfo {author} {\bibfnamefont {M.}~\bibnamefont {Reiher}},\
  }\href@noop {} {\bibfield  {journal} {\bibinfo  {journal} {Z. Phys. Chem.}\
  }\textbf {\bibinfo {volume} {224}},\ \bibinfo {pages} {583} (\bibinfo {year}
  {2010})}\BibitemShut {NoStop}%
\bibitem [{\citenamefont {Szalay}\ \emph {et~al.}(2015)\citenamefont {Szalay},
  \citenamefont {Pfeffer}, \citenamefont {Murg}, \citenamefont {Barcza},
  \citenamefont {Verstraete}, \citenamefont {Schneider},\ and\ \citenamefont
  {Legeza}}]{Ors_ijqc}%
  \BibitemOpen
  \bibfield  {author} {\bibinfo {author} {\bibfnamefont {S.}~\bibnamefont
  {Szalay}}, \bibinfo {author} {\bibfnamefont {M.}~\bibnamefont {Pfeffer}},
  \bibinfo {author} {\bibfnamefont {V.}~\bibnamefont {Murg}}, \bibinfo {author}
  {\bibfnamefont {G.}~\bibnamefont {Barcza}}, \bibinfo {author} {\bibfnamefont
  {F.}~\bibnamefont {Verstraete}}, \bibinfo {author} {\bibfnamefont
  {R.}~\bibnamefont {Schneider}}, \ and\ \bibinfo {author} {\bibfnamefont
  {{\"O}.}~\bibnamefont {Legeza}},\ }\href@noop {} {\bibfield  {journal}
  {\bibinfo  {journal} {Int.~J.~Quantum~Chem.}\ }\textbf {\bibinfo {volume}
  {115}},\ \bibinfo {pages} {1342} (\bibinfo {year} {2015})}\BibitemShut
  {NoStop}%
\bibitem [{\citenamefont {Wouters}\ and\ \citenamefont {{Van
  Neck}}(2014)}]{wouters-review}%
  \BibitemOpen
  \bibfield  {author} {\bibinfo {author} {\bibfnamefont {S.}~\bibnamefont
  {Wouters}}\ and\ \bibinfo {author} {\bibfnamefont {D.}~\bibnamefont {{Van
  Neck}}},\ }\href@noop {} {\bibfield  {journal} {\bibinfo  {journal} {Eur.
  Phys. J. D}\ }\textbf {\bibinfo {volume} {68}},\ \bibinfo {pages} {272}
  (\bibinfo {year} {2014})}\BibitemShut {NoStop}%
\bibitem [{\citenamefont {Yanai}\ \emph {et~al.}(2015)\citenamefont {Yanai},
  \citenamefont {Kurashige}, \citenamefont {Mizukami}, \citenamefont
  {Chalupsky}, \citenamefont {Lan},\ and\ \citenamefont
  {Saitow}}]{yanai-review}%
  \BibitemOpen
  \bibfield  {author} {\bibinfo {author} {\bibfnamefont {T.}~\bibnamefont
  {Yanai}}, \bibinfo {author} {\bibfnamefont {Y.}~\bibnamefont {Kurashige}},
  \bibinfo {author} {\bibfnamefont {W.}~\bibnamefont {Mizukami}}, \bibinfo
  {author} {\bibfnamefont {J.}~\bibnamefont {Chalupsky}}, \bibinfo {author}
  {\bibfnamefont {T.~N.}\ \bibnamefont {Lan}}, \ and\ \bibinfo {author}
  {\bibfnamefont {M.}~\bibnamefont {Saitow}},\ }\href@noop {} {\bibfield
  {journal} {\bibinfo  {journal} {Int.~J.~Quantum~Chem.}\ }\textbf {\bibinfo
  {volume} {115}},\ \bibinfo {pages} {283} (\bibinfo {year}
  {2015})}\BibitemShut {NoStop}%
\bibitem [{\citenamefont {Keller}\ \emph {et~al.}(2015)\citenamefont {Keller},
  \citenamefont {Boguslawski}, \citenamefont {Janowski}, \citenamefont
  {Reiher},\ and\ \citenamefont {Pulay}}]{dmrg-17}%
  \BibitemOpen
  \bibfield  {author} {\bibinfo {author} {\bibfnamefont {S.}~\bibnamefont
  {Keller}}, \bibinfo {author} {\bibfnamefont {K.}~\bibnamefont {Boguslawski}},
  \bibinfo {author} {\bibfnamefont {T.}~\bibnamefont {Janowski}}, \bibinfo
  {author} {\bibfnamefont {M.}~\bibnamefont {Reiher}}, \ and\ \bibinfo {author}
  {\bibfnamefont {P.}~\bibnamefont {Pulay}},\ }\href@noop {} {\bibfield
  {journal} {\bibinfo  {journal} {J. Chem. Phys.}\ }\textbf {\bibinfo {volume}
  {142}},\ \bibinfo {pages} {244104} (\bibinfo {year} {2015})}\BibitemShut
  {NoStop}%
\bibitem [{\citenamefont {Stein}\ and\ \citenamefont
  {Reiher}(2016)}]{Stein2016}%
  \BibitemOpen
  \bibfield  {author} {\bibinfo {author} {\bibfnamefont {C.~J.}\ \bibnamefont
  {Stein}}\ and\ \bibinfo {author} {\bibfnamefont {M.}~\bibnamefont {Reiher}},\
  }\href@noop {} {\bibfield  {journal} {\bibinfo  {journal}
  {J.~Chem.~Theory~Comput.}\ }\textbf {\bibinfo {volume} {12}},\ \bibinfo
  {pages} {1760} (\bibinfo {year} {2016})}\BibitemShut {NoStop}%
\bibitem [{\citenamefont {Kin-Lic}\ and\ \citenamefont
  {Sandeep}(2011)}]{dmrg-21}%
  \BibitemOpen
  \bibfield  {author} {\bibinfo {author} {\bibfnamefont {C.~G.}\ \bibnamefont
  {Kin-Lic}}\ and\ \bibinfo {author} {\bibfnamefont {S.}~\bibnamefont
  {Sandeep}},\ }\href@noop {} {\bibfield  {journal} {\bibinfo  {journal} {Annu.
  Rev. Phys. Chem.}\ }\textbf {\bibinfo {volume} {62}},\ \bibinfo {pages} {465}
  (\bibinfo {year} {2011})}\BibitemShut {NoStop}%
\bibitem [{\citenamefont {Legeza}\ \emph {et~al.}()\citenamefont {Legeza},
  \citenamefont {Veis},\ and\ \citenamefont {Mosoni}}]{dmrg_ors}%
  \BibitemOpen
  \bibfield  {author} {\bibinfo {author} {\bibfnamefont {{\"O}.}~\bibnamefont
  {Legeza}}, \bibinfo {author} {\bibfnamefont {L.}~\bibnamefont {Veis}}, \ and\
  \bibinfo {author} {\bibfnamefont {T.}~\bibnamefont {Mosoni}},\ }\href@noop {}
  {\enquote {\bibinfo {title} {\textsc{QC-DMRG-Budapest}, a program for quantum
  chemical {DMRG} calculations. { \rm Copyright 2000--2018, HAS RISSPO
  Budapest}},}\ }\BibitemShut {NoStop}%
\bibitem [{\citenamefont {Bullock}(1969)}]{uranyl-spectra-1}%
  \BibitemOpen
  \bibfield  {author} {\bibinfo {author} {\bibfnamefont {J.~I.}\ \bibnamefont
  {Bullock}},\ }\href@noop {} {\bibfield  {journal} {\bibinfo  {journal} {J.
  Chem. Soc. A}\ ,\ \bibinfo {pages} {781}} (\bibinfo {year}
  {1969})}\BibitemShut {NoStop}%
\bibitem [{\citenamefont {Matsika}\ and\ \citenamefont
  {Pitzer}(2001)}]{uranyl-spectra-2}%
  \BibitemOpen
  \bibfield  {author} {\bibinfo {author} {\bibfnamefont {S.}~\bibnamefont
  {Matsika}}\ and\ \bibinfo {author} {\bibfnamefont {R.~M.}\ \bibnamefont
  {Pitzer}},\ }\href@noop {} {\bibfield  {journal} {\bibinfo  {journal}
  {J.~Phys.~Chem.~A}\ }\textbf {\bibinfo {volume} {105}},\ \bibinfo {pages}
  {637} (\bibinfo {year} {2001})}\BibitemShut {NoStop}%
\bibitem [{\citenamefont {Ruip\'{e}rez}\ \emph {et~al.}(2009)\citenamefont
  {Ruip\'{e}rez}, \citenamefont {Danilo}, \citenamefont {R\'{e}al},
  \citenamefont {Flament}, \citenamefont {Vallet},\ and\ \citenamefont
  {Wahlgren}}]{uranyl-spectra-3}%
  \BibitemOpen
  \bibfield  {author} {\bibinfo {author} {\bibfnamefont {F.}~\bibnamefont
  {Ruip\'{e}rez}}, \bibinfo {author} {\bibfnamefont {C.}~\bibnamefont
  {Danilo}}, \bibinfo {author} {\bibfnamefont {F.}~\bibnamefont {R\'{e}al}},
  \bibinfo {author} {\bibfnamefont {J.-P.}\ \bibnamefont {Flament}}, \bibinfo
  {author} {\bibfnamefont {V.}~\bibnamefont {Vallet}}, \ and\ \bibinfo {author}
  {\bibfnamefont {U.}~\bibnamefont {Wahlgren}},\ }\href@noop {} {\bibfield
  {journal} {\bibinfo  {journal} {J.~Phys.~Chem.~A}\ }\textbf {\bibinfo
  {volume} {113}},\ \bibinfo {pages} {1420} (\bibinfo {year}
  {2009})}\BibitemShut {NoStop}%
\bibitem [{\citenamefont {Tecmer}\ \emph {et~al.}(2012)\citenamefont {Tecmer},
  \citenamefont {Bast}, \citenamefont {Ruud},\ and\ \citenamefont
  {Visscher}}]{uranyl-spectra-4}%
  \BibitemOpen
  \bibfield  {author} {\bibinfo {author} {\bibfnamefont {P.}~\bibnamefont
  {Tecmer}}, \bibinfo {author} {\bibfnamefont {R.}~\bibnamefont {Bast}},
  \bibinfo {author} {\bibfnamefont {K.}~\bibnamefont {Ruud}}, \ and\ \bibinfo
  {author} {\bibfnamefont {L.}~\bibnamefont {Visscher}},\ }\href@noop {}
  {\bibfield  {journal} {\bibinfo  {journal} {J.~Phys.~Chem.~A}\ }\textbf
  {\bibinfo {volume} {116}},\ \bibinfo {pages} {7397} (\bibinfo {year}
  {2012})}\BibitemShut {NoStop}%
\bibitem [{\citenamefont {Wang}\ \emph {et~al.}(1995)\citenamefont {Wang},
  \citenamefont {Becke},\ and\ \citenamefont {Smith}}]{Wang1995}%
  \BibitemOpen
  \bibfield  {author} {\bibinfo {author} {\bibfnamefont {J.}~\bibnamefont
  {Wang}}, \bibinfo {author} {\bibfnamefont {A.~D.}\ \bibnamefont {Becke}}, \
  and\ \bibinfo {author} {\bibfnamefont {V.~H.}\ \bibnamefont {Smith}},\
  }\href@noop {} {\bibfield  {journal} {\bibinfo  {journal} {J.~Chem.~Phys.}\
  }\textbf {\bibinfo {volume} {102}},\ \bibinfo {pages} {3477} (\bibinfo {year}
  {1995})}\BibitemShut {NoStop}%
\bibitem [{\citenamefont {Wittbrodt}\ and\ \citenamefont
  {Schlegel}(1996)}]{Wittbrodt1996}%
  \BibitemOpen
  \bibfield  {author} {\bibinfo {author} {\bibfnamefont {J.~M.}\ \bibnamefont
  {Wittbrodt}}\ and\ \bibinfo {author} {\bibfnamefont {H.~B.}\ \bibnamefont
  {Schlegel}},\ }\href@noop {} {\bibfield  {journal} {\bibinfo  {journal}
  {J.~Chem.~Phys.}\ }\textbf {\bibinfo {volume} {105}},\ \bibinfo {pages}
  {6574} (\bibinfo {year} {1996})}\BibitemShut {NoStop}%
\bibitem [{\citenamefont {Gr\"afenstein}\ and\ \citenamefont
  {Cremer}(2001)}]{grafenstein2001}%
  \BibitemOpen
  \bibfield  {author} {\bibinfo {author} {\bibfnamefont {J.}~\bibnamefont
  {Gr\"afenstein}}\ and\ \bibinfo {author} {\bibfnamefont {D.}~\bibnamefont
  {Cremer}},\ }\href@noop {} {\bibfield  {journal} {\bibinfo  {journal}
  {Mol.~Phys.}\ }\textbf {\bibinfo {volume} {99}},\ \bibinfo {pages} {981}
  (\bibinfo {year} {2001})}\BibitemShut {NoStop}%
\bibitem [{\citenamefont {Cohen}\ \emph {et~al.}(2007)\citenamefont {Cohen},
  \citenamefont {Tozer},\ and\ \citenamefont {Handy}}]{Cohen2007}%
  \BibitemOpen
  \bibfield  {author} {\bibinfo {author} {\bibfnamefont {A.~J.}\ \bibnamefont
  {Cohen}}, \bibinfo {author} {\bibfnamefont {D.~J.}\ \bibnamefont {Tozer}}, \
  and\ \bibinfo {author} {\bibfnamefont {N.~C.}\ \bibnamefont {Handy}},\
  }\href@noop {} {\bibfield  {journal} {\bibinfo  {journal} {J.~Chem.~Phys.}\
  }\textbf {\bibinfo {volume} {126}},\ \bibinfo {pages} {214104} (\bibinfo
  {year} {2007})}\BibitemShut {NoStop}%
\bibitem [{\citenamefont {Mulliken}(1955)}]{mulliken-pop-analys-1955}%
  \BibitemOpen
  \bibfield  {author} {\bibinfo {author} {\bibfnamefont {R.~S.}\ \bibnamefont
  {Mulliken}},\ }\href@noop {} {\bibfield  {journal} {\bibinfo  {journal}
  {J.~Chem.~Phys.}\ }\textbf {\bibinfo {volume} {23}},\ \bibinfo {pages} {1833}
  (\bibinfo {year} {1955})}\BibitemShut {NoStop}%
\bibitem [{\citenamefont {Szabo}\ \emph {et~al.}(1996)\citenamefont {Szabo},
  \citenamefont {Glaser},\ and\ \citenamefont {Grenthe}}]{Szabo1996}%
  \BibitemOpen
  \bibfield  {author} {\bibinfo {author} {\bibfnamefont {Z.}~\bibnamefont
  {Szabo}}, \bibinfo {author} {\bibfnamefont {J.}~\bibnamefont {Glaser}}, \
  and\ \bibinfo {author} {\bibfnamefont {I.}~\bibnamefont {Grenthe}},\
  }\href@noop {} {\bibfield  {journal} {\bibinfo  {journal} {Inorg. Chem.}\
  }\textbf {\bibinfo {volume} {35}},\ \bibinfo {pages} {2036} (\bibinfo {year}
  {1996})}\BibitemShut {NoStop}%
\bibitem [{\citenamefont {Allen}\ \emph {et~al.}(1997)\citenamefont {Allen},
  \citenamefont {Bucher}, \citenamefont {Shuh}, \citenamefont {Edelstein},\
  and\ \citenamefont {Reich}}]{allen1997}%
  \BibitemOpen
  \bibfield  {author} {\bibinfo {author} {\bibfnamefont {P.~G.}\ \bibnamefont
  {Allen}}, \bibinfo {author} {\bibfnamefont {J.~J.}\ \bibnamefont {Bucher}},
  \bibinfo {author} {\bibfnamefont {D.~K.}\ \bibnamefont {Shuh}}, \bibinfo
  {author} {\bibfnamefont {N.~M.}\ \bibnamefont {Edelstein}}, \ and\ \bibinfo
  {author} {\bibfnamefont {T.}~\bibnamefont {Reich}},\ }\href@noop {}
  {\bibfield  {journal} {\bibinfo  {journal} {Inorg. Chem.}\ }\textbf {\bibinfo
  {volume} {36}},\ \bibinfo {pages} {4676} (\bibinfo {year}
  {1997})}\BibitemShut {NoStop}%
\bibitem [{\citenamefont {Bardini}\ \emph {et~al.}(1998)\citenamefont
  {Bardini}, \citenamefont {Rubini},\ and\ \citenamefont
  {Madic}}]{Np-5-waters}%
  \BibitemOpen
  \bibfield  {author} {\bibinfo {author} {\bibfnamefont {N.}~\bibnamefont
  {Bardini}}, \bibinfo {author} {\bibfnamefont {P.}~\bibnamefont {Rubini}}, \
  and\ \bibinfo {author} {\bibfnamefont {C.}~\bibnamefont {Madic}},\
  }\href@noop {} {\bibfield  {journal} {\bibinfo  {journal} {Radioch. Acta}\
  }\textbf {\bibinfo {volume} {83}},\ \bibinfo {pages} {189–194} (\bibinfo
  {year} {1998})}\BibitemShut {NoStop}%
\bibitem [{\citenamefont {Hay}\ \emph {et~al.}(2000)\citenamefont {Hay},
  \citenamefont {Martin},\ and\ \citenamefont
  {Schreckenbach}}]{schreckenbach-2000}%
  \BibitemOpen
  \bibfield  {author} {\bibinfo {author} {\bibfnamefont {P.~J.}\ \bibnamefont
  {Hay}}, \bibinfo {author} {\bibfnamefont {R.~L.}\ \bibnamefont {Martin}}, \
  and\ \bibinfo {author} {\bibfnamefont {G.}~\bibnamefont {Schreckenbach}},\
  }\href@noop {} {\bibfield  {journal} {\bibinfo  {journal} {J.~Phys.~Chem.~A}\
  }\textbf {\bibinfo {volume} {104}},\ \bibinfo {pages} {6259} (\bibinfo {year}
  {2000})}\BibitemShut {NoStop}%
\bibitem [{\citenamefont {Barcza}\ \emph {et~al.}(2011)\citenamefont {Barcza},
  \citenamefont {Legeza}, \citenamefont {Marti},\ and\ \citenamefont
  {Reiher}}]{mutual_information_1}%
  \BibitemOpen
  \bibfield  {author} {\bibinfo {author} {\bibfnamefont {G.}~\bibnamefont
  {Barcza}}, \bibinfo {author} {\bibfnamefont {{\"O}.}~\bibnamefont {Legeza}},
  \bibinfo {author} {\bibfnamefont {K.~H.}\ \bibnamefont {Marti}}, \ and\
  \bibinfo {author} {\bibfnamefont {M.}~\bibnamefont {Reiher}},\ }\href@noop {}
  {\bibfield  {journal} {\bibinfo  {journal} {Phys. Rev. A}\ }\textbf {\bibinfo
  {volume} {83}},\ \bibinfo {pages} {012508} (\bibinfo {year}
  {2011})}\BibitemShut {NoStop}%
\bibitem [{\citenamefont {Rissler}\ \emph {et~al.}(2006)\citenamefont
  {Rissler}, \citenamefont {Noack},\ and\ \citenamefont {White}}]{dmrg-11}%
  \BibitemOpen
  \bibfield  {author} {\bibinfo {author} {\bibfnamefont {J.}~\bibnamefont
  {Rissler}}, \bibinfo {author} {\bibfnamefont {R.~M.}\ \bibnamefont {Noack}},
  \ and\ \bibinfo {author} {\bibfnamefont {S.~R.}\ \bibnamefont {White}},\
  }\href@noop {} {\bibfield  {journal} {\bibinfo  {journal} {Chem. Phys.}\
  }\textbf {\bibinfo {volume} {323}},\ \bibinfo {pages} {519} (\bibinfo {year}
  {2006})}\BibitemShut {NoStop}%
\bibitem [{\citenamefont {Boguslawski}\ \emph {et~al.}(2012)\citenamefont
  {Boguslawski}, \citenamefont {Tecmer}, \citenamefont {Legeza},\ and\
  \citenamefont {Reiher}}]{dmrg-13}%
  \BibitemOpen
  \bibfield  {author} {\bibinfo {author} {\bibfnamefont {K.}~\bibnamefont
  {Boguslawski}}, \bibinfo {author} {\bibfnamefont {P.}~\bibnamefont {Tecmer}},
  \bibinfo {author} {\bibfnamefont {{\"O}.}~\bibnamefont {Legeza}}, \ and\
  \bibinfo {author} {\bibfnamefont {M.}~\bibnamefont {Reiher}},\ }\href@noop {}
  {\bibfield  {journal} {\bibinfo  {journal} {J. Phys. Chem. Lett.}\ }\textbf
  {\bibinfo {volume} {3}},\ \bibinfo {pages} {3129} (\bibinfo {year}
  {2012})}\BibitemShut {NoStop}%
\bibitem [{\citenamefont {Mottet}\ \emph {et~al.}(2014)\citenamefont {Mottet},
  \citenamefont {Tecmer}, \citenamefont {Boguslawski}, \citenamefont {Legeza},\
  and\ \citenamefont {Reiher}}]{dmrg-19}%
  \BibitemOpen
  \bibfield  {author} {\bibinfo {author} {\bibfnamefont {M.}~\bibnamefont
  {Mottet}}, \bibinfo {author} {\bibfnamefont {P.}~\bibnamefont {Tecmer}},
  \bibinfo {author} {\bibfnamefont {K.}~\bibnamefont {Boguslawski}}, \bibinfo
  {author} {\bibfnamefont {{\"O}.}~\bibnamefont {Legeza}}, \ and\ \bibinfo
  {author} {\bibfnamefont {M.}~\bibnamefont {Reiher}},\ }\href@noop {}
  {\bibfield  {journal} {\bibinfo  {journal} {Phys. Chem. Chem. Phys.}\
  }\textbf {\bibinfo {volume} {16}},\ \bibinfo {pages} {8872} (\bibinfo {year}
  {2014})}\BibitemShut {NoStop}%
\bibitem [{\citenamefont {Duperrouzel}\ \emph {et~al.}(2015)\citenamefont
  {Duperrouzel}, \citenamefont {Tecmer}, \citenamefont {Boguslawski},
  \citenamefont {Barcza}, \citenamefont {Legeza},\ and\ \citenamefont
  {Ayers}}]{Corinne_2015}%
  \BibitemOpen
  \bibfield  {author} {\bibinfo {author} {\bibfnamefont {C.}~\bibnamefont
  {Duperrouzel}}, \bibinfo {author} {\bibfnamefont {P.}~\bibnamefont {Tecmer}},
  \bibinfo {author} {\bibfnamefont {K.}~\bibnamefont {Boguslawski}}, \bibinfo
  {author} {\bibfnamefont {G.}~\bibnamefont {Barcza}}, \bibinfo {author}
  {\bibfnamefont {{\"O}.}~\bibnamefont {Legeza}}, \ and\ \bibinfo {author}
  {\bibfnamefont {P.~W.}\ \bibnamefont {Ayers}},\ }\href@noop {} {\bibfield
  {journal} {\bibinfo  {journal} {Chem. Phys. Lett.}\ }\textbf {\bibinfo
  {volume} {621}},\ \bibinfo {pages} {160} (\bibinfo {year}
  {2015})}\BibitemShut {NoStop}%
\bibitem [{\citenamefont {Barcza}\ \emph {et~al.}(2014)\citenamefont {Barcza},
  \citenamefont {Noack}, \citenamefont {S{\'o}lyom},\ and\ \citenamefont
  {Legeza}}]{barcza2014entanglement}%
  \BibitemOpen
  \bibfield  {author} {\bibinfo {author} {\bibfnamefont {G.}~\bibnamefont
  {Barcza}}, \bibinfo {author} {\bibfnamefont {R.}~\bibnamefont {Noack}},
  \bibinfo {author} {\bibfnamefont {J.}~\bibnamefont {S{\'o}lyom}}, \ and\
  \bibinfo {author} {\bibfnamefont {{\"O}.}~\bibnamefont {Legeza}},\
  }\href@noop {} {\bibfield  {journal} {\bibinfo  {journal} {Phys.~Rev.~B}\
  }\textbf {\bibinfo {volume} {92}},\ \bibinfo {pages} {125140} (\bibinfo
  {year} {2014})}\BibitemShut {NoStop}%
\bibitem [{\citenamefont {Boguslawski}\ and\ \citenamefont
  {Tecmer}(2015)}]{boguslawski2015}%
  \BibitemOpen
  \bibfield  {author} {\bibinfo {author} {\bibfnamefont {K.}~\bibnamefont
  {Boguslawski}}\ and\ \bibinfo {author} {\bibfnamefont {P.}~\bibnamefont
  {Tecmer}},\ }\href@noop {} {\bibfield  {journal} {\bibinfo  {journal} {Int.
  J. Quantum Chem.}\ }\textbf {\bibinfo {volume} {115}},\ \bibinfo {pages}
  {1289} (\bibinfo {year} {2015})}\BibitemShut {NoStop}%
\bibitem [{\citenamefont {Boguslawski}\ and\ \citenamefont
  {Tecmer}(2017)}]{ijqc-eratum}%
  \BibitemOpen
  \bibfield  {author} {\bibinfo {author} {\bibfnamefont {K.}~\bibnamefont
  {Boguslawski}}\ and\ \bibinfo {author} {\bibfnamefont {P.}~\bibnamefont
  {Tecmer}},\ }\href@noop {} {\bibfield  {journal} {\bibinfo  {journal}
  {Int.~J.~Quantum~Chem.}\ }\textbf {\bibinfo {volume} {117}},\ \bibinfo
  {pages} {e25455} (\bibinfo {year} {2017})}\BibitemShut {NoStop}%
\bibitem [{\citenamefont {Gomes}\ \emph {et~al.}(2008)\citenamefont {Gomes},
  \citenamefont {Jacob},\ and\ \citenamefont {Visscher}}]{gomes2008}%
  \BibitemOpen
  \bibfield  {author} {\bibinfo {author} {\bibfnamefont {A.~S.~P.}\
  \bibnamefont {Gomes}}, \bibinfo {author} {\bibfnamefont {C.~R.}\ \bibnamefont
  {Jacob}}, \ and\ \bibinfo {author} {\bibfnamefont {L.}~\bibnamefont
  {Visscher}},\ }\href@noop {} {\bibfield  {journal} {\bibinfo  {journal}
  {Phys. Chem. Chem. Phys}\ }\textbf {\bibinfo {volume} {10}},\ \bibinfo
  {pages} {5353} (\bibinfo {year} {2008})}\BibitemShut {NoStop}%
\bibitem [{\citenamefont {Cornet}\ \emph {et~al.}(2009)\citenamefont {Cornet},
  \citenamefont {H\"{a}ller}, \citenamefont {Sarsfield}, \citenamefont
  {Collison}, \citenamefont {Helliwell}, \citenamefont {May},\ and\
  \citenamefont {Kaltsoyannis}}]{exp_cci_np5_np6}%
  \BibitemOpen
  \bibfield  {author} {\bibinfo {author} {\bibfnamefont {S.~M.}\ \bibnamefont
  {Cornet}}, \bibinfo {author} {\bibfnamefont {L.~J.~L.}\ \bibnamefont
  {H\"{a}ller}}, \bibinfo {author} {\bibfnamefont {M.~J.}\ \bibnamefont
  {Sarsfield}}, \bibinfo {author} {\bibfnamefont {D.}~\bibnamefont {Collison}},
  \bibinfo {author} {\bibfnamefont {M.}~\bibnamefont {Helliwell}}, \bibinfo
  {author} {\bibfnamefont {I.}~\bibnamefont {May}}, \ and\ \bibinfo {author}
  {\bibfnamefont {N.}~\bibnamefont {Kaltsoyannis}},\ }\href@noop {} {\bibfield
  {journal} {\bibinfo  {journal} {Chem. Commun.}\ }\textbf {\bibinfo {volume}
  {0}},\ \bibinfo {pages} {917} (\bibinfo {year} {2009})}\BibitemShut {NoStop}%
\bibitem [{\citenamefont {Veis}\ \emph {et~al.}(2016)\citenamefont {Veis},
  \citenamefont {Antal\'{i}k}, \citenamefont {Brabec}, \citenamefont {Neese},
  \citenamefont {\:{O}rs Legeza},\ and\ \citenamefont {Pittner}}]{dmrg-tcc}%
  \BibitemOpen
  \bibfield  {author} {\bibinfo {author} {\bibfnamefont {L.}~\bibnamefont
  {Veis}}, \bibinfo {author} {\bibfnamefont {A.}~\bibnamefont {Antal\'{i}k}},
  \bibinfo {author} {\bibfnamefont {J.}~\bibnamefont {Brabec}}, \bibinfo
  {author} {\bibfnamefont {F.}~\bibnamefont {Neese}}, \bibinfo {author}
  {\bibnamefont {\:{O}rs Legeza}}, \ and\ \bibinfo {author} {\bibfnamefont
  {J.}~\bibnamefont {Pittner}},\ }\href@noop {} {\bibfield  {journal} {\bibinfo
   {journal} {J. Phys. Chem. Lett.}\ }\textbf {\bibinfo {volume} {7}},\
  \bibinfo {pages} {4072} (\bibinfo {year} {2016})}\BibitemShut {NoStop}%
\bibitem [{\citenamefont {Clark}\ \emph {et~al.}(2004)\citenamefont {Clark},
  \citenamefont {Sonnenberg}, \citenamefont {Hay},\ and\ \citenamefont
  {Martin}}]{clark2004}%
  \BibitemOpen
  \bibfield  {author} {\bibinfo {author} {\bibfnamefont {A.~E.}\ \bibnamefont
  {Clark}}, \bibinfo {author} {\bibfnamefont {J.~L.}\ \bibnamefont
  {Sonnenberg}}, \bibinfo {author} {\bibfnamefont {P.~J.}\ \bibnamefont {Hay}},
  \ and\ \bibinfo {author} {\bibfnamefont {R.~L.}\ \bibnamefont {Martin}},\
  }\href@noop {} {\bibfield  {journal} {\bibinfo  {journal} {J. Chem. Phys.}\
  }\textbf {\bibinfo {volume} {121}},\ \bibinfo {pages} {2563} (\bibinfo {year}
  {2004})}\BibitemShut {NoStop}%
\bibitem [{\citenamefont {Knecht}\ \emph {et~al.}(2014)\citenamefont {Knecht},
  \citenamefont {Legeza},\ and\ \citenamefont {Reiher}}]{dmrg_relativistic_1}%
  \BibitemOpen
  \bibfield  {author} {\bibinfo {author} {\bibfnamefont {S.}~\bibnamefont
  {Knecht}}, \bibinfo {author} {\bibfnamefont {{\"O}.}~\bibnamefont {Legeza}},
  \ and\ \bibinfo {author} {\bibfnamefont {M.}~\bibnamefont {Reiher}},\
  }\href@noop {} {\bibfield  {journal} {\bibinfo  {journal} {J.~Chem.~Phys.}\
  }\textbf {\bibinfo {volume} {140}},\ \bibinfo {pages} {041101} (\bibinfo
  {year} {2014})}\BibitemShut {NoStop}%
\bibitem [{\citenamefont {Battaglia}\ \emph {et~al.}(2018)\citenamefont
  {Battaglia}, \citenamefont {Keller},\ and\ \citenamefont
  {Knecht}}]{dmrg_relativistic_2}%
  \BibitemOpen
  \bibfield  {author} {\bibinfo {author} {\bibfnamefont {S.}~\bibnamefont
  {Battaglia}}, \bibinfo {author} {\bibfnamefont {S.}~\bibnamefont {Keller}}, \
  and\ \bibinfo {author} {\bibfnamefont {S.}~\bibnamefont {Knecht}},\
  }\href@noop {} {\bibfield  {journal} {\bibinfo  {journal}
  {J.~Chem.~Theory~Comput.}\ }\textbf {\bibinfo {volume} {14}},\ \bibinfo
  {pages} {2353} (\bibinfo {year} {2018})}\BibitemShut {NoStop}%
\end{thebibliography}%
\end{document}